\documentclass[12pt]{article}
\usepackage{axodraw} 
\usepackage{graphicx}
\usepackage{amsmath} 
\usepackage{amssymb}
\usepackage{bm} 
\usepackage{hyperref}
\usepackage{sectsty} 

\sectionfont{\large}
\subsectionfont{\normalsize}
\subsubsectionfont{\normalsize}
\numberwithin{equation}{section} 

\renewenvironment{thebibliography}[1]{%
  \begin{oldthebibliography}{#1}%
    \setlength{\parskip}{0ex}%
    \setlength{\itemsep}{0ex}%
}%
{%
  \end{oldthebibliography}%
}

\def\d{\partial}
\def\tr{{\rm tr}}
\def\grad{{\bm \nabla}}
\def\v{\bm v}
\def\vs{v_s}
\def\x{{\bf x}}

\def\k{{\bf k}}

\def\D{{\cal D}}

\def\q{{\bf q}}
\def\L{{\cal L}}
\def\chiT{\chi_{\rm\scriptscriptstyle T}}
\def\chiE{\chi_{\rm\scriptscriptstyle E}}
\def\e{\epsilon}
\def\vpar{{\bm v}_{\scriptscriptstyle\parallel}}
\def\vperp{{\bm v}_{\scriptscriptstyle\perp}}

\def\lmfp{l_{\rm mfp}}
\def\coeff#1#2{{\textstyle {\frac {#1}{#2}}}}

\begin{document}

\title{\Large Lectures on hydrodynamic fluctuations in relativistic theories}
\author{\normalsize Pavel Kovtun \footnotemark\\
{\it\normalsize Department of Physics \& Astronomy,  University of Victoria,}\\
{\it\normalsize PO Box 3055 STN CSC, Victoria, BC, V8W 3P6, Canada}
}
\def\thefootnote{\fnsymbol{footnote}}
\footnotetext[1]{pkovtun@uvic.ca}
\def\thefootnote{\arabic{footnote}}

\date{\normalsize May, 2012}

\maketitle

\begin{abstract}
\noindent
These are pedagogical lecture notes on hydrodynamic fluctuations 
in normal relativistic fluids.
The lectures discuss correlation functions of conserved densities in thermal equilibrium, 
interactions of the hydrodynamic modes, an effective action for viscous fluids,
and the breakdown of the derivative expansion in hydrodynamics.
\end{abstract}

\setcounter{tocdepth}{2}
\tableofcontents

\section{Basic hydrodynamics}

\subsection{Introduction}
\noindent
One can think about hydrodynamics as an effective long-distance description
for a given classical or quantum many-body system at non-zero temperature.
These lectures will be mostly about relativistic hydrodynamics,
in other words, about hydrodynamics of fluids whose microscopic constituents
are constrained by Lorentz symmetry, as happens in relativistic quantum field theories.
Relativistic hydrodynamics is not limited of course to
the description of the collective motion of particles that move at speeds
close to the speed of actual light.
All that matters is Lorentz symmetry: the hydrodynamic equations
describing the quark-gluon plasma~\cite{Teaney:2009qa} will look identical to the hydrodynamic equations
describing relativistic quantum critical regions in quantum magnets \cite{SS}.

Besides its relevance to the real-world physical phenomena,
hydrodynamics is an interesting subject from a purely theoretical physics point of view.
It turns out that hydrodynamic fluctuations
(both in normal fluids and in superfluids)
can be thought of as gravitational fluctuations of black holes, 
and vice versa \cite{Policastro:2002se, Policastro:2002tn}.
In fact, this connection between gravity and hydrodynamics works beyond the linear response,
allowing one to derive how the full non-linear relativistic Navier-Stokes equations
are encoded in Einstein's equations of General Relativity.
This relation of the Einstein's equations to the equations of relativistic hydrodynamics 
is often referred to as the ``fluid-gravity correspondence''~\cite{Hubeny:2011hd}.

What does this teach one beyond being a
curious mathematical correspondence?
One answer comes from the gauge-gravity duality \cite{Horowitz:2006ct, Aharony:1999ti}
which states (among many other things) that the hydrodynamics of a black hole does
in fact represent the hydrodynamics of a specific quantum system,
whose Hamiltonian is explicitly known.
In particular, one can easily evaluate transport coefficients in these quantum systems
from black hole physics.
This is an important advance because the quantum systems in question happen to be
strongly interacting, and the conventional methods such as the diagrammatic
expansion or Monte-Carlo simulations become inadequate to compute the
transport coefficients.
Because these quantum-mechanical models share important common features
with the theory of strong interactions (QCD), the computations of transport coefficients
from black hole physics becomes more than a purely academic exercise.

An example of a fruitful interplay between hydrodynamics and black hole physics
came from the realization that the kinematic viscosity in the gauge-gravity duality
takes a universal model-independent value~\cite{Kovtun:2004de, Buchel:2004qq}, and that the 
Heisenberg uncertainty relation may prevent the existence of perfect fluids in nature~\cite{Kovtun:2004de}. 
The two main competitors for the most nearly-perfect fluid are
the ultracold Fermi gases and the quark-gluon plasma, see~\cite{Schafer:2009dj} for a review.

On the other hand, the relation between hydrodynamics
and black holes has motivated new research into relativistic hydrodynamics itself,
and its applications.
One example is the classification of two-derivative terms
(originally introduced in order to restore the causality of the standard one-derivative
relativistic generalization of the Navier-Stokes equations)
which only appeared recently~\cite{Baier:2007ix, Bhattacharyya:2008jc}.
Another example is the application of the relativistic hydrodynamics
to thermo-magnetic charge transport in high-temperature superconductors~\cite{Hartnoll:2007ih},
and in graphene~\cite{Muller-Sachdev}.
Another example is the realization that the standard relativistic hydrodynamics needs 
to be corrected in the presence of quantum anomalies~\cite{Son:2009tf}.
Another example is the modification of relativistic hydrodynamics 
due to parity violation~\cite{Jensen:2011xb}.
Another example is the systematic study of dissipative 
relativistic superfluid hydrodynamics~\cite{Bhattacharya:2011tr}.
All the above examples could have been understood many years ago
without any black hole physics. But they weren't. 
Instead, the studies were heavily motivated 
by the recent experiments studying the quark-gluon plasma~\cite{Romatschke:2009im},
by the gauge-gravity duality~\cite{McGreevy:2009xe, CasalderreySolana:2011us},
and by the progress in understanding quantum critical phenomena~\cite{Sachdev:2011cs}.

These lectures will largely focus on hydrodynamic fluctuations,
that is small, long-wave\-length fluctuations near thermal equilibrium.
These fluctuations are long-lived: their frequency $\omega(\k)$
vanishes as $\k\to0$, and they can propagate long distances,
as one can see in the example of sound waves,
whose dispersion relation is $\omega(\k) = \vs |\k| + O(\k^2)$.
Such hydrodynamic modes are solutions to the hydrodynamic equations.
At the same time, such modes are a danger to the very existence of hydrodynamics,
as they violate the assumption of local equilibration.
Thermal excitations in these long-lived modes will contribute 
to both charge and momentum transport.
For non-relativistic fluids, 
it has been known for a long time that this effect wipes off
the very notion of classical hydrodynamics in 2+1 dimensions~\cite{Forster:1977zz},
and makes two-derivative corrections to classical hydrodynamics 
in 3+1 dimensions problematic~\cite{DeSchepper19741}.%
\footnote{%
	In systems with many degrees of freedom, these fluctuation effects
	can be suppressed if the limit $N\to\infty$ is taken before
	before the hydrodynamic limit $\k\to0$, $\omega\to0$.
	In other words, the large-$N$ limit does not commute with the
	hydrodynamic limit in 2+1 dimensions.
	In the context of the gauge-gravity duality, 
	taking the limit $N\to\infty$ first is what allows the classical 2+1 dimensional
	hydrodynamics to emerge from 3+1 dimensional gravity~\cite{Kovtun:2003vj, CaronHuot:2009iq}.
}
Relativistic hydrodynamics is no different in this respect:
for small fluctuations near thermal equilibrium,
there is not a great difference between relativistic and 
non-relativistic hydrodynamics.
Thus the classical equations of second-order relativistic hydrodynamics
in 3+1 dimensions do not describe the fluid correctly 
in the hydrodynamic limit~\cite{Kovtun:2011np}.

These lecture notes have three main parts.
The first part is about the equations of hydrodynamics: 
we need to know what the equations are, before we solve them!
The equations can be found for example in the book by Landau and Lifshitz~\cite{LL6},
or in the book by Weinberg~\cite{weinberg:1972}.
The second part is about how the classical equations of hydrodynamics
determine the two-point correlation functions of conserved densities,
(energy density $T^{00}$, momentum density $T^{0i}$ and charge density $J^0$) in equilibrium.
This will be done in the linear response theory, deriving the Kubo formulas
for the shear viscosity, the bulk viscosity, and the charge conductivity.
An excellent pedagogical reference on the linear response theory 
is the paper by Kadanoff and Martin~\cite{KM}. 
We will also compute the correlation functions by a somewhat different method,
by studying the response of the hydrodynamic equations to the
external gauge field, and to the external metric.
The third part of the lecture notes is about the interactions
of the hydrodynamic modes.
In order to treat the interactions systematically, one may introduce microscopic
random stresses and random currents,
thus converting the classical hydrodynamic equations into stochastic equations.
Using a standard method, hydrodynamics can then be recast as a real-time ``quantum''
field theory, where temperature plays the role of $\hbar$,
and where correlation functions follow from a generating functional 
defined by a functional integral with a local effective action.

\subsection{Non-relativistic hydrodynamics}
\label{sec:NR-hydro}
\noindent
Let us start with non-relativistic hydrodynamics of normal fluids
in three spatial dimensions.
The discussion here will be a brief review;
we will move onto relativistic fluids shortly which will be discussed in more detail.
Hydrodynamic equations express conservation laws
of whatever is conserved, supplemented by assumptions about
the dissipative response, and thermodynamic information
such as the equation of state.
Let us first discuss ideal (i.e. non-dissipative) hydrodynamics of
a simple non-relativistic fluid \cite{LL6}.

\subsubsection*{Ideal fluids}
\noindent
In a non-relativistic system, particle number is conserved,
in other words particles are not created or destroyed.
This conservation of particle number is expressed in
hydrodynamics as conservation of mass, by the continuity equation
\begin{equation}
  \d_t \rho + \d_i(\rho v_i) = 0\,,
\label{eq:continuity-equation}
\end{equation}
where $\rho$ is local fluid density (mass per unit volume),
and $\v$ is local fluid velocity.
Another equation is the equation of motion of a fluid element,
$\rho(dv_i/dt)=-\d_i p$, where $p$ is pressure, and
$(-\d_i p)$ is the force per unit volume.
In terms of the velocity field $\v(t,\x)$ this gives the Euler equation,
$\rho(\d_t v_i + v_k\d_{k} v_i) = -\d_i p$.
Using the continuity equation (\ref{eq:continuity-equation}),
one can rewrite the Euler equation in the form of momentum conservation,
\begin{equation}
   \d_t(\rho v_i) + \d_{j} \Pi_{ij} = 0\,,
\label{eq:Euler-equation}
\end{equation}
where $\Pi_{ij} = p\delta_{ij} + \rho v_i v_j$.
So far we have five unknown functions ($\rho, \v, p$), and four equations.
The equilibrium equation of state will provide a relation between
$p$ and $\rho$, but in general will introduce
another unknown parameter such as temperature or entropy. 
This means that we need one more dynamical equation.
The extra equation can be taken as the condition of entropy conservation,
$d\tilde s/dt=\d_t \tilde s + v_k\d_k \tilde s=0$
where $\tilde s$ is entropy per unit mass.
Equivalently, $\d_t s +\d_i(s v_i)=0$, where $s$ is entropy
per unit volume.
Using the continuity equation, the Euler equation,
and basic thermodynamic identities, the condition
of entropy conservation can be rewritten as 
conservation of energy,
\begin{equation}
  \d_t \Big(\epsilon+\frac{\rho v^2}{2} \Big) 
  + \d_i \Big((w+\frac{\rho v^2}{2}) v_i \Big) = 0 \,,
\label{eq:energy-equation}
\end{equation}
where $\epsilon$ is (internal) energy per unit volume,
and $w=\epsilon+p$ is enthalpy per unit volume.
We now have six unknown functions $(\rho, \v, p, \epsilon)$,
and five dynamical equations
(\ref{eq:continuity-equation}){--}(\ref{eq:energy-equation}),
plus the equation of state which relates $\epsilon, p$ and $\rho$.
The above equations of ideal fluid dynamics express the conservation
of mass, momentum, and energy.

In writing down 
Eqs.~(\ref{eq:continuity-equation}){--}(\ref{eq:energy-equation}),
we have implicitly assumed that the fluid is homogeneous,
for example there is only one species of particles.
If there are several conserved ``particle numbers'',
the set of hydrodynamic equations needs to be expanded,
reflecting the fact that there are more conserved quantities.
See for example~\cite{LL6},~\S58.\\

\subsubsection*{Dissipative fluids}
\begin{figure}
\begin{center}
  \includegraphics[width=2.0in]{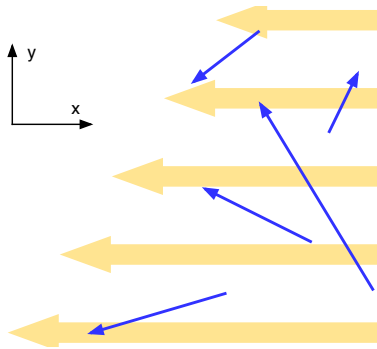}
\end{center}
  \caption{
  A stationary flow of an ideal fluid with a velocity gradient.
  Thin arrows represent particles which can transfer 
  $x$-momentum in the $y$-direction,
  eventually leading to the equilibration of the
  inhomogeneous velocity profile.
  }
  \label{fig:shear-flow}
\end{figure}

\noindent
Equations (\ref{eq:continuity-equation}){--}(\ref{eq:energy-equation}) of ideal hydrodynamics
ignore dissipation.
For example, sound waves derived from ideal hydrodynamics would propagate indefinitely without damping.
Physically, dissipation means that the fluid can move locally
not only because a fluid element is pushed around by other fluid elements,
but also because the microscopic movement of the
constituents of the fluid
can even out various inhomogeneities in velocity, temperature etc.
For example, consider the configuration
with constant $p$, $\rho$ and $\epsilon$, and with
\begin{equation}
  v_x=v_x(y)\,\quad
  v_y=v_z=0\,.
\label{eq:shear-flow}
\end{equation}
This describes homogeneous fluid which moves uniformly in the $x$
direction, but whose velocity depends on $y$, as shown in
Fig.~\ref{fig:shear-flow}.
For such configuration, $\d_i v_i=0$,
as well as $v_k\d_k v_i=0$, 
and therefore the flow (\ref{eq:shear-flow})
is a perfectly good stationary solution to the ideal hydro equations
(\ref{eq:continuity-equation}){--}(\ref{eq:energy-equation}).
However, this flow is unphysical
because the particles which comprise the fluid
can move in the $y$ direction, and this will lead to
momentum transfer between different layers in the fluid.
As a result, the velocity distribution $v_x(y)$
will tend to become more uniform,
and the configuration (\ref{eq:shear-flow}) will not remain
stationary.
It is precisely these kind of processes
that are taken into account by the dissipative terms
in the hydrodynamic equations.
The idea of dissipative hydrodynamics is to modify the
equations of the ideal hydrodynamics by adding extra terms
which are proportional to spatial gradients of the
hydrodynamic variables.
Moreover, one assumes that the gradients are small, so that
a ``gradient expansion'' can be constructed.
In hydrodynamics one always assumes that the
quantities of interest vary on scales which are large
compared to the microscopic length scale $\lmfp$,
so that $|\d_i\rho|\ll|\rho/\lmfp|$.
We will retain only those terms which are linear in spatial gradients.
In principle, one can write down dissipative terms which are
quadratic in spatial gradients, then cubic and so on.
One can think about hydrodynamics
as of an effective theory which is valid at long distance and time
scales, similar to the low-energy chiral theory in QCD.
Higher-order terms are suppressed by the powers of the cutoff
which in hydrodynamics is~$\lmfp$.

The equations of dissipative hydrodynamics can be written in the form~\cite{LL6}
\begin{subequations}
\label{eq:dissipative-NR-hydro}
\begin{eqnarray}
  &&   \d_t \rho + \d_i(\rho v_i) = 0\,,\\[5pt]
\label{eq:Navier-Stokes-eqn}
  &&   \d_t(\rho v_i) + \d_{j} \Pi_{ij} = 0\,,\\[5pt]
  &&   \d_t \Big(\epsilon+\frac{\rho v^2}{2} \Big) + \d_i j_i^\epsilon = 0 \,,
\end{eqnarray}
\end{subequations}
where the stress tensor is $\Pi_{ij} = p\delta_{ij} + \rho v_i v_j -\Sigma_{ij}$ with
\begin{equation*}
           \Sigma_{ij} = 
           \eta\,(\d_i v_j+\d_j v_i-\frac23\delta_{ij}\,\d_k v_k)+
           \zeta\delta_{ij} \d_k v_k\,,
\end{equation*}
and the energy current is
\begin{equation*}
   j_i^\epsilon = \left(w{+}\frac{\rho v^2}{2} \right)v_i
                   -\Sigma_{ij} v_j
                   -\kappa\,\d_{i} T  \,.
\end{equation*}
The coefficient $\eta$ is called shear viscosity,
$\zeta$ is called bulk viscosity, and $\kappa$ is called thermal conductivity.%
\footnote{%
     Often $\eta$ is called just ``viscosity'',
     and $\zeta$ is called ``second viscosity'' or ``volume viscosity''.
     The equation of momentum conservation (\ref{eq:Navier-Stokes-eqn})
     is called the Navier-Stokes equation.
     For a historical account of the formulation of the Navier-Stokes equation, see~\cite{Darrigol}.
}
The transport coefficients $\eta$, $\zeta$, and $\kappa$
should be thought of as input parameters for hydrodynamics:
like in any effective theory, they can not be evaluated in hydrodynamics itself, 
but rather need to be computed from the underlying short-distance physics,
for example using the Boltzmann equation~\cite{Chapman-Cowling}.
In a near-ideal gas, both $\eta$ and $\kappa$ are proportional to the mean-free path.

In writing down the equations~(\ref{eq:dissipative-NR-hydro}) of dissipative hydrodynamics,
a particular out-of-equilibrium definition of
the hydrodynamic variables has been chosen.
A different choice would lead to the hydrodynamic equations which look different
from (\ref{eq:dissipative-NR-hydro}).
We will discuss this issue when talking about relativistic hydrodynamics
in Section~\ref{sec:relativistic-hydro}.

One final comment concerns the terminology.
The words ``perfect fluid'', ``ideal fluid'', or ``ideal hydrodynamics'' usually refer to
non-dissipative hydrodynamics, i.e. the form of
the hydrodynamic equations with $\eta$, $\zeta$, and $\kappa$ set to zero.
However, ideal hydrodynamics is {\em not} the same as
hydrodynamics of the ideal gas.
In fact, ideal gas (i.e. gas without interactions between
the particles) has no hydrodynamics at all because
the thermal equilibrium state is unstable when there are no interactions.
This pathology of the ideal gas can be cured by introducing
small interactions between the particles, with interaction
strength determined by a small parameter~$\lambda$.
The equation of state in such a system is the same as the
equation of state of the ideal gas, up to small corrections
of order $\lambda$.
The mean free path, on the other hand, will be very large
(inversely proportional to $\lambda$), and the length scale
at which hydrodynamics is applicable must be even larger.
The shear viscosity and the thermal conductivity
in this slightly non-ideal gas are large,
because $\eta$ and $\kappa$ are directly proportional to the mean-free path.
By taking the limit $\lambda\to0$ one can see that
the ideal gas has {\em infinite} shear viscosity and thermal conductivity,
unlike the ideal fluid which has (by definition)
zero shear viscosity and thermal conductivity.

\subsubsection*{Hydrodynamic modes}
\noindent
Now that we have the equations of hydrodynamics,
we can study some simple solutions.
A fluid at rest, with constant $\rho{=}\bar{\rho}$,
$p{=}\bar{p}$, $\epsilon{=}\bar{\epsilon}$, and $\v{=}0$ is obviously
a solution.
Now let us look at small fluctuations around this solution:
$p{=}\bar{p}+\delta p$, $\epsilon{=}\bar{\epsilon}+\delta\epsilon$,
$\rho{=}\bar{\rho}+\delta\rho$, and small~$\v$.
``Small fluctuations'' means that $\delta p\ll \bar{p}$ etc,
but we need to say what does it mean to have small~$\v$.
We will see that the relevant condition is that $\v$
must be much smaller than the speed of sound.

Let us start with ideal hydrodynamics.
The linearized equations
(\ref{eq:continuity-equation}){--}(\ref{eq:energy-equation})
become
\begin{subequations}
\label{eq:linear-nr-ideal-hydro-eqs}
\begin{eqnarray}
  && \d_t \delta\rho + \bar{\rho}\,\d_i v_i=0\,,\\
  && \bar{\rho}\, \d_t v_i + \d_i \delta p=0\,,\\
  && \d_t\delta\epsilon + \bar{w}\, \d_i v_i =0\,,
\end{eqnarray}
\end{subequations}
where $\bar{w}=\bar{\epsilon}{+}\bar{p}$.
There are three unknown functions besides $\v$, and
we need to choose the variables:
given the equation of state, we can work with
$\delta\epsilon,\delta\rho$, or with $\delta p, \delta T$,
or with some other variables.
Let us choose $\delta\epsilon$ and $\delta\rho$
as the hydrodynamic variables, then we need an equation
of state of the form $p=p(\epsilon,\rho)$.
In practice these variables may not be the most convenient ones,
but they are natural from the effective field theory
point of view because they represent the densities of
conserved charges.
From the equation of state, we have
$\d_i \delta p = 
(\partial p/\partial\epsilon)_{\rho}\, \d_i \delta\epsilon + (\partial p/\partial\rho)_{\epsilon}\, \d_i \delta\rho$,
where the thermodynamic derivatives are determined by the equilibrium equation of state.

For fluctuations around an infinite homogeneous background,
we can take all variables proportional to $e^{i\k\cdot\x}$,
and decompose the velocity as $\v=\vpar+\vperp$,
where the parallel and perpendicular are with respect to the
wave vector $\k$. 
In particular, $\vpar=\k (\k{\cdot}\v)/\k^2$.
Then it is easy to show using
Eqs.~(\ref{eq:linear-nr-ideal-hydro-eqs})
that the transverse velocity component
becomes time independent and decouples.
On the other hand, the longitudinal component
satisfies the wave equation
$$
  \d_t^2 \vpar(\k,t) + \k^2\vs^2\, \vpar(\k,t) =0\,,
$$
where $\vs^2\equiv (\partial p/\partial\rho)_{\epsilon} + (\partial p/\partial\epsilon)_{\rho}\, {w}/{\rho}$.
This describes a hydrodynamic mode which propagates
with the velocity $\vs$.
As is evident from Eqs.~(\ref{eq:linear-nr-ideal-hydro-eqs}),
both mass density and energy density are part
of the same mode, and describe density fluctuations
which propagate with the velocity $\vs$.
This hydrodynamic mode is of course the sound wave,
and $\vs$ is the speed of sound.
The sound waves are longitudinal
because $\vperp$ is not a part of the propagating mode.
Also, Eqs.~(\ref{eq:linear-nr-ideal-hydro-eqs}) say that
$|\delta\rho/\bar{\rho}|=|\vpar/\vs|$ and $|\delta\epsilon/\bar{w}|=|\vpar/\vs|$,
so the magnitude of velocity fluctuations
must be much smaller than the velocity of sound
in order for the linearized approximation to be valid.
The eigenfrequencies of the system (\ref{eq:linear-nr-ideal-hydro-eqs})
are $\omega(\k)=\pm|\k|\vs$, plus two $\omega=0$ solutions,
one for $\vperp$, and one for $(\delta\rho,\vpar)$.
Once dissipation is included, the $\omega=0$ solutions
will become the transverse non-propagating shear mode
and the longitudinal non-propagating heat mode.

Let us now bring the speed of sound into a more familiar form.
To evaluate the thermodynamic derivatives, we need to choose
an equilibrium ensemble.
Let us work in the canonical ensemble with a fixed total
number of particles $N$.
Mass density is $\rho=mN/V$, where $m$ is the mass of one particle.
At fixed $N$, we have $TdS=dE+pdV$, as well as $dV/V=-d\rho/\rho$.
Also, the change in energy is
$dE/V=d(\epsilon V)/V=d\epsilon-\epsilon\, d\rho/\rho$, and thus
$$
  \frac{TdS}{V} = d\epsilon - \frac{w\,d\rho}{\rho}\,.
$$
This gives the variation of pressure in terms of $dS$ and $d\rho$:
$$
  dp = \left(\frac{\partial p}{\partial \epsilon}\right)_{\!\rho} d\epsilon + 
       \left(\frac{\partial p}{\partial\rho} \right)_{\!\epsilon} d\rho
     = \left(\frac{\partial p}{\partial \epsilon}\right)_{\!\rho} \frac{TdS}{V} 
       + \left[\left(\frac{\partial p}{\partial \epsilon}\right)_{\!\rho} \frac{w}{\rho} 
              + \left(\frac{\partial p}{\partial\rho} \right)_{\!\epsilon} \right] d\rho\,.
$$
The coefficient in the square brackets is precisely
$\vs^2$, and therefore
$$
  \vs^2 = \left( \frac{\d p}{\d\rho} \right)_{S,N}
$$
which is the familiar textbook expression.
For an ideal gas, $p=\rho T/m$, and also $p\propto\rho^\gamma$,
where $\gamma$ is the adiabatic index, about $1.4$ for air.
So the speed of sound in a non-relativistic ideal gas is
$$
  \vs = \sqrt{\frac{\gamma T}{m}}\,,
$$
proportional to the square root of temperature. 

The same analysis of small fluctuations around a static thermal
equilibrium state can be repeated for the dissipative hydrodynamics.
The dispersion relations $\omega(\k)$ for the hydrodynamic modes will get
imaginary parts proportional to $\k^2$ (at small $|\k|$) which
are determined by the dissipative coefficients $\eta$, $\zeta$
and $\kappa$.
As a result of the $O(\k^2)$ dissipation, 
low-frequency sound waves propagate farther than high-frequency sound waves.

\subsection{Relativistic hydrodynamics}
\label{sec:relativistic-hydro}
\noindent
Now let us move on to relativistic hydrodynamics.
As it is the main subject of these lectures, we will discuss it in
some more detail.
For simplicity, we will only discuss normal fluids (no superfluids),
and fluids without dynamical electromagnetic fields
(no magneto-hydrodynamics).
The speed of light is set to one.

\subsubsection*{Hydrodynamic variables}
\noindent
Just like in the non-relativistic case,
hydrodynamic equations express conservation laws
of whatever is conserved.
According to the Noether theorem, conservation laws are related
to continuous symmetries of the fundamental microscopic theory,
which imply the existence of conserved currents. 
In relativistic systems, the spacetime symmetries are given by
translations, rotations, and boosts. 
In addition, there may be other ``internal'' symmetries, 
such as the $U(1)$ baryon number symmetry.

The conserved current corresponding to the spacetime translation symmetry
is the energy-momentum tensor $T^{\mu\nu}$.
We will define the energy-momentum tensor by the variation
of the action with respect to the external metric~\cite{weinberg:1972},
in which case it is symmetric, $T^{\mu\nu} = T^{\nu\mu}$, 
and the conservation law is
\begin{subequations}
\label{eq:TJ-conservation}
\begin{equation}
\label{eq:Tmn-conservation}
  \partial_\mu T^{\mu\nu} = 0\,.
\end{equation}
The conserved currents corresponding to rotations and boosts are
${\cal M}^{\mu\nu\lambda} = x^\mu T^{\nu\lambda} - x^\nu T^{\mu\lambda}$,
which are identically conserved owing to the symmetry
of the energy-momentum tensor, $\partial_\lambda {\cal M}^{\mu\nu\lambda} = 0$.
In other words, given our definition of the energy-momentum tensor,
there is no extra conservation equation beyond (\ref{eq:Tmn-conservation})
expressing the symmetry under rotations and boosts.
Further, we will allow for a possibility of a conserved
current $J^\mu$ corresponding to a global $U(1)$ symmetry,%
\footnote{%
	For simplicity, we will assume a parity-symmetric
	microscopic theory, and take $J^\mu$ to be a vector current
	(charge density is parity-even).
	If parity is not a symmetry of the microscopic theory, or
	if $J^\mu$ is an axial current (charge density is parity-odd),
	there will be extra terms in the constitutive relations
	in addition to those we are going to discuss here~\cite{Bhattacharya:2011tr, Jensen:2011xb}.
}
\begin{equation}
\label{eq:Jm-conservation}
  \partial_\mu J^\mu = 0\,.
\end{equation}
\end{subequations}
In $d$ spatial dimensions, there are $d{+}1$ equations (\ref{eq:Tmn-conservation}),
and one equation (\ref{eq:Jm-conservation}), while there are
$(d{+}1)(d{+}2)/2$ components of $T^{\mu\nu}$, and $d{+}1$ components of $J^\mu$,
in other words there are more unknowns than there are equations.
The simplifying assumption of hydrodynamics is that
$T^{\mu\nu}$ and $J^{\mu}$ can be expressed as functions of
$d{+}2$ fields: a local temperature $T(x)$, a local fluid velocity $\v(x)$,
and a local chemical potential $\mu(x)$.
This makes the number of unknowns equal the number of equations.

The choice of hydrodynamic variables can be understood as follows.
The equilibrium state is characterized by the density operator $\hat{\varrho}$,
which is proportional to the exponential of the conserved charges~\cite{Israel-1981},
\begin{equation}
\label{eq:density-operator}
  \hat\varrho = \frac{1}{Z}\, e^{\,\beta_\mu P^\mu + \gamma N}\,.
\end{equation}
In this expression $Z=\tr\, e^{\,\beta_\mu P^\mu + \gamma N}$ is the partition function,
$P^\mu$ is the momentum operator, and $N$ is the operator of the conserved charge.
The density operator is characterized by
a timelike vector $\beta^\mu$ and a scalar $\gamma$,
which parametrize the manifold of thermal equilibrium states.
We will write these parameters as
$\beta^\mu = \beta u^\mu$, and $\gamma=\beta\mu$, where
the vector $u^\mu$ is normalized such that $u^2=-1$, and represents
the velocity of the fluid, $\beta=1/T$ is the inverse temperature,
and $\mu$ is the chemical potential.
In components, $u^\mu=(1{-}\v^2)^{-1/2}(1,\v)$,
where $\v$ is the spatial velocity.
Note that the equilibrium state breaks Lorentz symmetry
(specifically, boost invariance)
due to the presence of the preferred timelike vector $\beta^\mu$.

Hydrodynamic is concerned with states which deviate slightly
from the thermal equilibrium states (\ref{eq:density-operator})
specified by constant $\beta_\mu$ and $\gamma$.
Hence, one chooses to describe these slightly non-equilibrium states in terms of
slowly varying functions $\beta_\mu(x)$ and $\gamma(x)$, 
or equivalently in terms of slowly varying functions 
$u_\mu(x)$, $T(x)$, and $\mu(x)$.
As the microscopic theory is Lorentz invariant, 
and it is only the state (\ref{eq:density-operator}) that breaks the Lorentz symmetry,
the hydrodynamic equations formulated in terms of $\beta_\mu(x)$ and $\gamma(x)$ must be Lorentz covariant,
and the effective action for hydrodynamics must be Lorentz invariant.
We will return to the question of effective action in Section~\ref{sec:effective-action}.
In the following, we will focus on equations of relativistic hydrodynamics, 
and will use $u_\mu$, $T$, and $\mu$ as our hydrodynamic variables,
as is conventional. One could have used the variables $\beta_\mu$ and $\gamma$ instead.

\subsubsection*{Constitutive relations}
\noindent
For any timelike vector $u^\mu$, the energy-momentum tensor and the current
can be decomposed into components which are transverse and longitudinal
with respect to $u^\mu$, using the projector 
$\Delta^{\mu\nu} \equiv \eta^{\mu\nu} + u^\mu u^\nu$,
where $\eta_{\mu\nu}={\rm diag}\,(-1,1,...,1)$ is the flat-space metric.
We write the decomposition as \cite{PhysRev.58.919}
\begin{subequations}
\label{eq:TJ-decomposition}
\begin{eqnarray}
\label{eq:T1}
  && T^{\mu\nu} = {\cal E} u^\mu u^\nu + {\cal P} \Delta^{\mu\nu} +
     (q^\mu u^\nu + q^\nu u^\mu) + t^{\mu\nu}\,,\\
\label{eq:J1}
  && J^\mu = {\cal N} u^\mu + j^\mu\,,
\end{eqnarray}
\end{subequations}
where the coefficients ${\cal E}$, ${\cal P}$, and ${\cal N}$
are scalars, the vectors $q^\mu$ and $j^\mu$ are transverse, i.e.
$u_\mu q^\mu = u_\mu j^\mu = 0$, and the tensor
$t^{\mu\nu}$ is transverse, symmetric, and traceless.
Explicitly, the coefficients are:
\begin{subequations}
\label{eq:EPNqjt}
\begin{eqnarray}
  && {\cal E} = u_\mu u_\nu T^{\mu\nu}\,,\ \ \ \ 
     {\cal P} = \frac{1}{d} \Delta_{\mu\nu} T^{\mu\nu}\,,\ \ \ \ 
     {\cal N} = -u_\mu J^\mu\,,\\[5pt]
  && q_\mu = -\Delta_{\mu\alpha} u_\beta T^{\alpha\beta}\,,\ \ \ \ 
     j_\mu = \Delta_{\mu\nu} J^\nu\,,\\[5pt]
  && t_{\mu\nu} = \frac12\left(
     \Delta_{\mu\alpha}\Delta_{\nu\beta} + \Delta_{\nu\alpha}\Delta_{\mu\beta}
     -\frac{2}{d} \Delta_{\mu\nu} \Delta_{\alpha\beta}
     \right) T^{\alpha\beta}\,.
\end{eqnarray}
\end{subequations}
Equations (\ref{eq:TJ-decomposition}) are just identities
which hold locally in spacetime for any symmetric $T^{\mu\nu}(x)$
and any $J^\mu(x)$, given a chosen vector $u^\mu(x)$.
The assumption of hydrodynamics enters in expressing the
coefficients ${\cal E}$, ${\cal P}$, $q^\mu$, $t^{\mu\nu}$, ${\cal N}$, and $j^\mu$
in terms of the hydrodynamic variables $u_\mu$, $T$, and $\mu$.
For example, the coefficients ${\cal E}$, ${\cal P}$, and ${\cal N}$
will be functions of scalars
$T$, $\mu$, $\partial_\mu u^\mu$, $u^\mu \partial_\mu T$, $\partial^2{\!\mu}$,
$(\partial_\mu u_\nu)^2$ etc.
Similarly, the coefficients $j_\mu$ and $q_\mu$ will be functions
of transverse vectors
$\Delta_{\mu\nu} \partial^\nu T$, $\Delta_{\mu\nu} \partial^{\nu}\! \mu$ etc.
The expressions for $T^{\mu\nu}$ and $J^\mu$ in terms of the hydrodynamic variables
$T$, $u_\mu$, and $\mu$ are called the constitutive relations.

We will write down the constitutive relations in the 
derivative expansion, which is the expansion
in powers of derivatives of the hydrodynamic variables.
Because the deviations from thermal equilibrium are assumed small,
one expects that 
one-derivative terms will be smaller than zero-derivative terms, 
two-derivative terms will be smaller than one-derivative terms and so on.
This is analogous to the derivative expansion
in effective field theory, while the role of the cutoff
is played by the microscopic scale $\lmfp$.
At length scales shorter than $\lmfp$, hydrodynamics stops being
a sensible description of matter.
Ideal (non-dissipative) hydrodynamics corresponds to taking
into account only non-derivative terms of the hydro variables $T$, $u_\mu$, and $\mu$
in the expansion of $T^{\mu\nu}$ and $J^\mu$.
Conventional dissipative hydrodynamics corresponds to taking
into account both non-derivative and one-derivative terms of the hydro variables $T$, $u_\mu$, and $\mu$
in the expansion of $T^{\mu\nu}$ and $J^\mu$.

\subsubsection*{Zeroth-order hydrodynamics}
\noindent
Let us start with ideal hydrodynamics, i.e.~no derivatives of the hydro variables.
The transverse $q^\mu$, $t^{\mu\nu}$, and $j^\mu$
can only be built out of derivatives of the hydro variables,
hence there are no $q^\mu$, $t^{\mu\nu}$, and $j^\mu$ in the ideal hydrodynamics.
On the other hand, the coefficients ${\cal E}$, ${\cal P}$, and ${\cal N}$
can be functions of $T$ and $\mu$.
These coefficients have simple physical interpretation:
in static equilibrium, $T^{\mu\nu}={\rm diag}(\e, p,...,p)$, 
where $\e$ is the equilibrium energy density and $p$ is the equilibrium pressure,
while $J^\mu = (n,{\bf 0})$, where $n$ is the equilibrium charge density.
For a fluid which moves with constant velocity $u^\mu$,
the energy-momentum tensor and the current are obtained by performing the
corresponding Lorentz transformation, and one finds
\begin{eqnarray}
\label{eq:Tmunu-rel}
  &&  T^{\mu\nu}=\epsilon\, u^\mu u^\nu + p\,\Delta^{\mu\nu}\,,\\
\label{eq:Jmu-rel}
  &&  J^\mu = n u^\mu\,.
\end{eqnarray}
Ideal hydrodynamics corresponds to adopting this equilibrium form
of the energy-momentum tensor and the current, and promoting
$\epsilon$, $p$, $u^\mu$, and $n$ to slowly varying fields.
The constitutive relations in ideal relativistic hydrodynamics
thus have the form (\ref{eq:Tmunu-rel}), (\ref{eq:Jmu-rel}),
in other words we can identify 
${\cal E}(x)=\epsilon(x)$ with local energy density,
${\cal P}(x)=p(x)$ with local pressure,
${\cal N}(x)=n(x)$ with local charge density,
and $u^\mu(x)$ with local fluid velocity.
The equilibrium equation of state provides one with $p(T,\mu)$,
from which the energy density~$\epsilon$, entropy density~$s$,
and the charge density~$n$ can be found as 
$s = \partial p/\partial T$, $n = \partial p/\partial\mu$, 
$\epsilon = -p + Ts + \mu n$.

The longitudinal component of Eq.~(\ref{eq:Tmn-conservation}),
$u_\nu\d_\mu T^{\mu\nu}=0$, and current conservation give
\begin{subequations}
\label{eq:Tmunu-conserv}
\begin{eqnarray}
  && \d_\mu(w u^\mu) = u^\mu \d_\mu p\,,\\
  && \d_\mu (n u^\mu) =0\,,
\end{eqnarray}
\end{subequations}
where $w=\e+p$ is the density of enthalpy.
Using the thermodynamic relation $w=Ts+\mu n$ together with
$ d p = s\, dT + n\, d\mu$,
the conservation equations in
(\ref{eq:Tmunu-conserv}) combine to give
$$
  \d_\mu(s u^\mu) = 0\,.
$$
This is interpreted as the conservation of the ``entropy current'',
namely that (locally) the entropy does not increase
in ideal (non-dissipative) hydrodynamics.%
\footnote{%
   Even in zeroth-order hydrodynamics, entropy can increase
   in presence of shock waves which are discontinuities
   of the hydrodynamic flow~\cite{LL6}.
}

\subsubsection*{First-order hydrodynamics: frame choice}
\noindent
Let us now proceed to first-order hydrodynamics, 
i.e.~taking into account terms in the constitutive relations
with up to one derivative of the hydrodynamic variables.
This brings up a subtlety which was not apparent in the ideal hydrodynamics.
Namely, the notion of local temperature, local chemical potential,
and local fluid velocity are not uniquely defined out of equilibrium.
One may define many local temperature fields $T(x)$ which 
will differ from each other by gradients of the hydro variables,
and will approach the same constant value in equilibrium, when the gradients tend to zero.
The same applies to the chemical potential $\mu(x)$ and to the velocity $u^\mu(x)$.
In particular, this means that coefficients 
${\cal E}$, ${\cal P}$, and ${\cal N}$ in the 
decomposition (\ref{eq:TJ-decomposition}) are
\begin{subequations}
\label{eq:EPN-noneq}
\begin{eqnarray}
  {\cal E} &=& \epsilon(T,\mu) + f_{\cal E}(\partial T,\partial\mu, \partial u)\,,\\
  {\cal P} &=& p(T,\mu) + f_{\cal P}(\partial T,\partial\mu, \partial u)\,,\\
  {\cal N} &=& n(T,\mu) + f_{\cal N}(\partial T,\partial\mu, \partial u)\,,
\end{eqnarray}
\end{subequations}
where $\epsilon$, $p$, and $n$ are determined by the equation of state in equilibrium, and
the form of the out-of-equilibrium gradient corrections $f_{\cal E}$, $f_{\cal P}$, $f_{\cal N}$
depends on the definition of the local temperature, local chemical potential, and local velocity.
In hydrodynamics, redefinitions of the fields $T(x)$, $\mu(x)$, and $u^\mu(x)$ 
are often referred to as a choice of ``frame'', and we adopt this terminology here.
When thinking about frame choices, it is important to keep in mind that the parameters
$T(x)$, $\mu(x)$, and $u^\mu(x)$ have no first-principles microscopic definition out of equilibrium.
For example, there is no ``temperature operator'' whose expectation value in a given non-equilibrium state
would give a local temperature $T(x)$.
Rather, $T(x)$, $\mu(x)$, and $u^\mu(x)$ are merely auxiliary parameters used to 
parametrize $T^{\mu\nu}(x)$ and $J^{\mu}(x)$, which {\it do} have microscopic definitions.
The auxiliary parameters may be redefined at will, as long as the energy-momentum tensor
and the current remain unchanged under such redefinitions, see e.g.~\cite{Bhattacharya:2011tr}
for a recent discussion.

Consider a redefinition (or ``frame'' transformation)
\begin{eqnarray*}
   T(x) &\to& T'(x) = T(x) + \delta T(x)\,,\\
   \mu(x) &\to& \mu'(x) = \mu(x) + \delta \mu(x)\,,\\
   u^\mu(x) &\to& u^{\prime\mu}(x) = u^\mu(x) + \delta u^\mu(x)\,,
\end{eqnarray*}
where $\delta T$, $\delta\mu$ and $\delta u^\mu$ are first order in derivatives.
Note that $\delta u^\mu$ must be transverse, $u_\mu\, \delta u^\mu=0$,
due to the normalization condition $u^2=-1$.
Using the definitions (\ref{eq:EPNqjt}), and remembering that $q^\mu$, $j^\mu$ and $t^{\mu\nu}$
are first order in derivatives, while $T^{\mu\nu}$ and $J^{\mu}$ remain invariant, 
one finds to first order
\begin{subequations}
\label{eq:EPNqjt-transformations}
\begin{eqnarray}
\label{eq:EPN-transformations}
  && \delta {\cal E} = 0\,,\ \ \ \ \delta{\cal P} = 0\,,\ \ \ \ \delta{\cal N} = 0\,,\\
\label{eq:qj-transformations}
  && \delta q_\mu = -({\cal E} {+} {\cal P}) \delta u_\mu\,,\ \ \ \ 
     \delta j_\mu = -{\cal N} \delta u_\mu\,,\\
  && \delta t_{\mu\nu} = 0\,.
\end{eqnarray}
\end{subequations}
One can choose $\delta u_\mu$ such that $j_\mu=0$; 
this definition of the local fluid velocity which is adopted by Eckart~\cite{PhysRev.58.919},
is often referred to as ``Eckart frame'',
and implies no charge flow in the local rest frame of the fluid.
Or one can choose $\delta u_\mu$ such that  $q_\mu=0$; 
this definition of the local fluid velocity which is adopted in the book of Landau and Lifshitz~\cite{LL6},
is often referred to as ``Landau frame'',
and implies no energy flow in the local rest frame of the fluid.
Equation (\ref{eq:EPN-transformations}) says that 
$\epsilon(T,\mu) + f_{\cal E}(\partial T,\partial\mu, \partial u) = 
 \epsilon(T',\mu') + f'_{\cal E}(\partial T',\partial\mu', \partial u')$,
and similarly for ${\cal P}$ and ${\cal N}$,
which implies that 
\begin{eqnarray*}
  && f'_{\cal E} = f_{\cal E} 
     - \left(\frac{\partial\epsilon}{\partial T}\right)_{\!\mu} \delta T
     - \left(\frac{\partial\epsilon}{\partial \mu}\right)_{\!T} \delta \mu\,,\\
  && f'_{\cal P} = f_{\cal P} 
     - \left(\frac{\partial p}{\partial T}\right)_{\!\mu} \delta T
     - \left(\frac{\partial p}{\partial \mu}\right)_{\!T} \delta \mu\,,\\
  && f'_{\cal N} = f_{\cal N} 
     - \left(\frac{\partial n}{\partial T}\right)_{\!\mu} \delta T
     - \left(\frac{\partial n}{\partial \mu}\right)_{\!T} \delta \mu\,.
\end{eqnarray*}
This means that we can use the redefinitions of $T$ and $\mu$
to set two of the three functions $f'_{\cal E}$, $f'_{\cal P}$, $f'_{\cal N}$
to zero. 
It is conventional to set  $f'_{\cal E}$ and $f'_{\cal N}$ to zero,
which is to say that one can choose an off-equilibrium definition of $T$ and $\mu$
such that ${\cal E}=\epsilon$ and ${\cal N} = n$.

Other frames choices are possible: a generic frame would have both $q^\mu$ and $j^\mu$
non-zero, as well as $f_{\cal E}$, $f_{\cal P}$, $f_{\cal N}$ 
(defined by Eq.~(\ref{eq:EPN-noneq})) all non-zero.
As is evident from Eq.~(\ref{eq:qj-transformations}), the combination
$$
  \ell^\mu \equiv j^\mu  - \frac{n}{\epsilon+p} q^\mu
$$
is frame-invariant.
Using the above transformations of $f_{\cal E}$, $f_{\cal P}$, and $f_{\cal N}$,
it is easy to check that
$$
  f \equiv
  f_{\cal P} 
  - \left(\frac{\partial p}{\partial \epsilon}\right)_{\! n} \! f_{\cal E}
  - \left(\frac{\partial p}{\partial n}\right)_{\!\epsilon} \! f_{\cal N}
$$
is frame-invariant as well.
The dissipative terms may be shifted around in the constitutive relations
by frame transformations in a way that leaves $\ell^\mu$ and $f$ invariant.

\subsubsection*{First-order hydrodynamics: constitutive relations}
\noindent
We start with the Landau frame, 
where the velocity $u^\mu$ is chosen so that $q^\mu=0$,
and $T$ and $\mu$ are chosen so that ${\cal E}=\epsilon$ and ${\cal N}=n$.
The remaining coefficients ${\cal P}$, $t^{\mu\nu}$, and $j^\mu$
in Eqs.(\ref{eq:TJ-decomposition})
need to be expressed in terms of the hydrodynamic variables,
providing constitutive relations in first-order hydrodynamics.

There are three one-derivative scalars in addition to $T$ and $\mu$,
which can be built out of the hydro variables:
$u^\lambda \d_\lambda T$, $u^\lambda \d_\lambda \mu$, and $\d_\lambda u^\lambda$.
There are also three transverse vectors,
$\Delta^{\mu\nu}\d_\nu T$, $\Delta^{\mu\nu}\d_\nu \mu$,
and $\Delta^{\mu\nu} \dot{u}_\nu$, where $\dot{u}_\nu \equiv u^\lambda\d_\lambda u_\nu$.
Finally, there is one transverse traceless symmetric tensor,
$$
  \sigma^{\mu\nu}\equiv 
  \Delta^{\mu\alpha}\Delta^{\nu\beta}
  \left( \d_\alpha u_\beta + \d_\beta u_\alpha - \frac{2}{d} \eta_{\alpha\beta}\d_\mu u^\mu
  \right)\,.
$$
Let us look at the scalars first. 
We would write the derivative expansion as
$$
  {\cal P} = p + c_1 u^\lambda \d_\lambda T + c_2 u^\lambda \d_\lambda \mu + c_3 \d_\lambda u^\lambda +O(\d^2)\,,
$$
where $p$ is thermodynamic pressure in the local rest frame of the fluid,
$c_{1,2,3}$ are some coefficients, and $O(\d^2)$ denote the terms
which contain either second derivatives of the hydro variables, or are quadratic
in the first derivatives.
We can simplify our life by making use of the known zeroth-order hydro equations:
indeed, two scalar equations $u_\mu \d_\nu T^{\mu\nu}=0$ and $\d_\mu J^\mu=0$
provide two relations among the three one-derivative scalars.
We can use the zeroth-order hydro equations to eliminate two of the scalars
in the above expansion for ${\cal P}$, and the error we are making by
using the zeroth-order equations is only going to enter at $O(\d^2)$,
which we are neglecting anyway.
It is conventional to eliminate 
$u^\lambda \d_\lambda T$ and $u^\lambda \d_\lambda \mu$, and keep $\d_\lambda u^\lambda$
as the only independent scalar at first order.
Thus the expansion of ${\cal P}$ becomes,
$$
  {\cal P} = p - \zeta\, \partial_\lambda u^\lambda + O(\d^2)\,,
$$
where $\zeta$ is a coefficient which needs to be determined
from the microscopic theory.
The coefficient $\zeta$ is the bulk viscosity.

Similarly, we would write $j_\mu$ as a combination of the three
transverse one-derivative vectors. 
Again, we can use the zeroth order hydro equations:
there is one transverse vector equation 
$\Delta_{\lambda\nu}\d_\mu T^{\mu\nu}=0$,
and therefore only two of the three transverse vectors are independent.
We choose to eliminate $\Delta^{\mu\nu} \dot{u}_\nu$,
and therefore the expansion of $j^\mu$ becomes
$$
  j^\mu = -\sigma T \Delta^{\mu\nu} \partial_\nu (\mu/T) 
          + \chiT \Delta^{\mu\nu} \partial_\nu T + O(\d^2)\,,
$$
where $\sigma$ and $\chiT$ are coefficients which need to be determined
from the microscopic theory.
The coefficient $\sigma$ is the charge conductivity;
we will see later that the coefficient $\chiT$ must be zero.

Finally, $\sigma^{\mu\nu}$ is the only transverse traceless symmetric tensor, hence 
$$
  t^{\mu\nu} = - \eta\, \sigma^{\mu\nu} + O(\d^2)\,,
$$
where the coefficient $\eta$ is the shear viscosity.
Thus the constitutive relations in first-order
relativistic hydrodynamics in the Landau frame can be taken as:
\begin{subequations}
\label{eq:const-rel}
\begin{align}
  & T^{\mu\nu} = \epsilon u^\mu u^\nu + p \Delta^{\mu\nu}
     - \eta\, 
      \Delta^{\mu\alpha}\Delta^{\nu\beta}\!\!
      \left( \d_\alpha u_\beta + \d_\beta u_\alpha - 
      \frac{2}{d} \eta_{\alpha\beta}\d_\mu u^\mu
      \right)
     - \zeta \Delta^{\mu\nu} \d_\lambda u^\lambda + O(\partial^2)\,,\\[5pt]
  & J^\mu = n u^\mu 
     -\sigma T \Delta^{\mu\nu} \partial_\nu (\mu/T) 
     +\chiT \Delta^{\mu\nu} \partial_\nu T + O(\partial^2)\,.
\end{align}
\end{subequations}
We see that Lorentz covariance restricts the constitutive relations
up to four transport coefficients $\eta$, $\zeta$, $\sigma$, and $\chiT$.
We will see later that $\eta$, $\zeta$, and $\sigma$ must be non-negative,
while $\chiT$ has to vanish (even though a non-zero value of $\chiT$ is allowed by Lorentz symmetry).
The equation of state provides one with $p(T,\mu)$, from which one can find
$\epsilon(T,\mu)$ and $n(T,\mu)$.
The transport coefficients $\eta$, $\zeta$, and $\sigma$ depend on $T$ and $\mu$
in a way that is determined by the underlying microscopic theory.

In a generic unspecified frame, the constitutive relations for frame-invariant quantities
in first-order hydrodynamics take the form
$$
  t^{\mu\nu} = - \eta\, \sigma^{\mu\nu}\,,\ \ \ \ 
  f = -\zeta\, \partial_\lambda u^\lambda\,,\ \ \ \ 
  \ell^\mu = -\sigma T \Delta^{\mu\nu} \partial_\nu (\mu/T) +\chiT \Delta^{\mu\nu} \partial_\nu T \, .
$$
For example, one could choose a frame in which the bulk viscosity
appears as a non-equilibrium correction to charge density,
and the charge conductivity appears in the constitutive relations
for both the energy-momentum tensor and the current.
In the Eckart frame, the constitutive relations become
\begin{subequations}
\label{eq:const-rel-EF}
\begin{align}
  & T^{\mu\nu} = \epsilon u^\mu u^\nu + p \Delta^{\mu\nu}
     + (q^\mu u^\nu {+} q^\nu u^\mu)
     - \eta\, \sigma^{\mu\nu}
     - \zeta \Delta^{\mu\nu} \d_\lambda u^\lambda + O(\partial^2) \,,\\
  & J^\mu = n u^\mu + O(\partial^2)\,,
\end{align}
\end{subequations}
where 
$q^\mu = (\sigma T \Delta^{\mu\nu} \partial_\nu (\mu/T) 
        -\chiT \Delta^{\mu\nu} \partial_\nu T) (\epsilon{+}p)/n$.
The expression for $q^\mu$ can be rewritten in an equivalent form
by eliminating $\Delta^{\mu\nu}\partial_\nu \mu$ in favor of
$\Delta^{\mu\nu}\partial_\nu T$ and $\Delta^{\mu\nu} \dot{u}_\nu$,
using the equations of ideal hydrodynamics:
$$
  q^\mu = -\kappa\, \Delta^{\mu\nu}\! \left( T \dot{u}_\nu  + \partial_\nu T \right) 
          -\frac{\epsilon{+}p}{n} \chiT\, \Delta^{\mu\nu} \partial_\nu T  + O(\partial^2)\,,
$$
where $\kappa\equiv \sigma (\epsilon{+}p)^2/(n^2 T)$ is the heat conductivity.
It is this Eckart-frame form of the relativistic hydrodynamic equations
with $\chiT=0$ that becomes Eq.~(\ref{eq:dissipative-NR-hydro})
in the non-relativistic limit.

The coefficients $\eta$, $\zeta$, $\sigma$, and $\chiT$ can be viewed as parameters
in the effective theory that need to be matched to the microscopic theory.
This matching can be done in the linear response theory, as we will see in the next section.
The linear response theory will give explicit expressions for $\eta$, $\zeta$, and $\sigma$ 
in terms of correlations functions of the energy-momentum tensor and the current.
As the latter are independent of the ``frame'', this provides an alternative way to see that 
the transport coefficients $\eta$, $\zeta$, and $\sigma$ are frame invariant:
choosing a different frame may change the place where they appear in the constitutive relations,
but not their value. 

As this example of first-order hydrodynamics illustrates, Lorentz covariance alone is not sufficient
to enumerate the transport coefficients: the number of non-vanishing transport coefficients
does not coincide with the number of ``independent tensor structures" that one can write down 
consistent with Lorentz symmetry, up to a given order in the derivative expansion. 
Other ingredients are needed (such as the local form of the second law of thermodynamics,
or the linear response theory) in order to find the number of independent transport coefficients.
The same happens in second-order hydrodynamics~\cite{Bhattacharyya:2012ex, Banerjee:2012iz, Jensen:2012jh}.

\subsubsection*{First-order hydrodynamics: entropy current}
\noindent
Before moving on to the linear response theory,
let us briefly discuss the entropy current in first-order relativistic hydrodynamics.
In thermal equilibrium state with constant $u^\mu$, the entropy current is $S^\mu = s u^\mu$.
We saw that $\partial_\mu (s u^\mu) =0$ in zeroth order (ideal) hydrodynamics. 
One now assumes that in first-order hydrodynamics there exists a current
$$
  S^\mu = s u^\mu + \textrm{(gradient\ corrections)}\,,
$$
which satisfies $\partial_\mu S^\mu\geqslant0$
when the equations of hydrodynamics are satisfied.
The gradient corrections vanish in equilibrium, and are
built out of derivatives of $T$, $\mu$, and $u^\mu$.
The entropy current defined this way is in general not unique,
however, demanding that $\partial_\mu S^\mu{\geqslant}0$, one can obtain the constitutive relations
together with constraints on the transport coefficients~\cite{LL6,Hiscock:1985zz}.
An expression for the entropy current may be written down 
by using a covariant version of the thermodynamic relation 
$Ts = p + \epsilon -\mu n$ as~\cite{Israel-1981}
$$
  T S^\mu  = p u^\mu - T^{\mu\nu} u_\nu - \mu J^\mu.
$$
Using the decomposition of the energy-momentum tensor and the current (\ref{eq:TJ-decomposition}),
we find
\begin{equation}
\label{eq:entropy-current}
  S^\mu = \left[ s + \frac{1}{T} ({\cal E} {-}\epsilon) -\frac{\mu}{T}({\cal N} {-} n) \right] u^\mu
         +\frac{1}{T} q^\mu - \frac{\mu}{T} j^\mu\,,
\end{equation}
where ${\cal E}$, ${\cal N}$, $q^\mu$ and $j^\mu$ are defined by
(\ref{eq:TJ-decomposition}) and (\ref{eq:EPNqjt}).
Using frame transformations (\ref{eq:EPNqjt-transformations}) together with
$\delta p = p(T',\mu') - p(T,\mu) = s \delta T + n \delta \mu$,
it is straightforward to check that the entropy current (\ref{eq:entropy-current})
is frame invariant.
The expression for the entropy current simplifies in a frame with 
${\cal E} = \epsilon$, ${\cal N} = n$, and in addition one can use either
the Eckart frame condition $j^\mu=0$, or the Landau frame condition $q^\mu=0$. 
The positivity of entropy production $\partial_{\mu} S^\mu\geqslant0$ will dictate that 
\begin{equation}
\label{eq:eta-sign}
  \eta\geqslant 0\,,\ \ \ \ 
  \zeta\geqslant 0\,,\ \ \ \ 
  \sigma\geqslant 0\,,\ \ \ \ 
  \chiT=0\,.
\end{equation}
The canonical form (\ref{eq:entropy-current}) of the entropy current and the constraints
(\ref{eq:eta-sign}) arise by demanding the positivity of entropy production in curved
spacetime, $\nabla_{\!\mu} S^\mu = 0$, see Ref.~\cite{Bhattacharya:2011tr}.
We will see that the same conditions (\ref{eq:eta-sign})
can be obtained in the linear response theory, 
without reference to an entropy current.

\section{Hydrodynamic equations and correlation functions}
\label{sec:linear-response}
\noindent
If we wish to study small hydrodynamic fluctuations about a
static thermal equilibrium state, we can take the full non-linear
relativistic hydrodynamic equations (\ref{eq:TJ-conservation})
with constitutive relations (\ref{eq:const-rel}),
and linearize them around the solution $u^\mu=(1,{\bf 0})$, $T={\rm const}$,
$\mu={\rm const}$.
For most of this section, rather than working with $T$, $u^i$, and $\mu$,
we will work with their conjugate variables,
which are the energy density, momentum density, and charge density.
The reason is that the latter have a microscopic definition
given by the operators $T^{0\mu}(x)$ and $J^0(x)$.
For a pedagogical discussion of the linear response theory
in non-relativistic hydrodynamics, see the beautiful paper
by Kadanoff and Martin~\cite{KM}.

\subsection{Simple diffusion}
\label{sec:diffusion}
\noindent
Before moving on to the full linearized hydrodynamics,
we start with a simple example of the diffusion equation for a 
conserved density $n(t,\x)$,
\begin{equation}
\label{eq:diffusion-equation}
  \partial_t n - D\grad^2 n = 0\,,
\end{equation}
where $\grad^2\equiv\partial^i\partial_i$ is the spatial Laplacian.
This equation of course describes the diffusion of $n(t,\x)$ with
a diffusion coefficient $D$,
and gives a simple example of relaxation with 
$\omega(\k)\to0$ as $\k{\to}0$.
To proceed, we define the spatial Fourier transform by
$$
  n(t,\x) = \int\!\frac{d^dk}{(2\pi)^d}\, e^{i\k{\cdot}\x} n(t,\k)\,.
$$
We can now solve the equation,
\begin{equation}
\label{eq:ntk}
  n(t,\k) = e^{-D\k^2 t} n_0(\k)\,,
\end{equation}
where $n_0(\k)\equiv n(t{=}0,\k)$.
What this equation tells us is how the density relaxes at $t>0$ 
given an initial disturbance $n_0(\k)$.
We will be working later with functions of frequency, rather than time,
and so we would like to Fourier transform (\ref{eq:ntk}) in time as well.
However, $n(t,\k)$ in Eq.~(\ref{eq:ntk}) is only defined for $t\geqslant 0$, so
instead of the Fourier transform, we will use the Laplace transform in time, defined as
$$
  n(z,\k) = \int_0^\infty\!dt\, e^{izt} n(t,\k)\,,
$$
where $z$ must have a positive imaginary part for convergence.
Applying the Laplace transform to the diffusion equation (\ref{eq:diffusion-equation}),
we find the solution as
$$
  n(z,\k) = \frac{n_0(\k)}{-iz+D\k^2}\,.
$$
We can also write this in terms of the source $\mu(t,\x)$ for the charge density,
which is the chemical potential.
For small fluctuations, we have $n(t,\x) = \chi\,\mu(t,\x)$, where 
$\chi\equiv(\partial n/\partial\mu)_{\mu=0}$ is the static susceptibility.
Note that $\chi$ is a static thermodynamic quantity, 
not a dynamic response quantity.
We now write the solution to the diffusion equation in terms
of the initial value of the source as
\begin{equation}
\label{eq:nzk2}
  n(z,\k) = \frac{\chi\,\mu_0(\k)}{-iz+D\k^2}\,.
\end{equation}

We now would like to connect the solution to the diffusion equation
to correlation functions in our system.
To do so, we imagine the following process.
We turn on a source for the density of the form
$\mu(t,\x) = e^{\varepsilon t} \mu(\x) \theta(-t)$.
In other words, the source $\mu(\x)$ is adiabatically turned on
at $t=-\infty$, and then switched off at $t=0$.
We then watch the system evolve at time $t>0$.
The master formula for the whole of linear response theory
comes from first-order time-dependent perturbation theory in Quantum Mechanics.
For a system with a time-independent Hamiltonian $H$ 
and a Heisenberg-picture operator $A(t,\x)$
defined with respect to $H$,
we imagine adding a time-dependent contribution $\delta H(t)$ to the Hamiltonian.
Then the change in the expectation value of $A(t,\x)$ is
\begin{equation}
\label{eq:master-lr}
   \delta\langle A(t,\x)\rangle = -i\int_{-\infty}^t\!dt'\,
   \langle [A(t,\x), \delta H(t')] \rangle\,,
\end{equation}
to first order in $\delta H$.
In our case $\delta H$ is due to the external source $\mu(t,\x)$,
and has the form
$$
  \delta H = -\int\! d^dx\;\mu(t,\x)\, n(t,\x)
$$
[think grand canonical density operator, $H\to H{-}\mu Q$].
Now, from Eq.~(\ref{eq:master-lr}), the change in the density is
\begin{eqnarray*}
  \langle n(t,\x)\rangle & = & 
  i\int_{-\infty}^t\!\!dt'\; e^{\varepsilon t'} \theta(-t')
  \int\! d^dx'\; \mu(\x') \, \langle [n(t,\x), n(t',\x')] \rangle\\[5pt]
  & = &
  i\int_{-\infty}^0\!\!dt'\; e^{\varepsilon t'} \theta(t{-}t')
  \int\! d^dx'\; \mu(\x') \, \langle [n(t,\x), n(t',\x')] \rangle\,.
\end{eqnarray*}
This can be written more compactly if we introduce the retarded function
$$
  G^R_{nn}(t{-}t', \x{-}\x') \equiv -i\theta(t{-}t')\, \langle [n(t,\x), n(t',\x')] \rangle\,.
$$
Then the change in density induced by the external source $\mu(t,\x)$ becomes
\begin{equation}
\label{eq:nmugr}
  \langle n(t,\x)\rangle = 
  -\int_{-\infty}^0\!\!dt'\; e^{\varepsilon t'}
  \int\! d^dx'\; \mu(\x') \, G^R_{nn}(t{-}t', \x{-}\x')\,.
\end{equation}
Our goal now is to take this equation for $t>0$, 
find the Laplace transform $\langle n(z,\k)\rangle$, 
and compare with the prediction (\ref{eq:nzk2}) of the diffusion equation.
This will give us the retarded function.

We first Fourier transform Eq.~(\ref{eq:nmugr}) in space, which gives
\begin{equation}
\label{eq:nmugr2}
   \langle n(t,\k)\rangle = 
  -\int_{-\infty}^0\!\!dt'\; e^{\varepsilon t'}
  \mu(\k) \, G^R_{nn}(t{-}t', \k)\,.
\end{equation}
Note that $\mu(\k)$ is the Fourier transform of $\mu(\x)$, in other words 
it is the Fourier transform of the external source $\mu(t,\x)$ at $t{=}0$,
hence we will write $\mu(\k)=\mu_0(\k)$, to use the notation
of Eq.~(\ref{eq:nzk2}).
Next, we Fourier transform the retarded function in time,
$$
  G^R(t{-}t',\k) = \int_{-\infty}^{\infty}\!\frac{d\omega}{2\pi}\, G^R(\omega,\k)\,
  e^{-i\omega(t-t')}\,.
$$
Note that $G^R(t,\k)$ is only non-zero for $t>0$,
hence $G^R(\omega,\k)$ is an analytic function in the upper half-plane
of complex $\omega$. We can then analytically continue $G^R(\omega,\k)$
to lower half-plane.
Now the density induced by the external source becomes
$$
   \langle n(t,\k)\rangle = 
   - \mu_0(\k)\int\!\!\frac{d\omega}{2\pi}\;  G^R_{nn}(\omega,\k)
   \frac{e^{-i\omega t}}{i\omega + \varepsilon}\,.
$$
We multiply both sides by $e^{izt}$ (with ${\rm Im}\,z>0$),
and integrate over $t$ from $0$ to $\infty$,
which gives
$$
   \langle n(z,\k)\rangle = 
   - \mu_0(\k)\int\!\!\frac{d\omega}{2\pi}\;  
   \frac{G^R_{nn}(\omega,\k)}{(i\omega + \varepsilon)\, (i(\omega{-}z) + \varepsilon)}\,.
$$
To do the integral, we close the contour in the
upper-half plane where $G^R$ is analytic.
There are two poles inside the contour, 
at $\omega=i\varepsilon$, and $\omega=z+i\varepsilon$, thus
\begin{equation}
   \langle n(z,\k)\rangle = 
   - \mu_0(\k)  
   \frac{G^R_{nn}(z,\k) - G^R_{nn}(z{=}0,\k)}{iz}\,, 
\end{equation}
where the argument of $G^R$ is understood to be slightly
above the real axis.
Now comparing with equation (\ref{eq:nzk2}),
we find:
$$
  G^R_{n n}(z,\k) - G^R_{n n}(z{=}0,\k) = \frac{-iz\chi}{-iz+D\k^2}\,.
$$
The only missing piece is $G^R(z{=}0,\k)$.
It is easy to find if one looks at Eq.~(\ref{eq:nmugr2})
at $t{=}0$,
$$
  \langle n(t{=}0,\k)\rangle = 
  -\mu_0(\k) \int_0^\infty\! dt'\, e^{-\varepsilon t'} G^R_{n n}(t',\k)=
  -\mu_0(\k) G^R_{nn}(z{=}0,\k)\,.
$$
So in the small-$\k$ limit we can identify $G^R_{nn}(z{=}0,\k) = -\chi$,
which gives the retarded function
$$
  G^R_{nn}(z,\k) = \frac{\chi D\k^2}{iz - D\k^2}\,.
$$
This function is analytic in the upper-half plane
of complex $z$ as it should be.
We can define $G^R(\omega,\k)$ in the whole complex plane
as the analytic continuation of $G^R(z,\k)$ 
from the upper-half plane.
In the lower half-plane, the retarded function has a pole at 
$\omega=-iD\k^2$, corresponding to the diffusive mode.

Given the above expression for the retarded function,
we can deduce the standard Kubo formula for the diffusion constant,
$$
  D\chi = -\lim_{\omega\to 0}\lim_{k\to 0} \frac{\omega}{\k^2}{\rm Im\,} G^R_{nn}(\omega,\k)\,.
$$

\subsection{Canonical approach to hydrodynamic response functions}
\label{sec:generalization}
\noindent
The above example of diffusion allows for a simple generalization
when there are several fields present.
Let $\varphi_a(t,\x)$ be the set of hydrodynamic variables
which have microscopic operator definitions, 
such as the charge density $J^0$, momentum density $T^{0i}$ etc.
We add the sources~$\lambda_a$ to the Hamiltonian as
$$
  \delta H = -\int \!d^d x\; \lambda_a(t,\x) \,\varphi_a(t,\x)\,.
$$
The master relation (\ref{eq:master-lr}) now gives
$$
  \delta\langle\varphi_a(t,\x)\rangle = 
  -\int_{-\infty}^{\infty}\!\!dt' \int\!\!d^d x'\;
  G^R_{ab}(t{-}t', \x{-}\x')\, \lambda_b(t',\x')\,,
$$
or in terms of the Fourier components simply
\begin{equation}
\label{eq:GG}
  \delta\langle\varphi_a(\omega,\k)\rangle = -
  G^R_{ab}(\omega,\k)\, \lambda_b(\omega,\k)\,.
\end{equation}
As before, the retarded function is defined by
\begin{equation}
\label{eq:GR2}
  G^R_{ab}(t{-}t', \x{-}\x') \equiv 
  -i\theta(t{-}t')\, \langle [\varphi_a(t,\x), \varphi_b(t',\x')] \rangle\,.
\end{equation}
We can repeat the initial-value analysis just like we did
for diffusion.
In momentum space the variables $\varphi_a$ obey linear equations
\begin{equation}
\label{eq:phi-M}
  \partial_t \varphi_a(t,\k) + M_{ab}(\k)\varphi_b(t,\k) = 0\,,
\end{equation}
where the matrix $M_{ab}$ is determined by the equations that 
$\varphi_a$ satisfy, such as the equations of relativistic hydrodynamics.
Applying the Laplace transform, one finds
$$
  (-iz\delta_{ab} + M_{ab}) \varphi_b(z,\k) = \varphi_a^0(\k)\,,
$$
where $\varphi_a^0(\k)\equiv \varphi_a(t{=}0,\k)$.
The initial values of the hydrodynamic variables are related to the initial values
of the sources by the susceptibility matrix,
$\varphi_a^0(\k{\to}0) = \chi_{ab}\,\lambda_b^0(\k{\to}0)$,
where%
\footnote{%
	Note that the relation between the source and the field
	is local in space, but non-local in time, so it would be incorrect to write
	$\varphi_a(\omega,\k) = \chi_{ab}\,\lambda_b(\omega,\k)$.
}
$$
  \chi_{ab} = \left(\frac{\partial\varphi_a}{\partial\lambda_b}\right)
$$
is the static thermodynamic susceptibility.
As before, the retarded function at zero frequency and small momentum
is just minus the susceptibility,
$G^R_{ab}(z{=}0,\k{\to}0) = -\chi_{ab}$.
The solution to the hydrodynamic equations can now be written as
\begin{equation}
\label{eq:phi-hydro}
  \varphi_a(z,\k) = (K^{-1})_{ab}\, \chi_{bc}\,\lambda_c^0(\k)\,,
\end{equation}
where $K_{ab} = -iz\delta_{ab} + M_{ab}(\k)$.
On the other hand, $\varphi_a(z,\k)$ can be expressed
in terms of the retarded function, just like we did for diffusion:
\begin{equation}
\label{eq:phiGR}
  \varphi_a(z,\k) = -\frac{1}{iz} 
  \left( G^R_{ab}(z,\k) - G^R_{ab}(z{=}0,\k) \right) \lambda_b^0(\k)\,.
\end{equation}
Comparing the solution to the hydrodynamic equations (\ref{eq:phi-hydro})
with the expression (\ref{eq:phiGR}), we find the retarded function
\begin{equation}
\label{eq:GR-phi}
  G^R(z,\k) = -\left( {\bf 1} + iz K^{-1} \right)\chi\,,
\end{equation}
omitting the matrix indices.

\subsection{General properties of response functions}
\noindent
In addition to the retarded function (\ref{eq:GR2}),
other response functions are useful. 
For bosonic operators $\varphi_a$, $\varphi_b$, we define
\begin{eqnarray*}
  &&  G^R_{ab}(t{-}t', \x{-}\x') \equiv 
      -i\theta(t{-}t')\, \langle [\varphi_a(t,\x), \varphi_b(t',\x')] \rangle\,,\\[5pt]
  &&  G^A_{ab}(t{-}t', \x{-}\x') \equiv 
      i\theta(t'{-}t)\, \langle [\varphi_a(t,\x), \varphi_b(t',\x')] \rangle\,,\\[5pt]
  &&  G_{ab}(t{-}t', \x{-}\x') \equiv 
      \coeff12 \, \langle \{ \varphi_a(t,\x), \varphi_b(t',\x')\} \rangle\,,\\[5pt]
  &&  \rho_{ab}(t{-}t', \x{-}\x') \equiv 
      \langle [\varphi_a(t,\x), \varphi_b(t',\x')] \rangle\,.
\end{eqnarray*}
The first one is the retarded function, the second one is the advanced function,
the third one is the symmetrized function.
The Fourier transform of the commutator $\rho_{ab}(\omega,k)$ is called the spectral function.
The expectation values are taken in static thermal equilibrium in the grand canonical ensemble,
and the Heisenberg operators are defined with the Hamiltonian $H'=H-\mu Q$.
The above four functions are not independent, and can all be deduced
from the Euclidean time-ordered function, see e.g.~\cite{Fetter-Walecka}.

Thanks to the theta functions of time, 
$G^R(\omega,\k)$ is analytic in the upper half plane of complex $\omega$, while
$G^A(\omega,\k)$ is analytic in the lower half plane of complex $\omega$.
Translation invariance implies that 
\begin{eqnarray*}
  &&  G^R_{ab}(\omega,\k) = G^A_{ba}(-\omega,-\k)\,,\\[5pt]
  &&  \rho_{ab}(\omega,\k) = -\rho_{ba}(-\omega,-\k)\,.
\end{eqnarray*}
From the definition of $G^R$ and $G^A$ one finds
\begin{eqnarray*}
  && G^{R,A}_{ab}(\omega,\k) = \int\! \frac{d\omega'}{2\pi}\,
     \frac{\rho_{ab}(\omega',\k)}{\omega-\omega' \pm i\varepsilon}\,,
\end{eqnarray*}
with the upper sign for $G^R$ and the lower sign for $G^A$.
For Hermitian operators $\varphi_a$ and $\varphi_b$, the matrix
$\rho_{ab}(\omega,\k)$ is Hermitian.
The diagonal components $\rho_{aa}(\omega,\k)$ are then real,
and we find for real $\omega$
\begin{eqnarray*}
  && {\rm Re}\, G^R_{aa}(\omega,\k) = {\rm Re}\, G^A_{aa}(\omega,\k) = 
     {P}\! \int\!\frac{d\omega'}{2\pi}\, \frac{\rho_{aa}(\omega',\k)}{\omega-\omega'}\,,\\[5pt]
  && {\rm Im}\, G^R_{aa}(\omega,\k) = -{\rm Im}\, G^A_{aa}(\omega,\k) = -\coeff12 \rho_{aa}(\omega,\k)\,,
\end{eqnarray*}
where ${P}$ denotes the principal value of the integral.

Writing out the definitions of $G_{ab}(\omega,\k)$ and $\rho_{ab}(\omega,\k)$
in the basis of $H'$ eigenstates in the grand canonical ensemble
and inserting a complete set of states, one can easily show that
$$
  G_{ab}(\omega,\k) = \frac12\, \frac{1{+}e^{-\beta\omega}}{1{-}e^{-\beta\omega}}\, \rho_{ab}(\omega,\k)\,.
$$
In the hydrodynamic regime $\beta\omega\ll1$ this gives
$\rho_{ab}(\omega,\k) = \beta\omega G_{ab}(\omega,\k)$, or
\begin{equation}
\label{eq:fdt}
  G_{ab}(\omega,\k) = -\frac{2T}{\omega} {\rm Im}\, G^R_{ab}(\omega,\k)\,.
\end{equation}
For a Hermitian operator $\varphi_a(t,\x)$, the spectral decomposition implies
$G_{aa}(\omega,\k)\geqslant0$, or $\omega\rho_{aa}(\omega,\k)\geqslant0$, which gives
\begin{equation}
\label{eq:GR-positivity}
  -{\rm Im}\, G^R_{aa}(\omega,\k) \geqslant 0\ \ 
  {\rm for}\ \omega\geqslant0\,.
\end{equation}
Applied to hydrodynamic response functions,
this condition will imply that the transport coefficients
have a definite sign:
$\eta\geqslant0$, $\zeta\geqslant0$, and $\sigma\geqslant0$.

In addition, the retarded functions
have to be consistent with the symmetries of the theory,
such rotation invariance, 
parity, charge conjugation, time-reversal, and
various global symmetries.
Time-reversal turns out to be particularly useful in constraining 
the transport coefficients that appear in the constitutive
relations, as was pointed out by Onsager~\cite{PhysRev.37.405, PhysRev.38.2265}.
Hence the constraints on the transport coefficients due to 
the time-reversal covariance are called ``Onsager relations''.
The implications of time-reversal for the retarded function (\ref{eq:GR2})
are easy to derive. 
The anti-unitarity of the time-reversal operator $\Theta$ 
implies that~\cite{Sakurai}
$$
  \langle \beta| A |\alpha\rangle = 
  \langle \tilde\alpha| \Theta A^\dagger \Theta^{-1} |\tilde\beta\rangle
$$
for any linear operator $A$, where $|\tilde\alpha\rangle = \Theta|\alpha\rangle$
are the time-reversed states.
We now apply this to the retarded function (\ref{eq:GR2})
of Hermitian operators $\varphi_a$ which transform in a definite way under time-reversal,
$\Theta\, \varphi_a(t,\x)\Theta^{-1}=\eta_a \varphi_a(-t,\x)$,
where $\eta_a=\pm1$ is the time-reversal eigenvalue of $\varphi_a$.
The expectation value in (\ref{eq:GR2}) is taken in the grand canonical ensemble,
$$
  \langle \dots \rangle = \frac1Z {\rm tr}( e^{-\beta H + \beta\mu Q}\dots )\,,
$$
and the trace can be taken either in the basis of energy eigenstates $|n\rangle$,
or in the basis of the time-reversed states $|\tilde n\rangle$.
If the microscopic system is time-reversal invariant, i.e. $[H,\Theta]=0$,
then the space-time translation invariance implies that
$$
  G^R_{ab}(t, \x) = 
  -i\theta(t)\, \langle [\varphi_a(t,\x), \varphi_b(0)] \rangle = 
  -i\theta(t)\, \langle [\varphi_b(t,-\x), \varphi_a(0)] \rangle\,\eta_a \eta_b = 
  G^R_{ba}(t,-\x)\eta_a \eta_b\,.
$$
If the microscopic system does not have time-reversal invariance,
it may still be possible to use time-reversal covariance
to find a similar relation for the retarded function.
Namely, let us assume that the Hamiltonian depends on some
time-reversal breaking parameters $B$ such that
$\Theta H(B)\Theta^{-1} = H(-B)$.
For example, $B$ could be the external magnetic field, 
or it could be the mass for Dirac fermions in 2+1 dimensions.
The retarded function must then satisfy
\begin{equation}
\label{eq:OR}
G^R_{ab}(\omega,\k;B) = \eta_a \eta_b G^R_{ba}(\omega,-\k;-B)\,.
\end{equation}
This can be also written in the matrix form as
$
  G(\omega,\k;B) = S G^T(\omega,-\k;-B) S\,,
$
where $S\equiv{\rm diag} (\eta_1, \eta_2, \dots)$
is the matrix of time-reversal eigenvalues of the
hydro variables~$\varphi_a$, which satisfies $S^2{=}1$.
Equation (\ref{eq:OR}) is the basis for the Onsager relations.
Let us now apply it to our hydrodynamic correlation functions.

Taking $\omega{=}0$ and $\k{\to}0$ in (\ref{eq:OR})
we find for the static susceptibility matrix
\begin{equation}
\label{eq:OR-chi}
  S\chi(B)\,S =  \chi^T(-B)\,.
\end{equation}
Further, applying the relation (\ref{eq:OR})
to the hydrodynamic correlation function (\ref{eq:GR-phi}),
evaluated in the linear response theory, one finds
\begin{equation}
\label{eq:OR2}
  \chi(B) S M^T(-\k;-B) = M(\k;B) \chi(B) S\,.
\end{equation}
In these lectures, we will not consider any {\bf T}-breaking
parameters $B$; in fact the breaking of {\bf T} due to
either magnetic field, or the fermion mass in 2+1 dimensions would give rise to
extra terms in the constitutive relations (\ref{eq:const-rel}), see Ref.~\cite{Jensen:2011xb}.
The condition (\ref{eq:OR2}) may be viewed as a constraint on hydrodynamic constitutive
relations arising from the time-reversal covariance.
Since the matrix $M(\k)$ simply represents the constitutive relations
in linearized hydrodynamics, this tells us that the 
constitutive relations in hydrodynamics can not be arbitrary, 
but must be such that Eq.~(\ref{eq:OR2}) is satisfied.
As an example, we will see shortly that the 
constitutive relations~(\ref{eq:const-rel}) in relativistic hydrodynamics
do not satisfy Eq.~(\ref{eq:OR2}), unless $\chiT=0$.
In other words, time-reversal covariance
may be used to deduce that $\chiT=0$, without alluding to the entropy current argument.

\subsection{Retarded functions in relativistic hydrodynamics at $\mu=0$}
\label{sec:mu0}
\noindent
As a simple example of the above general formalism, let us consider
small hydrodynamic fluctuations about an equilibrium state with $\mu=0$,
in other words $\bar n=0$ in equilibrium.%
\footnote{
	As a reminder, a state in the canonical ensemble which has a fixed value of the total
    conserved charge (such as the number of particles) is {\em not} the same as the state in the
    grand canonical ensemble at $\mu{=}0$.
	Hydrodynamic fluctuations in a state with $\bar n=0$ is a perfectly sensible thing
	in relativistic hydrodynamics: $\bar n=0$ simply means that there is 
	an equal number of particles and antiparticles in thermal equilibrium.
	On the other hand, in non-relativistic systems there are no antiparticles,
	and $\bar n=0$ means that the equilibrium state is empty,
	with no matter that could flow.
}
As the Eckart frame is ill-defined for states with $\bar n=0$, we will use the Landau frame.
The hydrodynamic equations are thus given by the conservation laws (\ref{eq:TJ-conservation}),
supplemented by the constitutive relations~(\ref{eq:const-rel}).
We linearize the equations around the static equilibrium state
$v^i=0$, $T={\rm const}$, $\mu=0$, and choose the hydrodynamic variables to be
the fluctuation in the energy density $\delta\epsilon(t,\x)=\delta T^{00}$,
momentum density $\pi_i(t,\x)=T^{0i}$,
and charge density $n(t,\x)= J^0$.

First, let us look at the constitutive relation for the current in~(\ref{eq:const-rel}).
All terms in the constitutive relation must transform the same way
under charge conjugation {\bf C} as the current $J^\mu$ does, in other words
$\sigma$ is {\bf C}-even, and $\chiT$ is {\bf C}-odd.
Assuming a {\bf C}-invariant microscopic theory, the only source
of {\bf C} violation in hydrodynamics is the chemical potential $\mu$,
hence $\chiT \to0$ as $\mu\to0$.
In other words, when $\mu$ vanishes in equilibrium,
the $\chiT$~term does not contribute to the linearized hydrodynamic equations.
The fluctuation in charge density
$n(t,\x)$ then decouples from the fluctuations of energy and momentum densities,
and satisfies the diffusion equation which we already studied 
in Section~\ref{sec:diffusion}.
The diffusion constant is $D=\sigma/\chi$, where $\chi=(\partial n/\partial\mu)_{\mu=0}$
is the static charge susceptibility.
Hence, the two-point retarded function of $J^0(t,\x)$ 
is given by
\begin{equation}
\label{eq:GRnn}
  G^R_{nn}(\omega,\k) = \frac{D\chi\k^2}{i\omega - D\k^2}\,.
\end{equation}

The equations for $\delta\epsilon(t,\x)$ 
and $\pi_i(t,\x)$ come from the conservation of the
energy-momentum tensor.
We Fourier transform in space, and decompose 
$\pi_i = \pi_i^\parallel + \pi_i^\perp$, where parallel and perpendicular are
with respect to the direction of the spatial momentum.
Choosing $\k$ along $x$, we have the following set of equations
\begin{eqnarray*}
  && \partial_t\,\delta\epsilon + i k_x\pi_x = 0\,,\\
  && \partial_t \pi^\parallel + ik_x\vs^2\,\delta\epsilon + \gamma_s \k^2\pi_x = 0\,,\\
  && \partial_t \pi_i^\perp + \gamma_\eta \k^2 \pi_i^\perp = 0\,.
\end{eqnarray*}
Here $\vs^2 = \partial p/\partial\epsilon$,
$\gamma_\eta = \eta/(\bar\epsilon{+}\bar p)$,
$\gamma_s = (\frac{2d-2}{d}\eta+\zeta)/(\bar\epsilon{+}\bar p)$,
$\eta$ and $\zeta$ are shear and bulk viscosities,
and $d$ is the number of spatial dimensions.

The transverse momentum density $\pi_i^\perp$ obeys the
diffusion equation, with $\gamma_\eta$ playing the role of the diffusion constant.
This means that we should be able to use the results for the diffusion equation.
The source for the momentum density is the velocity, and it
appears in the Hamiltonian in the form
$$
  \delta H = -\int \!d^dx\; v_i(t,\x)\, \pi_i(t,\x)
$$
[think grand canonical density operator, 
$-\beta H\to \beta u_\mu P^\mu\approx -\beta(H{-}{\bf v}{\cdot}{\bf P})$].
The role of the susceptibility $\chi$ is played by
the enthalpy density $\bar w\equiv(\bar\epsilon{+}\bar p)$, 
because to linear order $\pi_i=\bar w v_i$.
The retarded correlation function of the transverse momentum density
is now easy to write down:
for $\k$ along $x$, we have
\begin{equation}
\label{eq:GR-piypiy}
  G^R_{\pi_y \pi_y}(\omega,\k) = 
  \frac{\bar w\gamma_\eta \k^2}{i\omega-\gamma_\eta \k^2}\,.
\end{equation}
Now let us look at the coupled equations for $\delta\epsilon$ and $\pi^\parallel$.
These equations describe sound waves,
and have the form of Eq.~(\ref{eq:phi-M}) with
$\varphi_a = (\delta\epsilon,\pi_x)$, and 
$$
  M_{ab} = \begin{pmatrix}
  0 & ik_x \\
  ik_x \vs^2 & \gamma_s \k^2
  \end{pmatrix}\,.
$$
To find the susceptibility matrix, we need to identify the sources $\lambda_a$.
A disturbance in energy density can be created by a disturbance in temperature,
$-\beta H\to -(\beta+\delta\beta)H$.
Noting that $\delta\beta/\beta=-\delta T/T$, we have
$$
  \delta H = -\int\!\!d^dx \left(
  \frac{\delta T(t,\x)}{T} \epsilon(t,\x) + v_i(t,\x) \pi_i(t,\x)
  \right)\,.
$$
Thus the sources corresponding to the fields $\varphi_a = (\delta\epsilon,\pi_x)$ 
are $\lambda_a = (\delta T/T, v_x)$.
The susceptibility matrix is therefore diagonal,
$$
  \chi_{ab} = \begin{pmatrix}
  c_v T & 0\\
  0 & \bar w
  \end{pmatrix}
$$
where 
$c_v = \partial\epsilon/\partial T = \langle H^2\rangle_{\rm conn}/(VT^2)$
is the specific heat.
At $\mu=0$, it can be easily related to the enthalpy and the speed of sound,
$\bar w/T = s = \frac{\partial p}{\partial T} = 
\frac{\partial p}{\partial\epsilon} \frac{\partial\epsilon}{\partial T}
=\vs^2 c_v$.
The matrix of {\bf T}-eigenvalues is $S={\rm diag}(1,-1)$,
and one can easily check that the relation (\ref{eq:OR2})
is satisfied, so the response functions will come out consistent
with the time-reversal invariance.
The retarded function is given by Eq.~(\ref{eq:GR-phi}),
$$
  G^R_{ab}(\omega,\k) = 
  \frac{\bar w}{\omega^2 - \k^2\vs^2 + i\omega\gamma_s \k^2}
  \begin{pmatrix}
  \k^2 & \omega k_x \\
  \omega k_x & \k^2 \vs^2 - i\omega\gamma_s \k^2
  \end{pmatrix}\,.
$$
In the plane of complex $\omega$, there are two 
poles, when the denominator vanishes.
In the limit of small momenta, the poles are at
$\omega = \pm |\k| \vs -i\gamma_s \k^2/2$,
corresponding to weakly damped sound waves.
We can combine the contributions from $\pi^\perp$ and $\pi^\parallel$
into one retarded function for momentum density,
\begin{equation}
\label{eq:GRpiipij}
  G^R_{\pi_i \pi_j}(\omega,\k) = 
  \left(\delta_{ij} - \frac{k_i k_j}{\k^2}\right)
  \frac{\eta \k^2}{i\omega-\gamma_\eta \k^2}
  +
  \frac{k_i k_j}{\k^2}\,
  \frac{\bar w (\k^2 \vs^2 - i\omega\gamma_s \k^2)}
  {\omega^2-\k^2\vs^2+i\omega\gamma_s \k^2}
  \,.
\end{equation}
For completeness, the other retarded functions are
\begin{eqnarray*}
  &&  G^R_{\epsilon \pi_i}(\omega,\k) = G^R_{\pi_i \epsilon}(\omega,\k) =
      \frac{\bar w\, \omega k_i}{\omega^2-\k^2\vs^2+i\omega\gamma_s \k^2}\,,\\[5pt]
  &&  G^R_{\epsilon\epsilon}(\omega,\k) = 
      \frac{\bar w\, \k^2}{\omega^2-\k^2\vs^2+i\omega\gamma_s \k^2}\,.
\end{eqnarray*}
Evaluating the imaginary parts of the retarded functions, we find
the Kubo formulas
\begin{subequations}
\label{eq:Kubo-formulas}
\begin{eqnarray}
\label{eq:Kubo-sigma-n}
  && \sigma = -\frac{\omega}{\k^2}\,{\rm Im}\, G^R_{nn}(\omega,\k{\to}0)\,,\\[5pt]
\label{eq:Kubo-eta-pi}
  && \eta = -\frac{\omega}{\k^2}\, \frac{1}{d{-}1} \left(\delta_{ij} - \frac{k_i k_j}{\k^2}\right)
     {\rm Im}\, G^R_{\pi_i \pi_j}(\omega,\k{\to}0)\,,\\[5pt]
  && \frac{2d{-}2}{d}\eta + \zeta = 
     -\frac{\omega^3}{\k^4}\,
     {\rm Im}\, G^R_{\epsilon\epsilon}(\omega,\k{\to}0) \,.
\end{eqnarray}
\end{subequations}
The positivity condition (\ref{eq:GR-positivity}) implies that
$\eta\geqslant0$, $\zeta\geqslant0$, and $\sigma\geqslant0$,
consistent with the requirement that small hydrodynamic fluctuations
decay (rather than grow) with time.
Using the relation (\ref{eq:fdt}), the above Kubo formulas can be equivalently expressed
in terms of either the symmetrized correlation functions $G_{ab}$, 
or the spectral functions $\rho_{ab}$.

The Kubo formulas can be also written in terms of correlation functions of
spatial currents, rather than charge densities.
To do so, we demand that the correlation functions of conserved currents satisfy 
\begin{eqnarray*}
  && k^\mu G_{\!J_\mu J_\nu}(\omega,\k) = 0\,,\\
  && k^\mu G_{ T_{\mu\nu} T_{\alpha\beta}}(\omega,\k) = 0\,.
\end{eqnarray*}
For the conductivity, we then find
\begin{subequations}
\label{eq:Kubo-currents}
\begin{equation}
\label{eq:Kubo-sigma}
  \sigma
  = \frac{1}{2Td}\, G_{\! J_i J_i}(\omega,\k{=}0)
  = \frac{1}{2T}\, G_{\! J_x J_x}(\omega,\k{=}0)\,.
\end{equation}
In writing down this expression, we have used 
the relation (\ref{eq:fdt}) between the symmetrized and retarded functions,
as well as time-reversal invariance which by Eq.~(\ref{eq:OR})
implies $G_{\! J_0 J_i} = G_{\! J_i J_0}$,
and rotation invariance which implies that at zero spatial momentum $G_{\! J_i J_j}$ 
must be proportional to $\delta_{ij}$.
A similar argument gives Kubo formulas for shear viscosity
\begin{equation}
\label{eq:Kubo-eta}
    \eta = 
    \frac{1}{2T} \frac{1}{d^2{+}d{-}2}\, H^{mi}_{nj}\, G_{T_{mi}, T_{nj}}
    = \frac{1}{2T} G_{T_{xy} T_{xy} }\,,
\end{equation}
and for the bulk viscosity
\begin{equation}
\label{eq:Kubo-zeta}
  \zeta = \frac{1}{2T} \frac{1}{d^2} \delta_{mi} \delta_{nj} \, G_{T_{mi}, T_{nj}}
    = \frac{1}{2Td} \left( G_{T_{xx} T_{xx} } {+} (d{-}1)G_{T_{xx} T_{yy} }\right)
    = \frac{1}{2T} \left( G_{T_{xx} T_{xx} } {-} \coeff{2d{-}2}{d} G_{T_{xy} T_{xy} }\right)\,,
\end{equation}
\end{subequations}
where 
$H^{ij}_{kl} = \frac12 \delta^i_k \delta^j_l + \frac12 \delta^i_l \delta^j_k -\frac1d \delta^{ij} \delta_{kl}$
is a projector onto symmetric traceless tensors, and all correlation functions
are evaluated at $\k{=}0$.
The positivity of bulk viscosity implies that zero-momentum correlation functions
of the stress tensor obey
$$
  G_{T_{xx} T_{xx}}(\omega,\k{=}0) \geqslant \frac{2d{-}2}{d}\, G_{T_{xy} T_{xy}}(\omega,\k{=}0)\,,
$$
as a consequence of rotation invariance.
Taking a closer look at the hydrodynamic retarded functions, we see that
\begin{eqnarray*}
  && \omega G^R_{\epsilon\epsilon}(\omega,\k) - k_j G^R_{\pi_j\epsilon}(\omega,\k) = 0\,,\\[5pt]
  && \omega G^R_{\epsilon\pi_i}(\omega,\k) - k_j G^R_{\pi_j \pi_i}(\omega,\k) = \bar w k_i\,,
\end{eqnarray*}
in other words there is a contact term in the right-hand side
of momentum conservation equation applied to our correlation functions.

Finally, let us look at the the speed of sound:
$$
  \vs^2 = \frac{(s/T)}{({\partial s}/{\partial T})}\,,
$$
as follows from the relation $s=\vs^2 c_v$.
While it is easy to see that $\vs^2\geqslant0$ because
both $s$ and $c_v$ are non-negative,
thermodynamic inequalities do not demand that $\vs^2\leqslant1$,
as one may expect in a consistent relativistic theory.
Demanding that $\vs^2\leqslant1$ produces a constraint on the equation of state.
For example, for the equation of state of the form 
$p(T) = c\, T^\alpha$, one must have
$\alpha>2$ in order for the speed of sound to be less than the speed of light.%
\footnote{
	More precisely, the effective ``speed of light'' which determines the Lorentz invariance
	of the microscopic theory. For example, the effective ``speed of light'' in graphene is
	about 300 times less than the velocity of electromagnetic waves in the vacuum~\cite{RevModPhys.81.109}.
}

\subsection{Retarded functions in relativistic hydrodynamics at $\mu\neq0$}
\label{sec:mu1}
\noindent
Now let us consider
linearized relativistic hydrodynamics at non-zero $\mu$ in equilibrium.
This means that the equilibrium state has non-zero charge density $\bar n$,
and as a result, the fluctuations of charge density will couple
to the fluctuations of energy and momentum density.
Again, the hydrodynamic equations are given by the conservation 
laws~(\ref{eq:TJ-conservation}),
supplemented by the constitutive relations~(\ref{eq:const-rel}),
in the Landau frame.
We linearize the equations around the static equilibrium state
$v^i=0$, $T={\rm const}$, $\mu={\rm const}$, and choose the hydrodynamic variables to be
the fluctuation in the energy density $\delta\epsilon(t,\x)=\delta T^{00}$,
momentum density $\pi_i(t,\x)=T^{0i}$,
and charge density $n(t,\x)= J^0$.
Choosing the spatial momentum in the $x$ direction,
the conservation of the energy-momentum tensor gives
\begin{eqnarray*}
  && \partial_t\delta\epsilon + i k_x\pi_x = 0\,,\\[5pt]
  && \partial_t \pi^\parallel + 
     ik_x\beta_1 \delta\epsilon + 
     ik_x\beta_2 \delta n +
     \gamma_s \k^2\pi_x = 0\,,\\[5pt]
  && \partial_t \pi_i^\perp + \gamma_\eta \k^2 \pi_i^\perp = 0\,,
\end{eqnarray*}
where 
$\beta_1=\left(\frac{\partial p}{\partial\epsilon}\right)_{\!n}$,
$\beta_2=\left(\frac{\partial p}{\partial n}\right)_{\!\epsilon}$.
The equation of current conservation gives
$$
  \partial_t\delta n + \frac{\bar n}{\bar w} ik_x \pi_x +
  \sigma\alpha_1 \k^2 \delta\epsilon + \sigma\alpha_2 \k^2 \delta n =0\,,
$$
where the coefficients
$\alpha_1$ and $\alpha_2$ are
$$
  \alpha_1 = \left(\frac{\partial\mu}{\partial\epsilon}\right)_{\!n}
            -\left(\frac{\mu}{T}+\frac{\chiT}{\sigma}\right)
             \left(\frac{\partial T}{\partial\epsilon}\right)_{\!n}\,,
            \ \ \ \ 
  \alpha_2 = \left(\frac{\partial\mu}{\partial n}\right)_{\!\epsilon}
            -\left(\frac{\mu}{T}+\frac{\chiT}{\sigma}\right)
             \left(\frac{\partial T}{\partial n}\right)_{\!\epsilon}\,.
$$
The equation for $\pi^\perp$ decouples, as before,
and gives rise to the same two-point function~(\ref{eq:GR-piypiy}).
The remaining equations
have the form of Eq.~(\ref{eq:phi-M}) with
$\varphi_a = (\delta\epsilon,\pi_x, \delta n)$, and 
$$
  M_{ab} = \begin{pmatrix}
  0 & ik_x & 0\\
  ik_x \beta_1 & \gamma_s \k^2 & ik_x\beta_2\\
  \sigma \alpha_1 \k^2 & ik_x \bar n/\bar w & \sigma \alpha_2 \k^2
  \end{pmatrix}\,.
$$
Again, to find the susceptibility matrix, we need to identify the sources $\lambda_a$.
For infinitesimal constant disturbances $\delta T$, $\delta\mu$, and $v_i$,
the Hamiltonian changes as
$$
  H \to H -\frac{\delta T}{T}\left( H{-}\mu Q\right) 
        -\delta\mu\, Q - {\bf v}{\cdot}{\bf P}
$$
For non-constant slowly varying sources, we take
$$
  \delta H = -\int\!\!d^dx \left(
  \frac{\delta T(t,\x)}{T} \Big(\epsilon(t,\x)-\mu\, n(t,\x)\Big)
  +\delta\mu(t,\x)\, n(t,x)
  + v_i(t,\x) \pi_i(t,\x)
  \right)\,.
$$
Thus we identify the sources corresponding to 
$\varphi_a = (\delta\epsilon,\pi_x, \delta n)$ as
$\lambda_a = (\delta T/T, v_x, \delta\mu - \frac{\mu}{T}\delta T)$,
and therefore the susceptibility matrix is
\begin{equation}
\label{eq:chi2}
  \chi_{ab} = 
  \begin{pmatrix}
  T \!\left(\frac{\partial\epsilon}{\partial T}\right)_{\mu/T} & 0 &
  \left(\frac{\partial\epsilon}{\partial \mu}\right)_{T} \\
  0 & \bar w & 0 \\
  T\! \left(\frac{\partial n}{\partial T}\right)_{\mu/T} & 0 & 
  \left(\frac{\partial n}{\partial \mu}\right)_{T}
  \end{pmatrix}\,.
\end{equation}
Note that 
$T \!\left(\frac{\partial\epsilon}{\partial T}\right)_{\mu/T} = 
T \!\left(\frac{\partial\epsilon}{\partial T}\right)_{\mu} +
  \mu\! \left(\frac{\partial\epsilon}{\partial \mu}\right)_{T}$,
and similarly for $n$.
The definition of the thermodynamic quantities in the
grand canonical ensemble implies that
$T(\partial n /\partial T)_{\mu/T} = (\partial\epsilon/\partial\mu)_T$,
in other words, the susceptibility matrix is symmetric, $\chi_{13}=\chi_{31}$,
consistent with (\ref{eq:OR-chi}).
Further, $\chi_{11}\geqslant0$ and $\chi_{33}\geqslant0$ because the
first is proportional to 
$\langle H^2\rangle_{\rm conn}$,
while the latter is proportional to 
$\langle N^2\rangle_{\rm conn}$.
In addition, ${\det}(\chi)\geqslant0$ 
because it is proportional to 
$\langle H^2\rangle_{\rm conn}\langle N^2\rangle_{\rm conn} - \langle HN\rangle^2_{\rm conn}$,
which is non-negative by the Schwarz inequality.
Thus the susceptibility matrix $\chi_{ab}$ is positive-definite.%
\footnote{%
		Note that only those $p(T,\mu)$ that follow from the partition function
		in the grand canonical ensemble represent legitimate equations of state.
		An arbitrary function $p(T,\mu)$ will give rise to
		a susceptibility matrix which is not necessarily positive-definite.
}

An exercise in thermodynamic derivatives shows that
\begin{eqnarray*}
  && \beta_1\chi_{11} + \beta_2\chi_{31} = \bar w\,,\\
  && \beta_1\chi_{13} + \beta_2\chi_{33} = \bar n\,,\\
  && \alpha_1\chi_{11} + \alpha_2\chi_{31} = -T\chiT/\sigma\,,\\
  && \alpha_1\chi_{13} + \alpha_2\chi_{33} = 1\,.
\end{eqnarray*}
The matrix of {\bf T}-eigenvalues is $S={\rm diag}(1,-1,1)$, and
it's a matter of simple algebra to check that 
the condition of time-reversal covariance (\ref{eq:OR2})
can only be satisfied if $\chiT=0$, which we take from now on.
The coefficients $\beta_{1,2}$ and $\alpha_{1,2}$
can be easily expressed in terms of the components of the susceptibility matrix $\chi_{ab}$.

The retarded functions can be found by Eq.~(\ref{eq:GR-phi});
in the limit $\omega{\to}0$ and $\k{\to}0$ one finds
\begin{eqnarray*}
  &&  G^R_{\epsilon\epsilon}(\omega,\k) = \frac{\bar w\, \k^2}{d(\omega,\k)} \left(
      \omega 
      +i\sigma\alpha_2 \k^2
      \right)\,,\\[5pt]
  &&  G^R_{\pi_x \pi_x}(\omega,\k) = \frac{\bar w\, \k^2}{d(\omega,\k)} \left(
      \omega \vs^2
      +i\sigma\k^2 (\alpha_2\beta_1{-}\alpha_1\beta_2)
      \right)\,,\\[5pt]
  &&  G^R_{nn}(\omega,\k) = \frac{\k^2}{d(\omega,\k)} \Big(
      \omega {\bar n^2}\!/{\bar w}
      +i\sigma\k^2 (\alpha_2\beta_1{-}\alpha_1\beta_2)\chi_{33}
      \Big)\,,\\[5pt]
  && G^R_{\epsilon \pi_x}(\omega,\k) = G^R_{\pi_x \epsilon}(\omega,\k) = 
     \frac{\bar w\, \omega k}{d(\omega,\k)}
     \left( \omega + i\sigma\alpha_2 \k^2 \right)\,,\\[5pt]
  && G^R_{\pi_x n}(\omega,\k) = G^R_{n \pi_x}(\omega,\k) =  
     \frac{\omega k }{d(\omega,\k)} 
     \left( \bar n \omega - i\sigma\alpha_1 \bar w \k^2\right)\,,\\[5pt]
  && G^R_{\epsilon n}(\omega,\k) = G^R_{n \epsilon}(\omega,\k) =  
     \frac{\k^2 }{d(\omega,\k)} 
     \left( \bar n \omega - i\sigma\alpha_1 \bar w \k^2\right)\,.
\end{eqnarray*}
where 
\begin{equation}
\label{eq:d-mu}
  d(\omega,\k) = \omega^3 + i\omega^2\k^2(\gamma_s{+}\sigma\alpha_2)
                -\omega\k^2 \vs^2 
                +i\sigma\k^4(\alpha_1\beta_2{-}\alpha_2\beta_1)\,,
\end{equation}
and $\vs^2\equiv \beta_1 + \bar n\beta_2/\bar w$.
Setting $d(\omega,\k)=0$ determines the eigenfrequencies of the system.
There are three modes, whose frequencies in the limit $\k\to0$ are
\begin{eqnarray*}
  && \omega = \pm k\vs - i\frac{\Gamma}{2}k^2\,,\\[5pt]
  && \omega = -iDk^2\,,
\end{eqnarray*}
where
$
  \Gamma = \gamma_s + {\sigma\beta_2}
  \left( \alpha_1+\frac{\bar n}{\bar w}\alpha_2 \right)\!/{\vs^2} \,,\ 
  D = {\sigma}\left( \alpha_2\beta_1{-}\alpha_1\beta_2 \right)\!/{\vs^2}\,.
$
These are the familiar sound and diffusive modes.
The expression for the speed of sound in terms
of the components of the susceptibility matrix (\ref{eq:chi2}) is
$$
  \vs^2 = 
  \frac{\bar{n}^2 \chi_{11} + \bar{w}^2\chi_{33} - 2 \bar{n}\bar{w}\chi_{13}}
  {{\det}(\chi)}\,.
$$
The numerator can be written as
$(\bar{n}\sqrt{\chi_{11}} - \bar{w}\sqrt{\chi_{33}})^2 + 
2\bar{n}\bar{w}(\sqrt{\chi_{11} \chi_{33}} - \chi_{13})$, and therefore
the positive-definiteness of the susceptibility matrix implies $\vs^2\geqslant0$.
By the same argument as in Sec.~\ref{sec:NR-hydro} one finds
$
  \vs^2 = \left({\partial P}/{\partial \epsilon}\right)_{S,N}\,.
$
Note that the speed of sound does not depend on the dissipative
transport coefficients,
and is only determined by the equation of state.
In a scale-invariant theory, the equation of state
$p(T,\mu)=T^{d+1}f(T/\mu)$ implies $p=\epsilon/d$,
which is equivalent to the tracelessness of 
the energy-momentum tensor (\ref{eq:Tmunu-rel}) in equilibrium.
Thus in a scale-invariant theory,
the speed of sound is $\vs=1/\sqrt{d}$, and is independent of temperature and chemical potential.%
\footnote{%
	In QCD,
	the equation of state at asymptotically high temperatures is 
	$p=\epsilon/3$, and therefore the speed of sound at asymptotically high temperatures is $1/\sqrt{3}$.
}
The expression for the diffusion constant in terms
of the components of the susceptibility matrix is
$$
  D = \frac{\sigma \, \bar{w}^2}
      {\bar{n}^2 \chi_{11} + \bar{w}^2\chi_{33} - 2 \bar{n}\bar{w}\chi_{13}}\,,
$$
and shows that $D$ is only positive for $\sigma$ positive.
The expression for the sound wave damping constant in terms
of the components of the susceptibility matrix is
$$
  \Gamma = \frac{1}{\bar{w}} \left(\frac{2d{-}2}{d}\eta+\zeta\right) 
  + \frac{\sigma\, \bar{w}}{{\det}(\chi)}\,
    \frac{(\bar{n}\chi_{11} {-} \bar{w}\chi_{13})^2}
         {\bar{n}^2 \chi_{11} + \bar{w}^2\chi_{33} - 2 \bar{n}\bar{w}\chi_{13}}\,.
$$
Evaluating the imaginary parts of the retarded functions, we find
the Kubo formulas~(\ref{eq:Kubo-formulas}) for the transport coefficients.
In other words, the Kubo formulas for $\eta$, $\zeta$, and $\sigma$
have the same form at $\mu\neq0$ as they do at $\mu=0$,
even though the values of the transport coefficients of course
depend on~$\mu$.
Again, the positivity condition (\ref{eq:GR-positivity}) implies that
$\eta\geqslant0$, $\zeta\geqslant0$, and $\sigma\geqslant0$,
consistent with the requirement that small hydrodynamic fluctuations
decay (rather than grow) with time.
Our hydrodynamic correlation functions satisfy
\begin{subequations}
\label{eq:WI-mu}
\begin{eqnarray}
  && \omega G^R_{\epsilon n} - k_j G^R_{\pi_j n} = 0\,,\\[5pt]
  && \omega G^R_{\epsilon\epsilon}(\omega,\k) - k_j G^R_{\pi_j\epsilon}(\omega,\k) = 0\,,\\[5pt]
  && \omega G^R_{\epsilon\pi_i}(\omega,\k) - k_j G^R_{\pi_j \pi_i}(\omega,\k) = \bar w k_i\,,
\end{eqnarray}
\end{subequations}
again with a contact term for momentum conservation in the right-hand side.

Similar to the sound in the $\mu=0$ fluid,
thermodynamics does not constrain the speed of sound to be 
less than the speed of light.
Again, demanding that $\vs^2\leqslant1$ produces a constraint 
on the equation of state $p(T,\mu)$,
$$
 \bar{n}^2 \chi_{11} + \bar{w}^2\chi_{33} - 2 \bar{n}\bar{w}\chi_{13}
 \leqslant {\det}(\chi)\,.
$$

\subsection{Variational approach to hydrodynamic response functions}
\label{sec:variational}
\noindent
The response functions we computed so far were evaluated using the standard 
linear response approach, as described for example in Ref.~\cite{KM}.
After introducing external sources for the conserved densities 
(energy density, momentum density, charge density),
the response functions can be evaluated starting from Eq.~(\ref{eq:master-lr})
in the canonical (operator) formalism.
The advantage of this approach is that the external sources can be introduced quite easily:
their coupling to conserved densities follows from equilibrium thermodynamics.
The disadvantage is that the response functions of spatial currents (charge current, stress tensor)
are not easily accessible. 
For example, the component of the spatial current which is transverse 
to the spatial momentum does not couple to the charge density, and hence the correlation function
$(\delta_{ij}-k_i k_j/\k^2) G_{J_i J_j}(\omega,\k)$ at non-zero $\k$
can not be determined by this method.

The general relation (\ref{eq:GG}) between the fields and the sources suggests that 
it would be more elegant to introduce sources for $J^\mu$ and $T^{\mu\nu}$
rather than just for the conserved densities.
The retarded functions $G^R_{ab}(\omega,\k)$ can then be evaluated 
by taking the variation of one-point functions with respect to the source.
The advantage of this approach is that it gives direct access to all response functions
$G^R_{J_\mu J_\nu}$, $G^R_{J_\mu T_{\alpha\beta}}$, and $G^R_{T_{\mu\nu} T_{\alpha\beta}}$.
The disadvantage is that it may not always be obvious how to couple the currents to external sources.
In relativistic hydrodynamics, Lorentz symmetry is of great help. 
In what follows, we will write the constitutive relations in the presence of background metric $g_{\mu\nu}$
and background gauge field $A_\mu$ which couples to the conserved current $J^\mu$.
When $u^\mu$, $T$, and $\mu$ satisfy the hydrodynamic equations,
the constitutive relations provide the one-point functions of $T_{\mu\nu}$ and $J_\mu$
in the presence of external sources.
Defining
\begin{equation}
 {\cal J}^\mu(x) \equiv \sqrt{-g}\, \langle J^\mu(x) \rangle_{A,g}\,,\ \ \ \ 
 {\cal T}^{\mu\nu}(x) \equiv \sqrt{-g}\, \langle T^{\mu\nu}(x) \rangle_{A,g}\,.
\end{equation}
and expanding the metric about the flat space as $g_{\mu\nu} = \eta_{\mu\nu} {+} h_{\mu\nu}$,
the retarded functions may be defined as
\begin{subequations}
\label{eq:GR-sources}
\begin{align}
	G^{\,R}_{J^\mu \! J^\nu}(x) & = -
	\left.\frac{\delta {\cal J}^{\mu}(x) }{\delta A_{\nu}(0)}\right|_{A=h=0}\,, &
	G^{\,R}_{T^{\mu\nu} \! J^\sigma}(x) & = -
	\left.\frac{\delta {\cal T}^{\mu\nu}(x)}{\delta A_{\sigma}(0)}\right|_{A=h=0}\,, \\[5pt]
	G^{\,R}_{J^\sigma T^{\mu\nu}}(x) & = - 2
	\left.\frac{\delta {\cal J}^{\sigma}(x) }{\delta h_{\mu\nu}(0)}\right|_{A=h=0}\,, &
	G^{\,R}_{T^{\sigma\tau} T^{\mu\nu}}(x) & = - 2
	\left.\frac{\delta {\cal T}^{\sigma\tau}(x) }{\delta h_{\mu\nu}(0)}\right|_{A=h=0}\,.
\end{align}
\end{subequations}
The factor of $\sqrt{-g}$ gives rise to contact terms, by virtue of 
$\delta\sqrt{-g} = \frac12 \sqrt{-g}\, g^{\mu\nu} \delta g_{\mu\nu}$.
In the Landau frame, the constitutive relations in the presence of external sources are
\begin{subequations}
\label{eq:const-rel-sources}
\begin{eqnarray}
 && T^{\mu\nu} = \epsilon u^\mu u^\nu + p \Delta^{\mu\nu} 
  - \eta\, 
    \Delta^{\mu\alpha}\Delta^{\nu\beta}\!
    \left( \nabla_{\!\alpha} u_\beta + \nabla_{\!\beta} u_\alpha - 
    \frac{2}{d}\, g_{\alpha\beta}\nabla_{\!\mu} u^\mu
    \right) 
  - \zeta \nabla_{\!\mu} u^\mu\,,\\[5pt]
 && J^\mu = n u^\mu + \sigma \Delta^{\mu\lambda} V_\lambda
    + \chiE \Delta^{\mu\lambda} E_\lambda
    + \chiT \Delta^{\mu\lambda} \nabla_{\!\lambda}T\,,
\end{eqnarray}
\end{subequations}
where $\Delta^{\mu\nu} = g^{\mu\nu} + u^\mu u^\nu$, $\nabla_{\!\alpha}$ is the covariant derivative,
$V_\lambda \equiv E_\lambda - T \Delta_{\lambda\rho} \nabla^\rho (\mu/T)$,
$E_\mu \equiv F_{\mu\nu} u^\nu$ is the electric field,
and $F_{\mu\nu}$ is the field strength of the external gauge field.
None of the hydrodynamic variables carry charge, hence there are no gauge-covariant derivatives.
The velocity satisfies $u_\mu u^\mu =-1$, and
consequently $u_\alpha \nabla_{\!\mu}u^\alpha = 0$ because the metric is covariantly constant.
In writing down the constitutive relations for the current we have grouped 
the three independent vectors
$\nabla_{\!\lambda} T$, $\nabla_{\!\lambda}\mu$ and $E_\lambda$
into combinations with coefficients $\sigma$, $\chiT$, and $\chiE$.
The coefficient $\sigma$ is the electrical conductivity, and 
the term $\sigma \Delta^{\mu\lambda} E_\lambda$ is just a relativistic version of the Ohm's law.
As for the coefficients $\chiT$ and $\chiE$ (which are allowed by Lorentz covariance),
they must vanish as a consequence of either the positivity of entropy production, or of consistent thermodynamics with external sources, see for example~\cite{Jensen:2011xb,Jensen:2012jh}.

With the external sources $A_\mu$ and $g_{\mu\nu}$,
the hydrodynamic equations take the form
\begin{subequations}
\label{eq:TJ-conservation-sources}
\begin{eqnarray}
  && \nabla_{\!\mu} T^{\mu\nu} = F^{\nu\lambda} J_\lambda\,,\\
  && \nabla_{\!\mu} J^\mu = 0\,.
\end{eqnarray}
\end{subequations}
In order to find the retarded functions, we 
linearize the hydrodynamic equations around the equilibrium state with constant 
$\bar\epsilon$, $\bar{p}$, and $\bar{n}$,
and solve the conservation equations (\ref{eq:TJ-conservation-sources})
with the constitutive relations (\ref{eq:const-rel-sources}),
expressing $\delta T$, $\delta\mu$, and $\delta u^\mu$ in terms of $A_\mu$ and $h_{\mu\nu}$
to linear order. 
Upon substituting the solution back into (\ref{eq:const-rel-sources}), the retarded functions
can be read off from the definition (\ref{eq:GR-sources}).

The retarded functions obtained in this variational approach will differ from 
their counterparts in Section~\ref{sec:mu1} by contact terms.
For example, taking the spatial momentum in the $x$-direction, 
we have for the transverse momentum and current densities
\begin{eqnarray}
\label{eq:T0yT0y-sources}
  &&  G^R_{T^{0y} T^{0y}}(\omega,\k) = \frac{\eta \k^2}{i\omega - \gamma_\eta\k^2} + \bar\epsilon\,,\\[5pt]
  &&  G^R_{J^y J^y}(\omega,\k) = 
      -i\omega\sigma + \frac{\bar{n}^{\, 2}}{\bar\epsilon+\bar{p}}\, \frac{i\omega}{i\omega - \gamma_\eta \k^2}\,,
\end{eqnarray}
where $\gamma_\eta \equiv \eta/(\bar{\epsilon} + \bar{p})$, as before.
Note that at non-zero charge density in equilibrium,
$G^R_{J^y J^y}$ has the same shear-mode singularity as $G^R_{T^{0y} T^{0y}}$,
due to the convective term in the current.
For the retarded functions of spatial currents at zero spatial momentum we find
\begin{subequations}
\begin{eqnarray}
  && G^R_{J^x \! J^x} = \frac{\bar{n}^{\, 2}}{\bar\epsilon + \bar p} - i\omega\sigma  + \dots \,,\\[5pt]
    \label{eq:TxyTxy-sources}
  && G^R_{T^{xy} T^{xy}} = \bar{p} - i\omega\eta + \dots\,,\\[5pt]
  && \frac{1}{d}\left( G^R_{T^{xx} T^{xx}} + (d{-}1) G^R_{T^{xx} T^{yy}}\right) = 
     -\frac{\bar p}{d} + (\bar\epsilon+\bar{p})\vs^2 - i\omega\zeta + \dots\,.
\end{eqnarray}
\end{subequations}
Remembering the relation (\ref{eq:fdt}) between $G_{ab}$ and $G^R_{ab}$,
this gives precisely the Kubo formulas (\ref{eq:Kubo-currents}).
The variational response functions satisfy
\begin{subequations}
\label{eq:WI-mu-sources}
\begin{eqnarray}
  && \omega G^R_{T^{00} J^0} - k_j G^R_{T^{0j} J^0} = 0\,,\\[5pt]
  && \omega G^R_{T^{00} T^{00}} - k_j G^R_{T^{0j} T^{00}} = -\bar{\epsilon}\,\omega \,,\\[5pt]
  && \omega G^R_{T^{00} T^{0i}} - k_j G^R_{T^{0j} T^{0i}} = \bar{p}\, k_i\,,
\end{eqnarray}
\end{subequations}
with contact terms in the right-hand side.
Comparing with the analogous expressions (\ref{eq:WI-mu}), we see that the contact terms differ
between the response functions evaluated in the canonical approach and in the variational approach.
For a discussion of contact terms in the variational approach, see Ref.~\cite{Herzog:2009xv}.

We conclude the section about correlation functions with a comment about the
choice of variables in the constitutive relations. 
When writing down the constitutive relations in (\ref{eq:const-rel}) or (\ref{eq:const-rel-sources}),
we chose to use the transverse vectors 
$\Delta^{\mu\lambda}\nabla_{\!\lambda} T$ and $\Delta^{\mu\lambda}\nabla_{\!\lambda} \mu$.
However, one could have chosen to work with a different set of transverse vectors,
say with $\Delta^{\mu\lambda}\nabla_{\!\lambda} T$ and $\Delta^{\mu\lambda}\dot{u}_\lambda$,
which is a choice adopted by Eckart~\cite{PhysRev.58.919, weinberg:1972}.
By using the zeroth-order hydrodynamic equations, we find
$$
  j^\mu = \sigma\! \left[E^\mu - T \Delta^{\mu\lambda}\nabla_{\!\lambda}\!\left(\frac{\mu}{T}\right) \right]
  = \sigma \frac{\epsilon + p}{n T}\, \Delta^{\mu\lambda}\left[T \dot{u}_\lambda + \nabla_\lambda T \right] 
    + O(\partial^2)\,.
$$
The latter form of $j^\mu$ is, however, not suitable for studying
hydrodynamic fluctuations in the ``neutral'' state with $\bar n = 0$, hence our preference 
for working with $\Delta^{\mu\lambda}\nabla_{\!\lambda} T$ and $\Delta^{\mu\lambda}\nabla_{\!\lambda} \mu$.

Another place where this choice of variables in the constitutive relations
becomes peculiar are response functions in a state with non-zero $\bar n$,
evaluated from Eckart-frame hydrodynamics.
Focussing on the response function of the transverse momentum density (the shear mode),
and using Eckart-frame constitutive relations (\ref{eq:const-rel-EF}) with
$q^\mu = -\sigma V^\mu \,(\epsilon{+}p)/{n}$ leads to expression~(\ref{eq:T0yT0y-sources}), as before.
On the other hand, using Eckart-frame constitutive relations (\ref{eq:const-rel-EF}) with
$q^\mu = -\kappa \Delta^{\mu\lambda}\left[T \dot{u}_\lambda + \nabla_\lambda T \right]$ leads to
\begin{equation}
\label{eq:T0yT0y-sources-2}
  G^R_{T^{0y} T^{0y}}(\omega,\k) = 
  \frac{\eta \k^2(1+i\omega/\omega_0)}
  {i\omega (1+i\omega/\omega_0) - \gamma_\eta\k^2} + \bar\epsilon\,,
\end{equation}
where $\omega_0 \equiv \bar{n}^2/(\sigma \bar{w})$, 
and we have used the relation between heat and charge conductivities
$\kappa = \sigma (\epsilon{+}p)^2/(n^2 T)$.
There is now a $O(\omega^2)$ term in the denominator of the response function,
however, in the hydrodynamic regime, $\omega/\omega_0 \ll 1$,
and this term should be ignored.%
\footnote{%
	Second-order terms in the constitutive relations will only contribute
	$O(\partial^3)$ terms such as $O(\omega\k^2)$, $O(\omega^3)$
	to the denominator of (\ref{eq:T0yT0y-sources-2}).
}
If one naively proceeds with the analysis of the singularities
of the response function (\ref{eq:T0yT0y-sources-2}), one finds two poles:
the normal shear pole at $\omega_1 = -i\gamma_\eta \k^2 + O(\k^4)$,
and the second pole at $\omega_2 = i \omega_0 + O(\k^2)$.
The second pole is in the upper half-plane of complex $\omega$, and,
if taken seriously, would represent an instability of the thermal equilibrium state~\cite{Hiscock:1985zz}.
The instability is, however, fictitious, as the second pole is outside of the
validity regime of hydrodynamics:
the corresponding time scale $\sigma (\bar{\epsilon} {+} \bar{p})/\bar{n}^2$
is microscopic, not macroscopic (in a gas, this time scale would be determined by the mean-free time
between particle collisions).
Hydrodynamics is only a sensible effective description at time scales much longer
than such microscopic times, and the correct way to interpret the response function (\ref{eq:T0yT0y-sources-2})
is to expand $(\omega{-}\omega_2)$ in the denominator in powers of $\omega$.
This gives
$$
  G^R_{T^{0y} T^{0y}}(\omega,\k) = 
  \frac{\eta \k^2}{i\omega - \gamma_\eta\k^2 +O (\k^4)} 
  \left[1 + O(\k^2) + O(\omega\k^2) + O(\omega^2) \right]
  + \bar\epsilon\,,
$$
and shows that the contribution of the term $\omega/\omega_0$ in (\ref{eq:T0yT0y-sources-2})
is of the same order as contributions from higher-order hydrodynamics.
Therefore, this term should not be retained in response functions
evaluated using first-order hydrodynamics. 
In first-order hydrodynamics, the response function of 
the transverse momentum density is given by Eq.~(\ref{eq:T0yT0y-sources}),
and only has a pole in the lower half-plane of complex $\omega$, as it should.
It happens quite generally that certain terms in response functions evaluated from first-order hydrodynamics
may be ``contaminated'' by higher-derivative corrections. 
Thus, some care needs to be taken
in order to determine the response functions reliably, up to a given order
in the derivative expansion.

\section{Interactions of hydrodynamic modes}
\label{sec:interactions}
\noindent
In the last section, we looked at two-point correlation functions
of $T^{\mu\nu}$ and $J^\mu$ in thermal equilibrium state,
evaluated from linearized hydrodynamics.
What happens to these correlation functions if the non-linearities
in the hydrodynamic equations are taken into account?
While the linear equations describe freely propagating 
hydrodynamic modes, the non-linearities are responsible for the
interactions among the modes. 
These interactions will modify the equilibrium correlation functions
of $T^{\mu\nu}$ and $J^\mu$, and in particular
will modify the transport coefficients $\eta$, $\zeta$, and $\sigma$
compared to their naive ``bare'' values in linearized hydrodynamics.
\begin{figure}
\begin{center}
\includegraphics[width=5cm]{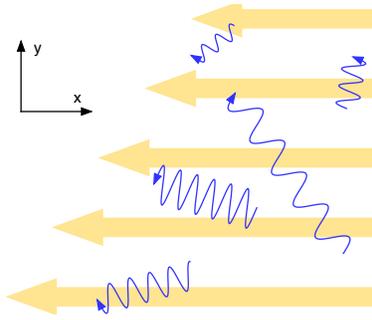}
\end{center}
\caption{
  A fluid flow with an inhomogeneous velocity profile $v_x(y)$
  corresponding to the shear mode,
  represented by thick arrows.
  Thin wiggly arrows represent sound and shear modes generated 
  by thermal fluctuations in fluid elements in local thermal equilibrium.
  These collective excitations can transfer 
  $x$-momentum in the $y$-direction,
  thus contributing to shear viscosity.
}
\label{fig:shear-wave-2}
\end{figure}

The modification of viscosity compared to its naive value can be understood as follows.
An inhomogeneous velocity profile $v_x(y)$ of the shear mode
will eventually equilibrate due to particles transferring $x$-momentum
between the layers of fluid, as shown in Fig.~\ref{fig:shear-flow}.
However, momentum can also be transferred by collective excitations
such as sound and shear modes, with wavelength shorter than the 
length scale of the inhomogeneity of the flow, as illustrated in 
Fig.~\ref{fig:shear-wave-2}.
Such collective excitations will be present even in thermal equilibrium
due to fluctuations, hence the shear viscosity measured for example
from equilibrium two-point correlation functions will receive contributions
from the collective modes. 
To estimate the contribution to $\eta$ from the sound waves, 
one can think about the short-wavelength sound waves as particles,
whose ``mean free path'' (propagation distance)
is proportional to $1/(\gamma_s k^2)$,
as is evident from the sound wave dispersion relation 
$\omega = \pm k \vs -i\gamma_s k^2/2$.
Hence the contribution to $\eta$ from sound waves
is proportional to $\int\! d^dk /(\gamma_s k^2)$,
in other words is inversely proportional to $\eta$ itself. 
The momentum integral is UV divergent at large $k$
and needs to be cut off at some $k_{\rm max}$,
which is the maximum value of $k$ at which the hydrodynamic
picture of collective excitations is still applicable. 

In this section, we will give a simple derivation of 
how two-point correlation functions of conserved currents are modified
by the non-linear terms in the hydrodynamic equations, and estimate
the corresponding fluctuation correction to the shear viscosity.
We will also see that hydrodynamic fluctuations 
will lead to the breakdown of the derivative expansion in hydrodynamics.
Put differently, second-order hydrodynamics (taken as a set of classical
partial differential equations) does not correctly describe the
relaxation of $T^{\mu\nu}$ and $J^\mu$ in the hydrodynamic regime.
In the context of non-relativistic hydrodynamics, the problem
with the derivative expansion was known for a long time~\cite{DeSchepper19741},
and relativistic hydrodynamics is no different in this respect. 
It is important to emphasize that the physics of hydrodynamic fluctuations
is only partially contained in the equations of hydrodynamics
if the latter are taken merely as classical partial differential equations. 
This is because the classical equations of hydrodynamics ignore
the excitation of collective modes by thermal fluctuations: 
the classical equations contain only the ``dissipation'' part
of the fluctuation-dissipation theorem~\cite{LL9}.

\subsection{Correlation functions of conserved currents}
\label{sec:loops}
\noindent
Following Ref.~\cite{Arnold:1997gh, Kovtun:2003vj}, we will evaluate
the effect of the non-linear terms in the hydrodynamic equations
on the correlation functions of spatial currents $J^i$ and stresses $T^{ij}$
in the hydrodynamic regime $\omega\to0$ and $\k\to0$.
For simplicity, we will do this in a scale-invariant theory,
in the equilibrium state with $\mu=0$. 
Choosing the hydrodynamic variables as $\delta\epsilon=T^{00}-\bar\epsilon$,
$\pi^i = T^{0i}$, and $n=J^0$, the constitutive relations in the Landau frame 
take the following form to quadratic order in fluctuations:
\begin{eqnarray*}
  && T^{ij} = \left(\bar{p} + \vs^2 \delta\epsilon \right) \delta^{ij} 
              - \gamma_\eta \!\left( \partial^i \pi^j + \partial^j \pi^i - 
                {\textstyle\frac2d}\, \delta^{ij} \partial_k \pi^k  \right) 
              + \frac{1}{\bar{w}}H^{ij}_{kl}\, \pi^k \pi^l + \dots\,,\\
  && J^i = -D\,\partial^i n + \frac{1}{\bar{w}} n \pi^i + \dots \,.
\end{eqnarray*}
Here $\bar{w}=(\bar{\epsilon} + \bar{p})$ is the equilibrium enthalpy density,
$D=\sigma/\chi$ is the diffusion constant for charge density, and
$\gamma_\eta=\eta/{\bar w}$ is the diffusion constant for the transverse momentum density.
The speed of sound is $\vs^2=1/d$,
and the bulk viscosity vanishes due to scale invariance.
The projector onto traceless symmetric tensors is 
$H^{ij}_{kl} = \frac12 \delta^i_k \delta^j_l + \frac12 \delta^i_l \delta^j_k -\frac1d \delta^{ij} \delta_{kl}$.
The dots denote the terms which are suppressed either due to
containing higher powers of the hydrodynamic variables,
or due to containing extra derivatives of the hydrodynamic variables.
The contributions to the equilibrium correlation functions from the quadratic terms 
in the constitutive relations thus have the form
\begin{eqnarray*}
  && G_{J_i J_k}^{(2)}(t,\x) = \frac{1}{\bar{w}^{\,2}}
     \langle n(t,\x)\, \pi^i(t,\x)\, n(0)\, \pi^k(0)\rangle\,,\\
  && G_{T_{ij} T_{kl}}^{(2)}(t,\x) = \frac{1}{\bar{w}^{\,2}}
     H^{ij}_{mn} H^{kl}_{pq}\,\langle\pi^m(t,\x)\, \pi^n(t,\x)\,\pi^p(0)\,\pi^q(0)\rangle\,.
\end{eqnarray*}
Assuming that the small equilibrium fluctuations are Gaussian and therefore
can be factorized, we have for the quadratic contributions to the
connected correlation functions
\begin{eqnarray*}
  && G_{J_i J_k}^{(2)}(\omega,\k) = \frac{1}{\bar{w}^{\,2}}
     \int\! \frac{d\omega'}{2\pi} \frac{d^dk'}{(2\pi)^d}\,
     G_{nn}^0(\omega',\k')\, G_{\pi_i \pi_k}^0(\omega{-}\omega',\k{-}\k')\,,\\[5pt]
  && G_{T_{ij} T_{kl}}^{(2)}(\omega,\k) = \frac{2}{\bar{w}^{\,2}}
     H^{ij}_{mn} H^{kl}_{pq}\,
     \int\! \frac{d\omega'}{2\pi} \frac{d^dk'}{(2\pi)^d}\,
     G_{\pi_m \pi_p}^0(\omega',\k')\, G_{\pi_n \pi_q}^0(\omega{-}\omega',\k{-}\k')\,.
\end{eqnarray*}
Here $G_{nn}^0(\omega,\k)$ and $G_{\pi_i \pi_j}^0(\omega,\k)$
are the symmetrized (or unordered) correlation functions,
obtained in the linear response theory.
Explicitly, relations~(\ref{eq:GRnn}), (\ref{eq:GRpiipij}), and (\ref{eq:fdt}) give
\begin{subequations}
\begin{eqnarray}
\label{eq:Gnn0}
  && G_{nn}^0(\omega,\k) = \frac{2TD\chi\k^2}{\omega^2 + (D\k^2)^2}\,,\\[5pt]
  && G_{\pi_i \pi_j}^0 (\omega,\k) = 
     \left( \delta_{ij} - \frac{k_i k_j}{\k^2} \right)
     \frac{2T\gamma_\eta \bar{w}\,\k^2}{\omega^2 + (\gamma_\eta \k^2)^2} + 
     \frac{k_i k_j}{\k^2} 
     \frac{2T \gamma_s \bar{w}\, \k^2 \omega^2}{(\omega^2-\k^2\vs^2)^2 + (\omega\gamma_s \k^2)^2}\,.
\label{eq:Gpipi0}
\end{eqnarray}
\end{subequations}
Let us first look at the current-current correlation function $G_{J_i J_k}^{(2)}(\omega,\k)$. 
There are two contributions, one of the ``shear-shear'' type, 
and one of the ``shear-sound'' type. 
After performing the frequency integrals, one is left with momentum integrals
which are linearly divergent in $d=3$, and logarithmically divergent in $d=2$.
Introducing the high-momentum cutoff $\Lambda$, one finds at zero momentum 
and small frequency in $d=3$
\begin{eqnarray*}
  && G_{J_i J_k}^{(2)}(\omega,\k{=}0) = 
     O\!\left( \Lambda \right) - 
     \delta_{ij} \frac23
     \frac{ (2T\chi) (2T \bar{w})}{4\pi \bar{w}^{\,2}}
     \frac{|\omega|^{1/2} }{[2(D{+}\gamma_\eta)]^{3/2}}
     + \dots
\end{eqnarray*}
where $\dots$ denote the terms analytic in $\omega$.
Only the ``shear-shear'' contribution gives rise to the non-analytic $|\omega|^{1/2}$ term.
Using the Kubo formula~(\ref{eq:Kubo-sigma}),
the $O(\Lambda)$ term can be interpreted as the contribution of shear waves to 
d.c. conductivity, while the $|\omega|^{1/2}$ term is the cutoff-independent contribution
of shear waves to $\sigma(\omega)$.
The cutoff $\Lambda$ is the momentum scale at which hydrodynamics ceases being a good 
effective description.
Its value can be estimated from the condition that 
one-derivative terms in the hydrodynamic expansion are comparable to zero-derivative terms,
$\Lambda \gamma_\eta\sim1$.
The non-analytic $|\omega|^{1/2}$ term corresponds to the $t^{-3/2}$ fall-off of the correlation function
in real time.
This phenomenon of ``long-time tails'' in correlation functions of conserved currents
has been known for a long time \cite{Pomeau:1974hg},
and, as is evident from the derivation, is not specific to relativistic hydrodynamics.
In $d=2$, one finds at zero momentum and small frequency
\begin{eqnarray*}
  && G_{J_i J_k}^{(2)}(\omega,\k{=}0) = 
     O\!\left( \ln \Lambda \right) + 
     \delta_{ij} \frac12
     \frac{ (2T\chi) (2T \bar{w})}{8\pi \bar{w}^{\,2} (D{+}\gamma_\eta)}
     \ln \frac{(D{+}\gamma_\eta) \Lambda^2}{|\omega|}
     + O(\omega^2 \ln(\Lambda^2/\omega^2))
\end{eqnarray*}
Again, the $O(\ln\Lambda)$ term can be interpreted as the contribution of shear waves to 
d.c. conductivity, while the $\ln(\Lambda^2/|\omega|)$ term can be interpreted 
in the sense of the renormalization group as the 
logarithmic contribution of shear waves to the ``running conductivity''.

For the correlation function of two stress tensors one finds similar expressions,
\begin{eqnarray*}
   & 
{\displaystyle
   G_{T_{ij} T_{kl}}^{(2)}(\omega,\k{=}0) = 
   O(\Lambda) - 
   H^{ij}_{kl}\, \frac{T^2}{60\pi}\left[ 7+ \left(\coeff32\right)^{3/2}\right]
   \frac{|\omega|^{1/2}}{\gamma_\eta^{3/2}} + \dots\,,
}
   &
   d=3\\[5pt]
   &
{\displaystyle 
   G_{T_{ij} T_{kl}}^{(2)}(\omega,\k{=}0) = 
   O(\ln \Lambda) + 
   H^{ij}_{kl}\, \frac{T^2}{4\pi \gamma_\eta} \ln\frac{\gamma_\eta \Lambda^2}{|\omega|} + \dots\,,
}
   &
   d=2
\end{eqnarray*}
Again, using the Kubo formula (\ref{eq:Kubo-eta}), the cutoff-dependent terms
can be interpreted as contributions of shear and sound waves to shear viscosity. 

\begin{figure}
\begin{center}
\includegraphics[width=8cm]{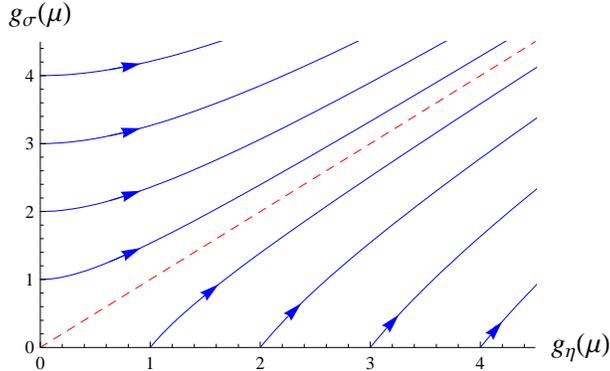}
\end{center}
\caption{
  The flow diagram for the differential equations (\ref{eq:rg-2d}), pictured
  in terms of the dimensionless variables $g_\eta = \eta/s$ and $g_\sigma = \sigma T/\chi$, where
  $\eta$ is the shear viscosity, $\sigma$ is the charge conductivity, $s$ is the
  equilibrium density of entropy, and $\chi$ is the equilibrium charge susceptibility.
  The arrows indicate the direction of decreasing $\mu$ (towards lower frequency).
  The dashed line is a straight line with a unit slope,
  indicating the infrared asymptotics as $\mu{\to}0$.
}
\label{fig:rg-flow-2des}
\end{figure}

In $d{=}2$ spatial dimensions, one can define the ``running conductivity''
$\sigma(\mu)\equiv\sigma(\omega{=}\mu)$, and ``running viscosity'' $\eta(\mu)\equiv\eta(\omega{=}\mu)$.
Demanding that the correlation functions do not depend on $\mu$ gives the following
renormalization group equations for $\sigma(\mu)$ and $\eta(\mu)$
\begin{subequations}
\label{eq:sigma-eta-RG}
\begin{eqnarray}
\label{eq:sigma-RG}
  && \mu \frac{\partial\sigma}{\partial\mu} 
     = -\frac{T}{8\pi} \frac{\bar\chi^{\, 2}}{\sigma \bar w + \eta\bar\chi}\,,\\[5pt]
\label{eq:eta-RG}
  && \mu \frac{\partial\eta}{\partial\mu}
     = -\frac{T}{16\pi} \frac{\bar w}{\eta}\,,
\end{eqnarray}
\end{subequations}
where $\bar\chi=(\partial \bar n/\partial\mu)_{\mu=0}$ is the equilibrium charge susceptibility,
and $\bar w = \bar\epsilon{+}\bar p$ is the equilibrium enthalpy density
[the sliding scale $\mu$ in Eq.~(\ref{eq:sigma-eta-RG}) is not to be confused with the chemical potential].
We can rewrite these equations in terms of dimensionless quantities
$g_\eta\equiv \eta/s$ and $g_\sigma\equiv \sigma T/\bar\chi$:
\begin{subequations}
\label{eq:rg-2d}
\begin{eqnarray}
  && \mu \frac{\partial g_\sigma}{\partial\mu} 
     = -\frac{1}{8\pi c} \frac{1}{g_\sigma + g_\eta}\,,\\[5pt]
  && \mu \frac{\partial g_\eta}{\partial\mu}
     = -\frac{1}{16\pi c} \frac{1}{g_\eta}\,,
\end{eqnarray}
\end{subequations}
where the constant $c$ is defined by the equilibrium entropy density as $s = c\, T^2$,
in other words $c$ measures the number of degrees of freedom in the system.
Note that the right-hand side of these renormalization-group equations is small in the large-$N$
limit, and the running of $\sigma$ and $\eta$ disappears as $N{\to}\infty$.
The flow diagram for equations~(\ref{eq:rg-2d}) is shown in Fig.~\ref{fig:rg-flow-2des}.
Both $\eta$ and $\sigma$ grow in the infrared,
as happens quite generally in two-dimensional hydrodynamics~\cite{KLS-1984}.
In the limit $\mu\to0$ the solutions asymptote to $g_\eta = g_\sigma$, in other words
\begin{equation}
\label{eq:etasigma2d}
  \frac{\eta}{s} = \frac{\sigma T}{\chi}.
\end{equation}
This shows that in the extreme low-frequency limit,
$\eta$ and $\sigma$ cease being independent transport coefficients in $d=2$.

\subsection{Breakdown of second-order hydrodynamics}
\label{sec:second-order}
\noindent
The non-analytic behavior of correlation functions of conserved currents at small frequency 
(proportional to $|\omega|^{1/2}$ in $d{=}3$ and $\ln |\omega|$ in $d{=}2$)
has implications for the derivative expansion in hydrodynamics.
As indicated in section \ref{sec:relativistic-hydro},
the derivative expansion is constructed by expanding the 
energy-momentum tensor and the current
order by order in derivatives of the hydrodynamic variables $T$, $\mu$, and $u^\mu$.
For example, the shear viscosity arises as the coefficient of 
the one-derivative tensor $\sigma^{\mu\nu}$ in the expansion of $T^{\mu\nu}$.
At second order in derivatives, there are other contributions to $T^{\mu\nu}$,
for example $\sigma^{\mu\nu} \partial_\lambda u^\lambda$.
References~\cite{Baier:2007ix, Bhattacharyya:2008jc, Romatschke:2009kr, Bhattacharyya:2012ex}
classify possible second-order terms in relativistic hydrodynamics.

The second-order terms will modify the linear-response correlation functions of $T_{\mu\nu}$ and $J_\mu$
evaluated in Sec.~\ref{sec:linear-response}.
Repeating the calculation in Section~\ref{sec:variational} changes Eq.~(\ref{eq:TxyTxy-sources})
to \cite{Baier:2007ix}
\begin{equation}
\label{eq:TxyTxy-sources-2}
  G^R_{T_{xy} T_{xy}} (\omega,\k{=}0) = \bar p - i\omega\eta + \eta \tau_\Pi \omega^2 + \dots
\end{equation}
where $\tau_\Pi$ is the Israel-Stewart stress relaxation time~\cite{Israel-Stewart}, and the dots
denote contributions from other second-order transport coefficients.
The derivative expansion is based on the assumption that higher-order terms
are smaller than the lower-order terms, and indeed this is true in Eq.~(\ref{eq:TxyTxy-sources-2}),
in the hydrodynamic limit $\omega{\to}0$.

We saw that in $d{=}3$ spatial dimensions, hydrodynamic fluctuations give rise 
to a non-analytic $\omega^{3/2}$ contribution to $G^R_{T_{xy} T_{xy}}(\omega,\k{=}0)$.
In the hydrodynamic limit, this is smaller than the first-order contribution $O(\omega)$,
but is {\it larger} than the second-order contribution $O(\omega^2)$.
Such non-analytic terms can not be represented by a local derivative expansion,
which means that the derivative expansion is 3+1 dimensional hydrodynamics breaks down at second order. 
This problem with the derivative expansion has been known for a long time~\cite{DeSchepper19741},
and is not specific to relativistic hydrodynamics.
See Ref.~\cite{Kovtun:2011np} for a discussion of this problem in the context of the quark-gluon plasma
in heavy-ion collisions.

We emphasize that this problem with the derivative expansion in hydrodynamics
is not an intrinsic problem with hydrodynamic equations themselves.
The equations of hydrodynamics
can be studied as partial differential equations on their own right at second
(or higher) order in the derivative expansion. 
The problem is that these partial differential equations do not necessarily capture
the correct behavior of fluids in the hydrodynamic regime.
Viewing hydrodynamics merely as a set of partial differential equations
misses information about hydrodynamic fluctuations which render these equations physically invalid as $\omega{\to}0$.
A complete description of fluids should supplement the hydrodynamic equations with
a prescription of how to deal with fluctuations.
This is the subject of the next section.

\section{Hydrodynamics as a field theory}
\label{sec:effective-action}
\noindent
So far, we looked at two methods of computing
correlation functions of conserved densities
in the hydrodynamic regime:
the standard linear-response formulation in 
Sections~\ref{sec:mu0}, \ref{sec:mu1},
and the variational formulation in Section~\ref{sec:variational}.
There is another way to look at equilibrium real-time correlation
functions, where the dissipation is attributed
to small-scale random stresses and random currents
in thermal equilibrium~\cite{LL-1957, LL9}.
This formalism has an added advantage that 
the interactions of the hydrodynamic modes can be treated in a
systematic way~\cite{Martin:1973zz},
so that both the linear response results of Section~\ref{sec:linear-response},
and the interaction results of Section~\ref{sec:interactions}
are included within the same framework.
See for example~\cite{Hohenberg:1977ym, PhysRevB.18.353, Khalatnikov:1983ak}.
The stochastic equations involving random stresses and random currents
can be converted to a functional integral,
which allows one to view hydrodynamics not just as a low-energy effective description,
but as a low-energy effective {\it field theory}.
We will start in Sections~\ref{sec:diff} and~\ref{sec:hydro}
with equilibrium correlation functions
derived from linearized relativistic hydrodynamics, which can be viewed as
propagators for the hydrodynamic perturbation theory.
Then in Section~\ref{sec:FNS}, we will discuss interactions between the
hydrodynamic modes in a simplified model of Ref.~\cite{Forster:1977zz}.

\subsection{Simple diffusion}
\label{sec:diff}
\noindent
Just like we did in Section~\ref{sec:linear-response},
it is instructive to look at a simple example
of charge diffusion before studying the full hydrodynamics.
Beyond being an interesting example by itself, it will also serve
as a basic building block for the future perturbative treatment
of fluctuations.

For the simple diffusive process studied in Section~\ref{sec:diffusion},
we have found for the retarded function $G^R_{nn}(\omega,\k)$
and the symmetrized function $G_{nn}(\omega,\k)$
\begin{equation}
\label{eq:diff-corr}
  G^R_{nn}(\omega,\k) = \frac{D\chi\,\k^2}{i\omega - D\k^2}\,,\ \ \ \ 
  G_{nn}(\omega,\k) = \frac{2T D \chi\,\k^2}{\omega^2 + (D\k^2)^2}\,.
\end{equation}
Here $D$ is the diffusion constant, and
$\chi$ is the static charge susceptibility,
$\chi \equiv (\partial n/\partial\mu)_{\mu=0}$.
The two functions are not independent, but rather are related by the
fluctuation-dissipation theorem (\ref{eq:fdt}).
One can ask the following question: given the hydrodynamic correlation functions
(\ref{eq:diff-corr}) of the charge density $n(t,\x)$,
what is the effective action $S_{\rm eff}[n]$ that would produce them? 
Let us now show how one can reproduce correlation functions (\ref{eq:diff-corr})
from a functional integral formalism, where the charge density
$n(t,\x)$ plays the role of a ``quantum field''.

\subsubsection*{Stochastic model for diffusion}
\noindent
The diffusion equation (\ref{eq:diffusion-equation})
is a consequence of the current conservation $\partial_\mu J^\mu=0$,
where the charge density is $J^0=n$, and the constitutive relation
for the spatial current is $J_k = -D\partial_k n$.
The diffusion equation is only an effective description
at long distance and time scales.
In order to model the microscopic effects, we introduce random currents
$r_k$ in equilibrium, so that the linearized constitutive relation 
for the current takes the form
$
  J_k = - D\partial_k n + r_k\,.
$
We take the microscopic currents to only have Gaussian short-distance correlations,
so that
\begin{equation}
\label{eq:current-noise}
  \langle r_i(\x,t)\, r_k(\x',t') \rangle = C\delta_{ik} 
  \delta(\x{-}\x') \delta(t{-}t')\,.
\end{equation}
where $C=2TD\chi$ determines the strength of the ``noise''.
The diffusion equation becomes a Langevin-type equation:
\begin{equation}
\label{eq:diff-eom}
  -\partial_t n + D\grad^2 n + \theta(\x,t) = 0\,,
\end{equation}
with a random Gaussian field $\theta = -\partial_k r_k$,
\begin{equation}
\label{eq:diff-noise}
  \langle \theta(\x,t)\, \theta(\x',t') \rangle = 
  -2T\chi\,D\grad^2 \delta(\x{-}\x') \delta(t{-}t')\,.
\end{equation}
The angular brackets in (\ref{eq:diff-noise})
denote the average over the noise, in other words
$$
  \langle \dots \rangle = \int\!\!\D\theta\, e^{-W_n[\theta]} \dots\,.
$$
For the chosen noise correlations given by Eq.~(\ref{eq:diff-noise}),
the weight factor is
$$
  W_n[\theta] = \frac12 \int\!\! dt\,d^d\!x\,d^d\!x'\; \theta(\x,t) D(\x,\x') \theta(\x',t)\,,
$$
where $D(\x,\x')$ satisfies 
$$
  -C \grad^2 D(\x,\x') =  \delta(\x{-}\x')\,.
$$

\subsubsection*{Effective action for diffusion}
\noindent
One way to compute the correlation functions is to solve
Eq.~(\ref{eq:diff-eom}) and express the density in terms of the noise $n=n_\theta(\x,t)$,
then take the product of the solutions $n_\theta(\x_1,t_1) n_\theta(\x_2,t_2) ...$,
and average it over the noise.
Alternatively, we can pick out the solutions to Eq.~(\ref{eq:diff-eom})
by integrating over all possible $n(\x,t)$, with a functional delta function
which enforces Eq.~(\ref{eq:diff-eom}):
$$
  \langle n(\x_1,t_1)\, n(\x_2,t_2) ... \rangle = \int\!\!\D n\,
  \langle \delta({\rm e.o.m.}) \rangle\, J\,
  n(\x_1,t_1)\, n(\x_2,t_2) ...
$$
where ``e.o.m.'' stands for the left-hand side of the ``equation of motion'' (\ref{eq:diff-eom}), and 
$J={\det}(\delta({\rm e.o.m.})/\delta n)$ is the Jacobian.
For a linear equation such as the diffusion equation,
the Jacobian does not depend on the fields, and can be absorbed into
the integration measure.
We represent the functional delta function by introducing an auxiliary field
$\rho(\x,t)$, namely
$\delta({\rm e.o.m.}) = \int \D\rho\, \exp\left( i\int ({\rm e.o.m.})\rho \right)$.
Now the integral over the noise $\theta(\x,t)$ can be performed
because it is Gaussian,
giving the following representation of the correlation functions:
$$
  \langle n(\x_1,t_1)\, n(\x_2,t_2) ... \rangle = \int\!\!\D n\, \D\rho\;
  e^{-S_{\rm eff}[\rho,n]} n(\x_1,t_1)\, n(\x_2,t_2) ...\,,
$$
where the action is
\begin{equation}
\label{eq:diff-action-1}
  S_{\rm eff}[\rho,n] = \int\!\!dt\,d^d x \left[
                \frac{C}{2} (\partial_i\rho)^2
              - i\rho\left( \partial_t n  - D\grad^2 n \right)
              \right]\,.
\end{equation}
This gives the functional integral representation of correlation functions
for diffusion.
We can also integrate out the auxiliary field $\rho$ to end up
with the action which depends on $n(\x,t)$ only:
$$
  \langle n(\x_1,t_1)\, n(\x_2,t_2) ... \rangle = \int\!\!\D n\;
  e^{-S_{\rm eff}[n]} n(\x_1,t_1)\, n(\x_2,t_2) ...\,.
$$
The effective action for the density fluctuations is
\begin{equation}
\label{eq:diff-action-2}
  S_{\rm eff}[n] = \frac12 \int\!\!dt\,d^d\!x\,d^d\!x'\;
  B(t,\x) D(\x,\x') B(t,\x')\,,
\end{equation}
where $B(t,\x) \equiv (\partial_t n  - D\grad^2 n)$\,.
This effective action does not involve any auxiliary fields,
it is both real and time-reversal invariant,
however it is non-local in space.
One can use either one of the actions
(\ref{eq:diff-action-1}) or (\ref{eq:diff-action-2})
to compute correlation functions.
We emphasize that the effective action (\ref{eq:diff-action-2}) 
describing diffusion of a conserved density
is not meant to be the classical action from which one obtains
the ``equation of motion'' (\ref{eq:diffusion-equation}) via the
standard Euler-Lagrange variational method.
Rather, it is the effective action that is to be used
in the generating functional for real-time correlation functions
of the conserved charge density $n(t,\x)$ in thermal equilibrium.

\subsubsection*{Correlation functions}
\noindent
What are the correlation functions that we compute in our
effective theory (\ref{eq:diff-action-1})?
In other words, what do they correspond to in the operator formalism?
It is natural to identify $\langle n(\x_1,t_1)\, n(\x_2,t_2) \rangle$
evaluated in the effective theory (\ref{eq:diff-action-1})
with the symmetrized (or unordered) density-density correlation function:
$$
  G_{n n}(t_1{-}t_2,\x_1{-}\x_2) = \langle n(\x_1,t_1)\, n(\x_2,t_2) \rangle\,.
$$
This is consistent with the fact that in the path integral formulation,
the order of the fields does not matter, in other words
$\langle n(\x_1,t_1)\, n(\x_2,t_2) \rangle = \langle n(\x_2,t_2)\, n(\x_1,t_1) \rangle$.
With this identification, it also important that the effective action
(\ref{eq:diff-action-2}) is time-reversal invariant.
This is because time-reversal invariant microscopic dynamics must give rise to
time-reversal invariant unordered density-density function.
However, the action (\ref{eq:diff-action-1}) need not be time-reversal invariant
because it also contains information about the mixed $\langle n\rho\rangle$
correlation functions, and those are not time reversal invariant.

Finally, note that $n(t,\x)$ is a real field, and its correlation functions
come from the real action, Eq.~(\ref{eq:diff-action-2}).
Therefore, $G_{n n}(t,\x)$ is real, or
$G_{n n}(\omega,\k)^* = G_{n n}(-\omega,-\k)$.
Time-reversal invariance of the action (\ref{eq:diff-action-2})
combined with rotation invariance gives
$G_{nn}(\omega,\k) = G_{n n}(-\omega,-\k)$,
in other words $G_{n n}(\omega,\k)$ must be real.

Let us evaluate the correlation functions in the effective theory (\ref{eq:diff-action-1}).
For a quadratic action with several real fields $\phi^a(\x,t)$,
we can Fourier transform the fields and write the action as
$$
  S = \frac12 \int_{\omega,\k} \phi^{a\ *}_{\omega,\k}\, P_{a b}(\omega,\k) \phi^b_{\omega,\k}\,.
$$
The two-point correlation functions are then given by
$$
  G_{a b}(\omega,\k) = (P^{-1})_{a b}\,.
$$
We apply this to the effective action (\ref{eq:diff-action-1}) and read off
the correlation functions:
\begin{eqnarray*}
  && G_{n n}(\omega,\k) = \frac{2T\chi D\k^2}{\omega^2 + (D\k^2)^2}\,, \\[5pt]
  && G_{n\rho}(\omega,\k) = \frac{-1}{\omega + i D\k^2}\,, \\[5pt]
  && G_{\rho n}(\omega,\k) = \frac{1}{\omega - i D\k^2}\,, \\ [5pt]
  && G_{\rho\rho}(\omega,\k) = 0\,.
\end{eqnarray*}
The $G_{n n}$ correlation function has precisely the expected form
for the linear response density-density correlation function~(\ref{eq:diff-corr}).
The mixed correlation functions
$G_{n\rho}(t,\x) = \langle n(t,\x) \rho(0)\rangle$ and
$G_{\rho n}(t,\x) = \langle \rho(t,\x) n(0)\rangle$
look almost like the retarded and advanced functions, and are not independent from $G_{n n}$.
To see this, we perform a variable change in the functional integral as
$\rho(t,\x) = \psi(t,\x) + i\int\! d^d x'\, D(\x,\x') B(\x',t)$,
and one finds
$$
  \langle n(t_1,\x_1) \rho(t_2,\x_2) \rangle =
  i\int\!\! d^d x'\, D(\x_2,\x') \langle n(t_1,\x_1) B(t_2,\x') \rangle\,.
$$
The Fourier transform of this relation is
\begin{equation}
\label{eq:diff-FDT-2}
  G_{n\rho}(\omega,\k) = i\frac{i\omega + D\k^2}{C\k^2} G_{nn}(\omega,\k)\,.
\end{equation}
Thus we identify $-iD\chi\k^2 G_{n\rho}(\omega,\k)$ with $G^R_{nn}(\omega,\k)$,
and $-iD\chi\k^2 G_{\rho n}(\omega,\k)$ with $G^A_{nn}(\omega,\k)$.
The relation (\ref{eq:diff-FDT-2}) now becomes the fluctuation-dissipation theorem
(\ref{eq:fdt}).

\subsection{Linearized hydrodynamics}
\label{sec:hydro}
\noindent
Now that we have worked out correlation functions for the linear diffusion process,
we can repeat the calculations for linearized hydrodynamics.
For simplicity, let us look at relativistic hydrodynamics 
at zero chemical potential, studied in Section~\ref{sec:mu0}.

\subsubsection*{Stochastic model for linearized hydrodynamics}
\noindent
The hydrodynamic equations are only an effective description
at long distance and time scales.
In order to model the microscopic effects, we introduce random stresses
$\tau_{ij}$ in equilibrium, so that the linearized constitutive relation 
for the stress tensor takes the form
$$
  T_{ij} = \delta_{ij}\! \left( \bar p + \vs^2 \delta\epsilon \right)
  -\gamma_\eta\! \left( \partial_i\pi_j + \partial_j\pi_i - \frac{2}{d}\delta_{ij}\partial_k\pi_k \right)
  -\gamma_\zeta \delta_{ij} \partial_k\pi_k + \tau_{ij}\,.
$$
As before, $\gamma_\eta\equiv\eta/\bar{w}$, $\gamma_\zeta\equiv \zeta/\bar{w}$, and
$\bar{w}\equiv \bar\epsilon {+} \bar p$ is the equilibrium enthalpy density,
which plays the role of the static susceptibility for momentum density fluctuations.
We take the microscopic stresses to have Gaussian short-distance correlations,
$$
  \langle \tau_{ij}(\x,t) \tau_{kl}(\x',t')\rangle = 
  2T\bar{w} \left[
  \gamma_\eta\! \left(\delta_{ik}\delta_{jl} + \delta_{il} \delta_{jk} -
               \frac2d \delta_{ij}\delta_{kl}\right)
  +\gamma_\zeta \delta_{ij}\delta_{kl}
  \right]\!
  \delta(\x{-}\x')\delta(t{-}t')\,.
$$
The hydrodynamic equations become
\begin{eqnarray}
\label{eq:hydro-e}
  && \partial_t \epsilon = -\partial_k \pi_k\,,\\ [5pt]
\label{eq:hydro-pi}
  && \partial_t \pi_i = -\vs^2 \partial_i\epsilon + {\cal M}_{ij}\pi_j + \xi_i(t,\x)\,,
\end{eqnarray}
where
$\xi_i = -\partial_j \tau_{ij}$ is the momentum density noise, 
${\cal M}_{ij}\equiv \gamma_\eta (\grad^2\delta_{ij} - \partial_i\partial_j) + \gamma_s \partial_i\partial_j$,
and $\gamma_s\equiv \gamma_\zeta + \frac{2d-2}{d}\gamma_\eta$ as before.
Note that there are no noise terms in the equation of energy conservation.
The noise $\xi_i(t,\x)$ has local Gaussian correlations,
$$
  \langle \xi_i(t,\x) \xi_j(t',\x') \rangle  = 
  C_{ij} \delta(\x{-}\x') \delta(t{-}t')\,,
$$
where $C_{ij} = -2T\bar{w} {\cal M}_{ij}$.
The average over the noise $\xi_i$ is performed with the weight
$$
  W_\pi[\xi] = \frac12 \int\!\! dt\,d^d\!x\,d^d\!x'\; \xi_i(\x,t) D_{ij}(\x,\x') \xi_j(\x',t)\,,
$$
where $D_{ij}(\x,\x')$ satisfies
$$
  C_{ij} D_{jk}(\x,\x') = \delta_{ik} \delta(\x{-}\x')\,.
$$

\subsubsection*{Effective action for linearized hydrodynamics}
\noindent
As before, we can represent correlation functions with a functional integral
over the hydro variables, which in this case are $\epsilon$ and $\pi^i$.
Again, we start with
$$
  \langle \epsilon(t_1,\x_1) \dots \rangle  =
  \int\!\! \D\epsilon\,\D{\bm\pi} \langle \delta({\rm e.o.m.}) \rangle \,\epsilon(t_1,\x_1) \dots
$$
where the delta function enforces the hydro equations
(\ref{eq:hydro-e}) and (\ref{eq:hydro-pi}).
We introduce an auxiliary field $\beta$ for energy conservation,
and auxiliary field $\lambda_i$ for momentum conservation.
Performing the integral over the noise,
we find the following representation of the correlation functions:
$$
   \langle \pi_k(t_1,\x_1) \pi_l(t_2,\x_2) \dots \rangle  =
   \int\!\! \D\epsilon\,\D{\bm\pi}\,\D\beta\,\D{\bm\lambda}\;
   e^{-S_{\rm eff}[\epsilon,\pi,\beta,\lambda]}\;
   \pi_k(t_1,\x_1) \pi_l(t_2,\x_2)  \dots
$$
where the effective action is
\begin{equation}
\label{eq:hydro-action-1}
  S_{\rm eff}[\epsilon,\pi,\beta,\lambda] =
  \int\!\!dt\,d^dx\, \left(
  \frac12 \lambda_i C_{ij}\lambda_j - i\lambda_k F_k
  -i\beta(\partial_t\epsilon{+}\partial_k\pi_k)
  \right)\,.
\end{equation}
Here $F_i\equiv \partial_t\pi_i + \vs^2\partial_i\epsilon - {\cal M}_{ij}\pi_j$.
Again, we can integrate out the auxiliary field $\lambda_i$,
which gives a representation of correlation functions
with a non-local action,
$$
   \langle  \pi_k(t_1,\x_1) \pi_l(t_2,\x_2) \dots \rangle  =
   \int\!\! \D\epsilon\,\D{\bm\pi}\,\D\beta\;
   e^{-S_{\rm eff}[\epsilon,\pi,\beta]}\;
   \pi_k(t_1,\x_1) \pi_l(t_2,\x_2) \dots
$$
where
\begin{equation}
\label{eq:hydro-action-2}
    S_{\rm eff}[\epsilon,\pi,\beta] = \frac12 \int\!\!dt\,d^d\!x\,d^d\!x'\;
    F_i(t,\x) D_{ij}(\x,\x') F_j(t,\x')
    - i \int\!\!dt\,d^d\!x\; \beta(\partial_t\epsilon {+} \partial_k\pi_k)\,.
\end{equation}
Again, we identify two-point functions of the hydro variables
evaluated with the action (\ref{eq:hydro-action-2}) with the
unordered (or symmetrized) correlation functions of the corresponding
operators in the microscopic theory.
The fact that the noise strength $C_{ij}$ is proportional to ${\cal M}_{ij}$
ensures that the effective action (\ref{eq:hydro-action-2})
is time-reversal invariant.

\subsubsection*{Correlation functions}
\noindent
We can evaluate the correlation functions either from the action
(\ref{eq:hydro-action-1}) or from the action (\ref{eq:hydro-action-2}).
Let's work them out using the action (\ref{eq:hydro-action-2}).
To do so, we express the action in terms of the Fourier components
of the fields, and express the fluctuation in energy density as 
$\delta\epsilon = k_i\pi_i/\omega$.
The function $F_i(\omega,\k)$ can now be expressed in terms of the
momentum density alone,
$$
  F_i(\omega,\k) = S_{ij}\pi_j(\omega,\k)\,,\ \ \ \
  {\rm where\ }
  S_{ij} = \Delta_\eta \!\left( \delta_{ij} - \frac{k_i k_j}{\k^2}\right)
  + \Delta_s \frac{k_i k_j}{\k^2}\,,
$$
where $\Delta_\eta \equiv (-i\omega +\gamma_\eta\k^2)$, and
$\Delta_s \equiv (-i\omega + i\vs^2 \frac{\k^2}{\omega} + \gamma_s\k^2)$.
Setting $\Delta_\eta$ to zero gives the dispersion relation
of the shear mode, while setting $\Delta_s$ to zero gives the dispersion relation
of the sound mode.
The action is
$$
  S_{\rm eff}[{\bm\pi}] = \frac12\int_{\omega,\k}
  \pi_l (S_{li} D_{ij} S^*_{jm}) \pi^*_m\,,
$$
so the correlation function is
$$
  G_{\pi_i \pi_j}(\omega,\k) = (S^{*\,-1} D^{-1} S^{-1})_{ij}\,.
$$
This can be straightforwardly evaluated:
the inverse of $S_{ij}$ is
$$
  (S^{-1})_{ij} = \frac{1}{\Delta_\eta} \left( \delta_{ij} - \frac{k_i k_j}{\k^2}\right)
  + \frac{1}{\Delta_s} \frac{k_i k_j}{\k^2}\,.
$$
and the correlation function becomes
$$
  G_{\pi_i\pi_j}(\omega,\k) =
   \left( \delta_{ij} - \frac{k_i k_j}{\k^2}\right)
   \frac{2T\gamma_\eta  \bar{w}\, \k^2}{\omega^2 + (\gamma_\eta\k^2)^2}
  +\frac{k_i k_j}{\k^2}
   \frac{2\gamma_s \bar{w} \, T\,\k^2\omega^2}{(\omega^2 -\vs^2 \k^2)^2 + (\gamma_s\k^2\omega)^2}\,,
$$
in agreement with the linear-response result (\ref{eq:Gpipi0}) found in Section~\ref{sec:mu0}.
Similarly to the relation (\ref{eq:diff-FDT-2}) between $G_{n\rho}$ and $G_{nn}$ 
for diffusion, one can derive a relation between
$G_{\pi_i \lambda_j}$ and $G_{\pi_i \pi_j}$ in linearized hydrodynamics.
Changing the functional integration variable from $\lambda_i$ to $\phi_i$ as
$\lambda_i(t,\x) = \phi_i(t,\x) + i\int \!d^dx' D_{ij}(\x,\x') F_j(t,\x')$
one finds
$$
  \langle \pi_i(t_1,\x_1)\, \lambda_j(t_2,\x_2) \rangle = 
  i\int\!\! d^dx'\, D_{jk}(\x_2,\x')\,
  \langle \pi_i(t_1,\x_1) F_k(t_2,\x') \rangle\,.
$$
After using
$\omega G_{\pi_i \epsilon} = k_l G_{\pi_i \pi_l}$
together with the above expression for $G_{\pi_i \pi_j}$,
this gives 
$$
  G_{\pi_i \lambda_j}(\omega,\k) = 
  \left( \delta_{ij} - \frac{k_i k_j}{\k^2}\right)
  \frac{-1}{\omega+i\gamma_\eta \k^2} + 
  \frac{k_i k_j}{\k^2}
  \frac{-\omega}{\omega^2 - \vs^2 \k^2 +  i\omega \gamma_s \k^2}\,.
$$

\subsection{One-loop corrections}
\label{sec:FNS}
\noindent
So far we looked at fluctuations in linearized hydrodynamics.
One could extend the analysis of Section~\ref{sec:hydro}
to the full non-linear relativistic hydrodynamics with the constitutive relations~(\ref{eq:const-rel}),
however the resulting effective action is messy and not illuminating.
In order to study the interactions of the hydrodynamic modes,
it is instructive to look at a toy model which is technically simpler,
yet still retains the essential features of the full hydrodynamics.
We will thus consider a toy model, with the following simplifying assumptions:
{\it i)} we only expand the hydrodynamic equations to quadratic order
in fluctuations, keeping only non-derivative quadratic terms, and
{\it ii)} we impose the ``incompressibility'' condition $\partial_i\pi_i=0$.
The latter amounts to ignoring the ``high-energy'' sound mode
which has $\omega\sim k$, while keeping the ``low-energy''
shear and diffusive modes which have $\omega\sim k^2$.

\subsubsection*{Stochastic model for hydrodynamics}
\noindent
We look at hydrodynamic fluctuations in an equilibrium state with $\mu=0$.
Imposing the condition $\partial_k\pi_k=0$ eliminates the energy density,
and leaves $n$ and $\pi_k$ as the only hydro variables.
Keeping only non-derivative quadratic terms,
the constitutive relations for $J_k$ and $T_{ij}$ become:
\begin{eqnarray*}
  && J_k = -D\partial_k n + \frac{n\pi_k}{\bar{w}} + r_k\,,\\
  && T_{ij} = p\delta_{ij} -\gamma_\eta(\partial_i\pi_j + \partial_j\pi_i)
     + \frac{\pi_i\pi_j}{\bar{w}} + \tau_{ij}\,,
\end{eqnarray*}
where $r_k$ is the current noise, $\tau_{ij}$ is the stress noise,
and $\gamma_\eta\equiv\eta/\bar{w}$.
The pressure $p$ needs to be expanded to quadratic order
in $n$ and $\pi_k$ as well, but these terms will turn out to be unimportant,
as we will see shortly.
The hydrodynamic equations give the following stochastic model:
\begin{subequations}
\label{eq:FNS-model}
\begin{eqnarray}
  &&  \partial_t n  = D\grad^2 n -\frac{1}{\bar{w}} \pi_k \partial_k n + \theta(t,\x)\,,\\[5pt]
\label{eq:FNS-2}
  &&  \partial_t \pi_i = -\partial_i p
                         + \gamma_\eta \grad^2 \pi_i
                         - \frac{1}{\bar{w}} \pi_k\partial_k\pi_i + \xi_i(t,\x)\,.
\end{eqnarray}
\end{subequations}
From the discussion of diffusion in Section~\ref{sec:diff},
we take the short-distance noise as
\begin{eqnarray*}
  && \langle \theta(t,\x) \theta(t',\x') \rangle = -C_\sigma \grad^2 \delta(\x{-}\x') \delta(t{-}t')\,,\\ [5pt]
  && \langle \xi_i(t,\x) \xi_j(t',\x') \rangle = -C_\eta \, \delta_{ij}
     \grad^2 \delta(\x{-}\x') \delta(t{-}t')\,,\\[5pt]
  && \langle \theta(t,\x) \xi_i(t',\x')\rangle=0\,,
\end{eqnarray*}
with the noise strength constants
$C_\sigma = 2TD\chi = 2T\sigma$, and $C_\eta=2T\gamma_\eta w = 2T\eta$.
The convective term $\pi_k \partial_k n$
couples the diffusive mode and the shear mode,
while the convective term $\pi_k\partial_k\pi_i$
gives rise to self-interactions of the shear mode.
The model (\ref{eq:FNS-model}) was studied by Forster, Nelson, and Stephen 
in~\cite{Forster:1977zz}.
Note that Eq.~(\ref{eq:FNS-2}) is just the non-relativistic 
incompressible Navier-Stokes equation with a random external ``stirring force'' $\xi_i$.
For a treatment which includes the sound mode in 
non-relativistic hydrodynamics, see~\cite{PhysRevE.55.403}.

The symmetry properties of the correlation functions in the model (\ref{eq:FNS-model})
are the same as in the linearized hydrodynamics.
The equations of the model are real, so the correlation functions
of $n(t,\x)$ and $\pi_i(t,\x)$ are real. 
The correlation functions are also invariant under space-time translations,
and transform covariantly under rotations.
As the order of the fields does not matter when averaging over the noise, we have
$G_{nn}(t,\x) = G_{nn}(-t,-\x)$, and
$G_{\pi_i \pi_j}(t,\x) = G_{\pi_j \pi_i}(-t,-\x)$.
Combined with rotation symmetry, this implies that
$G_{nn}(\omega,\k)$ and $G_{\pi_i \pi_j}(\omega,\k)$ are real
(for real $\omega$ and $\k$).

\subsubsection*{Effective action for hydrodynamics}
\noindent
We follow the same procedure as in Sections \ref{sec:diff} and \ref{sec:hydro}
to derive the effective action for the stochastic model~(\ref{eq:FNS-model}).
As before, there is an auxiliary field $\rho(t,\x)$
for charge density fluctuations, and an auxiliary field $\lambda_i(t,\x)$
for momentum density fluctuations, which we choose to satisfy
$\partial_i\lambda_i=0$, to match the condition $\partial_i\pi_i=0$.
This condition implies that the pressure drops out of the effective action
because the corresponding term in the effective action
$\int\!\lambda_i\,\partial_i p$ is a total derivative.
This gives the following representation for correlation functions:
$$
  \langle n(t_1,\x_1) n(t_2,\x_2)\dots\rangle =
  \int\!\!\D n\,\D{\bm\pi}\,\D\rho\,\D{\bm\lambda}\;
  |J|\,
  e^{-S_{\rm eff}[n,\pi,\rho,\lambda]}\;
  n(t_1,\x_1) n(t_2,\x_2)\dots\,,
$$
where
\begin{equation}
\label{eq:nonlin-hydro-action-1}
  S_{\rm eff}[n,\pi,\rho,\lambda] = \int\!\!dt\,d^dx\;
  \left(
  -\frac{C_\sigma}{2} \rho\grad^2\rho - \frac{C_\eta}{2}\lambda_i\grad^2\lambda_i
  -i\rho B - i\lambda_i F_i
  \right)\,.
\end{equation}
Here $B \equiv (\partial_t n - D\grad^2 n + \pi_k\partial_k n/{\bar{w}})$,
and $F_i \equiv (\partial_t\pi_i - \gamma_\eta \grad^2\pi_i + \pi_k\partial_k\pi_i/\bar{w} )$.
The conditions $\partial_i\pi_i=0$ and $\partial_i\lambda_i=0$ are implied.
Again, one could integrate out the auxiliary fields $\rho$ and $\lambda_i$
in Eq.~(\ref{eq:nonlin-hydro-action-1}) to arrive at a real
(though non-local) effective action.

The effective action (\ref{eq:nonlin-hydro-action-1}) is not real.
Rather, it is invariant under complex conjugation combined with
$\rho\to-\rho$ and $\lambda_i\to-\lambda_i$.
This symmetry implies that the ``mixed'' correlation functions
$G_{n\rho}(t,\x)$, $G_{\rho n}(t,\x)$,
$G_{\pi_i\lambda_j}(t,\x)$, and $G_{\lambda_i\pi_j}(t,\x)$
are purely imaginary.
Combined with translation invariance, this implies
\begin{eqnarray}
\label{eq:nonlin-hydro-Gnrho-Grhon}
  && G_{n\rho}(\omega,\k) = -G_{\rho n}(\omega,\k)^* \,,\\
  && G_{\pi_i \lambda_j}(\omega,\k) = -G_{\lambda_j \pi_i}(\omega,\k)^*\,.
\end{eqnarray}
Again, we can show that $G_{n\rho}$ is related to $G_{nn}$,
by repeating the same argument which leads to
Eq.~(\ref{eq:diff-FDT-2}), and the extra term
$\langle  n \pi_k \partial_k n \rangle$ vanishes by parity.
A similar argument gives the relation between $G_{\pi\lambda}$ and $G_{\pi\pi}$,
because the extra term $\langle \pi_i\pi_k\partial_k\pi_j\rangle$ vanishes by parity.
Remembering that $G_{nn}(\omega,\k)$  and $G_{\pi_i\pi_j}(\omega,\k)$ are real, we have
\begin{eqnarray}
\label{eq:nonlin-hydro-FDT-1}
  &&  G_{nn}(\omega,\k) = 2T\chi\, {\rm Im\,} G_{n\rho}(\omega,\k)\,,\\
  &&  G_{\pi_i\pi_j}(\omega,\k) = 2 T \bar{w}\, {\rm Im\,} G_{\pi_i\lambda_j}(\omega,\k)\,.
\end{eqnarray}
This means that in order to evaluate the correlation functions
$G_{nn}$ and $G_{\pi_i\pi_j}$, we can evaluate $G_{n\rho}$ and $G_{\pi_i\lambda_j}$,
and take their imaginary parts.
It will be simpler to do exactly that in practice.

The above functional integral representation of correlation functions
also contains the Jacobian $J$ which arises from the variable change to
the fields $n$ and ${\bm\pi}$.
In general, if there are several fields $\varphi_a$ 
satisfying equations $E_a[\varphi]=0$,
then the Jacobian is $J=\det(\delta E_a/\delta\varphi_b)$.
Note that in our model the noise terms only appear additively in the 
stochastic equations, therefore the Jacobian does not depend on the noise,
and the noise average can be easily performed leading to (\ref{eq:nonlin-hydro-action-1}).
We take our fields as $\varphi_a=\{n,\pi_i\}$, and we take
as their equations of motion $E_a[\varphi]$ the corresponding stochastic
equations for $n$ and $\pi_i$.
Note that the Jacobian is real because both the fields $\varphi_a$
and their equations of motion $E_a[\varphi]$ are real.
We can exponentiate the Jacobian by introducing the anticommuting ghost fields
$\psi_a$, $\bar\psi_a$ as was done by Parisi and Sourlas in \cite{Parisi:1979ka},
$$
  J = \int\!\!\D\bar\psi\D\psi\;e^{-S_g}\,,\ \ \ \
  S_g = \int\!\!dt\,d^d x\;\bar\psi_a \frac{\delta E_a[\varphi]}{\delta\varphi_b} \psi_b\,.
$$
In our model we will have both the density ghosts $\psi_{\rm n}$, $\bar\psi_{\rm n}$,
and the momentum ghosts $\psi_i$, $\bar\psi_i$.
For the momentum ghosts, we impose the ``incompressibility'' condition
$\partial_i\psi_i=0$, $\partial_i\bar\psi_i = 0$.
We can now include the ghosts in the effective action
(\ref{eq:nonlin-hydro-action-1}), and write it as
\begin{equation}
\label{eq:nonlin-hydro-action-2}
  S_{\rm eff} = \int\!dt\,d^dx \left( \L^{(2)} + \L^{(int)}  \right)\,,
\end{equation}
where $\L^{(2)}$ is quadratic in the fields, and $\L^{(int)}$
contains the interactions,
\begin{eqnarray}
     \L^{(2)} & = &
     -\frac{C_\sigma}{2} \rho\grad^2\rho - \frac{C_\eta}{2}\lambda_i\grad^2\lambda_i
     -i\rho (\partial_t n - D\grad^2 n ) - i\lambda_i (\partial_t\pi_i - \gamma_\eta \grad^2 \pi_i)\nonumber\\[5pt]
     && 
     +\,\bar\psi_i(\partial_t - \gamma_\eta \grad^2)\psi_i 
     +\,\bar\psi_{\rm n}(\partial_t - D\grad^2)\psi_{\rm n}  \,,
\end{eqnarray}

\begin{eqnarray}
     \L^{(int)} & = &
     -\frac{i}{\bar{w}} \rho\pi_j\partial_j n 
     - \frac{i}{\bar{w}} \lambda_i \pi_j \partial_j\pi_i\nonumber\\[5pt]
     &&
     +\frac{1}{\bar{w}}\bar\psi_i (\partial_k\pi_i)\psi_k
     +\frac{1}{\bar{w}}\bar\psi_i\pi_k\partial_k\psi_i 
     +\frac{1}{\bar{w}}\bar\psi_{\rm n} (\partial_i n)\psi_i 
     +\frac{1}{\bar{w}}\bar\psi_{\rm n} \pi_k\partial_k\psi_{\rm n}
     \,.
\end{eqnarray}
The conditions $\partial_i\pi_i=0$, $\partial_i\lambda_i=0$, 
$\partial_i\psi_i=0$, $\partial_i\bar\psi_i = 0$ can be imposed
by additional Lagrange multipliers.
These extra auxiliary fields only appear in $\L^{(2)}$, but
not in $\L^{(int)}$, hence their only effect is to modify
the propagators, but not the vertices.
The action above is suitable for calculating correlation functions
in the hydrodynamic perturbation theory.

\begin{figure}

\begin{center}
\begin{tabular}{p{5cm} p{5cm}}
 \begin{picture}(100,100)(0,0)
 \DashLine(10,50)(50,50){2}
 \Photon(50,50)(80,80){2}{8}
 \Line(50,50)(80,20)
 \Vertex(50,50){2}
 \put(30,55){$k_j$}
 \put(48,37){$j$}
 \put(75,50){ ${\displaystyle \frac{1}{\bar{w}} }$ }
 \end{picture}
 &
 \begin{picture}(100,100)(0,0)
 \Gluon(10,50)(50,50){2}{8}
 \Line(50,50)(80,80)
 \Line(50,50)(80,20)
 \Vertex(50,50){2}
 \put(62,77){$k_j$}
 \put(30,37){$i$}
 \put(52,57){$i$}
 \put(55,30){$j$}
 \put(75,50){ ${\displaystyle \frac{1}{\bar{w}} }$ }
 \end{picture}
\end{tabular}
\end{center}

\caption{
    The vertices for the effective action (\ref{eq:nonlin-hydro-action-1}).
    There is an overall sign depending on
    whether the momentum is flowing in or out of the vertex.
}
\label{fig:nonlin-hydro-vertices}
\end{figure}
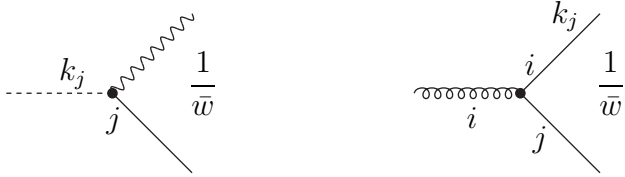

The action (\ref{eq:nonlin-hydro-action-2}) 
is invariant under space-time translations, spatial rotations,
parity, and charge conjugation.
The action (\ref{eq:nonlin-hydro-action-2}) is not
invariant under time-reversal, nor should it be.
The action is also invariant under Galilean symmetry under which
\begin{eqnarray*}
  && \pi_i(t,\x) \to \pi_i(t,\x {-} \v t) + \bar{w} v_i\,,\\
  && n(t,\x) \to n(t,\x{-}\v t)\,,
\end{eqnarray*}
and all other fields transforming the same way as $n(t,\x)$.
Further, the action (\ref{eq:nonlin-hydro-action-2}) is invariant under
the symmetry generated by
\begin{eqnarray*}
   && \delta n = \bar\xi\, \psi_{\rm n}\,,\ \ \ \ \ \delta\bar\psi_{\rm n} = i\bar\xi\,\rho\,,\\
   && \delta\pi_i = \bar\xi\, \psi_i\,,\ \ \ \ \delta\bar\psi_i = i\bar\xi\,\lambda_i\,,
\end{eqnarray*}
where $\bar\xi$ is an infinitesimal anticommuting parameter.
This symmetry is quite generic in stochastic 
systems~\cite{Parisi:1979ka, Feigelman:1983ac}.
It is an analogue of the BRST symmetry, and implies the relations
\begin{eqnarray*}
  && G_{\psi_{\rm n} \bar\psi_{\rm n}}(t,\x) = -i\, G_{n\rho}(t,\x)\,,\\
  && G_{\psi_i \bar\psi_j}(t,\x) = -i\, G_{\pi_i\lambda_j}(t,\x)\,.
\end{eqnarray*}
For the diagram technique,
we read off the propagators from $\L^{(2)}$, and the vertices from $\L^{(int)}$.
The propagators were already calculated in Sections~\ref{sec:diff} -- \ref{sec:hydro}, and
we summarize them here, associating a line with each one:
\begin{center}
\begin{tabular}{l p{5cm}}
 ${\displaystyle G^0_{nn}(\omega,\k) = \frac{C_\sigma\k^2}{\omega^2 + (D\k^2)^2} }$ &
 $\quad\quad$
 \begin{picture}(100,3)(0,0)
 \DashLine(0,0)(100,0){2}
 \end{picture}
 \\[15pt]
 ${\displaystyle G^0_{n\rho}(\omega,\k) = \frac{-1}{\omega + i D\k^2}}$ &
 $\quad\quad$
 \begin{picture}(100,3)(0,0)
 \DashLine(0,0)(50,0){2}
 \Photon(50,0)(100,0){2}{8}
 \end{picture}
 \\[15pt]
 ${\displaystyle G^0_{\rho n}(\omega,\k) = \frac{1}{\omega - i D\k^2}}$ &
 $\quad\quad$
 \begin{picture}(100,3)(0,0)
 \Photon(0,0)(50,0){2}{8}
 \DashLine(50,0)(100,0){2}
 \end{picture}
 \\ [15pt]
 ${\displaystyle G^0_{\rho\rho}(\omega,\k) = 0 }$ &
 $\quad\quad$
 \begin{picture}(100,3)(0,0)
 \Photon(0,0)(100,0){2}{16}
 \end{picture}
 \\ [15pt]
%
  ${\displaystyle
 G^0_{\pi_i\pi_j}(\omega,\k)
 = \frac{C_\eta \k^2}{\omega^2 + (\gamma_\eta \k^2)^2} \left(\delta_{ij} - \frac{k_i k_j}{\k^2} \right)
 }$ &
 $\quad\quad$
 \begin{picture}(100,10)(0,5)
 \Line(0,10)(100,10)
 \put(0,0){\small\it i}
 \put(95,0){\small\it j}
 \end{picture}
 \\[15pt]
 ${\displaystyle
 G^0_{\pi_i\lambda_j}(\omega,\k)
 = \frac{-1}{\omega + i\gamma_\eta \k^2} \left(\delta_{ij} - \frac{k_i k_j}{\k^2} \right)
 }$ &
 $\quad\quad$
 \begin{picture}(100,10)(0,5)
 \Line(0,10)(50,10)
 \Gluon(50,10)(100,10){2}{8}
 \put(0,0){\small\it i}
 \put(95,0){\small\it j}
 \end{picture}
 \\[15pt]
 ${\displaystyle
 G^0_{\lambda_i \pi_j}(\omega,\k)
 = \frac{1}{\omega - i\gamma_\eta \k^2} \left(\delta_{ij} - \frac{k_i k_j}{\k^2} \right)
 }$ &
 $\quad\quad$
 \begin{picture}(100,10)(0,5)
 \Gluon(0,10)(50,10){2}{8}
 \Line(50,10)(100,10)
 \put(0,0){\small\it i}
 \put(95,0){\small\it j}
 \end{picture}
 \\ [15pt]
 ${\displaystyle G^0_{\lambda_i\lambda_j}(\omega,\k) =0 }$ &
 $\quad\quad$
 \begin{picture}(100,10)(0,5)
 \Gluon(0,10)(100,10){2}{16}
 \put(0,0){\small\it i}
 \put(95,0){\small\it j}
 \end{picture}
\end{tabular}
\end{center}
The vertices are shown in Figure~\ref{fig:nonlin-hydro-vertices}.
There are also ghost propagators and ghost vertices which we didn't write down,
but they can be easily read off from the action~(\ref{eq:nonlin-hydro-action-2}).

\subsubsection*{Correction to the momentum density propagator}
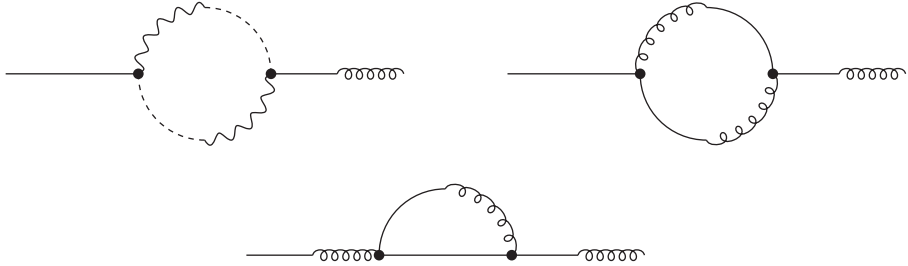
\begin{figure}
\begin{center}
\begin{tabular}{p{5cm} c p{5cm}}
\begin{picture}(150,50)(0,-20)
  \Line(0,0)(50,0)
  \DashCArc(75,0)(25,180,270){2}
  \PhotonArc(75,0)(25,270,360){2}{5}
  \DashCArc(75,0)(25,0,90){2}
  \PhotonArc(75,0)(25,90,180){2}{5}
  \Line(100,0)(125,0)
  \Gluon(125,0)(150,0){2}{5}
  \Vertex(50,0){2}
  \Vertex(100,0){2}
\end{picture}
  &
  $\quad\quad$
  &
\begin{picture}(150,50)(0,-20)
  \Line(0,0)(50,0)
  \CArc(75,0)(25,180,270)
  \GlueArc(75,0)(25,270,360){2}{5}
  \CArc(75,0)(25,0,90)
  \GlueArc(75,0)(25,90,180){2}{5}
  \Line(100,0)(125,0)
  \Gluon(125,0)(150,0){2}{5}
  \Vertex(50,0){2}
  \Vertex(100,0){2}
\end{picture}
\end{tabular}
\end{center}
\begin{center}
\begin{picture}(150,50)(0,-20)
  \Line(0,0)(25,0)
  \Gluon(25,0)(50,0){2}{5}
  \Line(50,0)(100,0)
  \Line(100,0)(125,0)
  \Gluon(125,0)(150,0){2}{5}
  \CArc(75,0)(25,90,180)
  \GlueArc(75,0)(25,0,90){2}{5}
  \Vertex(50,0){2}
  \Vertex(100,0){2}
\end{picture}
\end{center}
\caption{
    Three connected one-loop diagrams potentially contributing to $G_{\pi\lambda}$.
    In addition to the shown diagrams, there are two more diagrams with ghost loops
    which are completely analogous to the diagrams in the first row.
}
\label{fig:G-pi-lambda-1-loop}
\end{figure}

\noindent
At one loop, there are five connected diagrams potentially contributing to
$G_{\pi_i \lambda_j}$, three of which are shown in Figure~\ref{fig:G-pi-lambda-1-loop},
and the remaining two contain ghost loops.
The first two diagrams shown in the figure vanish.
To see this, let $\omega$ and $\k$ be the external frequency and momentum,
and $z$ and $\q$ the frequency and momentum running in the loop.
The first diagram will have $G_{n\rho}$ and $G_{\rho n}$ in the loop,
and is proportional to 
$$
  \int \frac{dz}{2\pi}\; \frac{-1}{z+iD\q^2}\; \frac{1}{(\omega{-}z) - iD(\k{-}\q)^2}\,.
$$
In the plane of complex $z$, both poles are in the lower half plane,
and as a result the integral is zero.
Exactly the same argument applies to the second diagram in 
Figure~\ref{fig:G-pi-lambda-1-loop}, and to the diagrams with ghost loops.
Thus it is only the last diagram in Figure~\ref{fig:G-pi-lambda-1-loop} which contributes.
Let us call the amputated part of the last diagram $\Sigma_{mn}(\omega,\k)$, i.e.
we write the connected one-loop contribution in momentum space as
$$
  G_{\pi_i\lambda_j} = 
  G_{\pi_i\lambda_j}^0
  + G_{\pi_i\lambda_m}^0 \Sigma_{mn}\, G_{\pi_n\lambda_j}^0\,.
$$
By rotation invariance, $\Sigma_{mn}$ must have the form 
$\Sigma_{mn} = \delta_{mn} \Sigma_1 + k_m k_n \Sigma_2$.
Because $G_{\pi_n\lambda_j}^0$ is transverse, it is only 
$\Sigma_1$ that contributes.
By summing all such one-loop diagrams, we find the correlation function as
$
  G = G^0 (1-\Sigma_1 G^0)^{-1}\,,
$
which gives
$$
  G_{\pi_i\lambda_j}(\omega,\k) = 
  \frac{-1}{\omega + i\gamma_\eta \k^2 + \Sigma_1(\omega,\k)}
  \left( \delta_{ij} - \frac{k_i k_j}{\k^2} \right)\,.
$$
If $\Sigma_1(\omega,\k)$ is proportional to $\k^2$ for small external momenta,
its imaginary part can be interpreted as a correction to $\gamma_\eta$,
or equivalently as a correction to viscosity.
The correlation function of momentum density then becomes
\begin{equation}
\label{eq:Gpipi-1-loop}
  G_{\pi_i\pi_j}(\omega,\k{\to}0) = 
  \frac{2T\bar{w}\,\gamma_\eta(\omega)\,\k^2}{\omega^2 + \Big(\gamma_\eta(\omega)\,\k^2\Big)^2} 
  \left(\delta_{ij} - \frac{k_i k_j}{\k^2}\right)\,,
\end{equation}
where $\gamma_\eta(\omega) = \gamma_\eta + {\rm Im}\,\Sigma_1(\omega,\k{\to}0)/\k^2$.
Explicitly, the loop integral is
\begin{equation}
\label{eq:loop-pi}
  \Sigma_1(\omega,\k{\to}0) = -\frac{\k^2}{s d} \int\!\! \frac{d^d q}{(2\pi)^d}
  \left(
  \frac{d^2{-}d{-}2}{\omega + 2i\gamma_\eta q^2} +
  \frac{2}{d{+}2} \frac{2i\gamma_\eta q^2}{(\omega + 2i\gamma_\eta q^2)^2}
  \right)\,,
\end{equation}
where $s=\bar{w}/T$ is the equilibrium density of entropy.
In the large-$N$ limit, both the viscosity and the entropy density grow with $N$,
so that $\gamma_\eta=O(1)$,
and $1/s$ can be treated as a small parameter, suppressing the
fluctuation corrections.
In $d=3$ spatial dimensions the integral (\ref{eq:loop-pi}) is linearly divergent at large momenta.
Introducing a large-momentum cutoff, we find
$$
  {\rm Re}\, \Sigma_1(\omega,\k{\to}0) = -\frac{23\,\k^2}{30\pi s}
  \frac{\sqrt{|\omega|}\, {\rm sign}(\omega)}{(4\gamma_\eta)^{3/2}}\,,\ \ \ \ 
  {\rm Im}\, \Sigma_1(\omega,\k{\to}0) = C_\Lambda \k^2 - \frac{23\,\k^2}{30\pi s}
  \frac{\sqrt{|\omega|} } {(4\gamma_\eta)^{3/2}}\,.
$$
The constant $C_\Lambda$ is proportional to the cutoff
and to $1/s$, but is $\omega$ and $\k$ independent.
The imaginary part of $\Sigma_1(\omega,\k)$ can be interpreted
as a correction to $\gamma_\eta$, producing a frequency-dependent 
viscous damping coefficient
$\gamma_\eta(\omega) = \gamma_\eta + C_\Lambda -\frac{23}{30\pi s}\sqrt{|\omega|}\, (4\gamma_\eta)^{-3/2}$.
We define the ``renormalized'' viscous damping constant as
$$
  \gamma_\eta^{\rm ren} = \gamma_\eta + C_\Lambda\,.
$$
We need to express the physically measurable $\gamma_\eta(\omega)$
in terms of finite $\gamma_\eta^{\rm ren}$.
Following the standard renormalization procedure,
to this order in $1/s$ we replace $\gamma_\eta$
with $\gamma_\eta^{\rm ren}$, which gives
\begin{equation}
\label{eq:Gamma-of-omega-3dim}
  \gamma_\eta(\omega) = \gamma_\eta^{\rm ren} - 
  \frac{23}{30\pi s} \frac{ \sqrt{|\omega|} } {(4\gamma_\eta^{\rm ren})^{3/2} }\,.
\end{equation}
The renormalized viscous damping constant in $d=3$ is given by the usual Kubo formula~(\ref{eq:Kubo-eta-pi}),
as $\omega\to0$ (recall that $\eta=\gamma_\eta \, \bar{w}$). 
For example, taking the spatial momentum along $z$, we have
\begin{equation}
\label{eq:nonlin-hydro-Kubo-Gamma}
  \gamma_\eta^{\rm ren} =
  \frac{1}{2T \bar{w}}
  \lim_{\omega\to0} \lim_{\k\to0} \frac{\omega^2}{\k^2}
  G_{\pi_x \pi_x}(\omega,\k)\,.
\end{equation}

In $d{=}2$ spatial dimensions, the integral (\ref{eq:loop-pi})
is logarithmically divergent at large momenta.
Introducing a large-momentum cutoff, we find
$$
  {\rm Re}\, \Sigma_1(\omega,\k{\to}0) = -\frac{\k^2}{32s} \frac{{\rm sign}(\omega)}{2\gamma_\eta}\,,
  \ \ \ \ 
  {\rm Im}\, \Sigma_1(\omega,\k{\to}0) = \frac{\k^2}{16\pi s} \frac{\ln(\Lambda/|\omega|)}{2\gamma_\eta}\,.
$$
Again, the imaginary part of $\Sigma_1(\omega,\k)$ can be interpreted 
as a correction to $\gamma_\eta$, producing a frequency-dependent viscous damping constant,
$\gamma_\eta(\omega) = \gamma_\eta + \frac{1}{16\pi s} \frac{1}{2\gamma_\eta} \ln(\Lambda/|\omega|)$.
If we interpret the logarithmic divergence in the renormalization-group sense,
we can define the ``renormalized'' damping constant $\gamma_\eta^{\rm ren}$ as 
$\gamma_\eta(\omega)$ evaluated at some
frequency scale $\omega=\mu$,
$$
  \gamma_\eta^{\rm ren} = \gamma_\eta + \frac{1}{16\pi s} \frac{1}{2\gamma_\eta} \ln\frac{\Lambda}{\mu}\,.
$$
To this order in $1/s$ we replace $1/\gamma_\eta$ in the right-hand side
with $1/\gamma_\eta^{\rm ren}$, and express the ``bare'' parameter $\gamma_\eta$
in terms of $\gamma_\eta^{\rm ren}$.
The ``bare'' parameter $\gamma_\eta$ knows nothing about
the arbitrary frequency scale $\mu$ at which we choose to define
the viscosity.
Hence $\partial \gamma_\eta/\partial\mu=0$, 
which implies
\begin{equation}
\label{eq:RG-eta}
  \mu\frac{\partial \gamma_\eta^{\rm ren}}{\partial\mu} = -\frac{1}{32\pi s} \frac{1}{\gamma_\eta^{\rm ren}}\,.
\end{equation}
In other words, in order for the frequency-dependent viscosity coefficient
$\gamma_\eta(\omega)$ in the correlation function~(\ref{eq:Gpipi-1-loop})
to be independent of the arbitrary scale $\mu$, 
the renormalized viscosity must depend on the arbitrary frequency scale
$\mu$, at which we choose to measure it,
so that $\gamma_\eta^{\rm ren}(\mu)$ grows
in the infrared, proportional to $\sqrt{\ln(1/\mu)}$.
The viscous damping coefficient $\gamma_\eta\equiv\eta/\bar{w}$ is
$$
  \gamma_\eta(\omega) = \gamma_\eta^{\rm ren}(\mu) + \frac{1}{32\pi s}\frac{1}{\gamma_\eta^{\rm ren}(\mu)}
  \ln\frac{\mu}{\omega}\,,
$$
and the correlation function of momentum density (\ref{eq:Gpipi-1-loop})
does not depend on $\mu$.
The correction to $\gamma_\eta$ differs from the one computed in Section~\ref{sec:loops}
because the present ``incompressible'' model ignores sound waves.

\subsubsection*{Correction to the charge density propagator}
\begin{figure}
\begin{center}

\begin{tabular}{p{5cm}}
\begin{picture}(150,50)(0,-20)
  \DashLine(0,0)(25,0){2}
  \Photon(25,0)(50,0){2}{5}
  \Line(50,0)(100,0)
  \DashLine(100,0)(125,0){2}
  \Photon(125,0)(150,0){2}{5}
  \DashCArc(75,0)(25,90,180){2}
  \PhotonArc(75,0)(25,0,90){2}{5}
  \Vertex(50,0){2}
  \Vertex(100,0){2}
\end{picture}
\end{tabular}
\end{center}
\caption{
    The connected one-loop diagram contributing to $G_{n\rho}$.
}
\label{fig:Gnrho-Grhon-1-loop}
\end{figure}
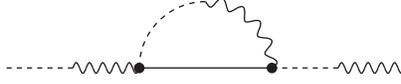

\noindent
Let us now calculate the correction to the charge density correlation function.
At one loop, there is only one connected diagram that contributes to $G_{n\rho}$,
as shown in Figure~\ref{fig:Gnrho-Grhon-1-loop}.
Let us call the amputated part of the $G_{n\rho}$ diagram $\Sigma(\omega,\k)$.
By summing all such diagrams, we find the corrected correlation function:
\begin{equation}
\label{eq:nonlin-hydro-Gnrho}
  G_{n\rho}(\omega,\k) = \frac{-1}{\omega+iD\k^2 +\Sigma(\omega,\k)}\,.
\end{equation}
For small external momenta, $\Sigma(\omega,\k)$ is proportional to $\k^2$,
and we obtain the correlation function of charge density as
\begin{equation}
\label{eq:Gnn-1-loop}
  G_{nn}(\omega,\k{\to}0) = 
  \frac{2T\chi\, D(\omega)\,\k^2}{\omega^2 + \Big(D(\omega)\,\k^2\Big)^2}\,,
\end{equation}
where $D(\omega) = D + {\rm Im}\,\Sigma(\omega,\k{\to}0)/\k^2$.
Explicitly, $\Sigma(\omega,\k)$ is given by
\begin{equation}
\label{eq:nonlin-hydro-Sigma}
  \Sigma(\omega,\k) =
  \frac{1}{\bar{w}^{\,2}} \int\frac{dz}{2\pi}\frac{d^dq}{(2\pi)^d}\;
  G^{0}_{\pi_i\pi_j}(z,\q)\;
  G^{0}_{n\rho}(\omega{-}z, \k{-}\q) \;(k_i{-}q_i) k_j\,.
\end{equation}
In the expression (\ref{eq:nonlin-hydro-Sigma}),
the $q_i$ factor does not contribute because $G^{0}_{\pi_i\pi_j}$ is transverse.
The frequency integral can be evaluated by closing the contour in the lower half-plane.
The remaining momentum integral simplifies when $\k\to0$,
and one finds
\begin{equation}
\label{eq:loop-n}
  \Sigma(\omega,\k{\to}0) = -\frac{\k^2}{s} \frac{d{-}1}{d}
  \int\!\!\frac{d^dq}{(2\pi)^d}\;
  \frac{1}{\omega + i(\gamma_\eta {+} D) \q^2}\,.
\end{equation}
In $d=3$ spatial dimensions, the integral (\ref{eq:loop-n})
is linearly divergent at large momenta.
Introducing a large-momentum cutoff, we find
$$
 {\rm Re}\,\Sigma(\omega,\k{\to}0) =
 -\frac{\k^2}{3\pi s}
 \frac{\sqrt{|\omega|}\, {\rm sign}(\omega) }{[2(\gamma_\eta{+}D)]^{3/2}}\,,
 \ \ \ \ 
 {\rm Im}\,\Sigma(\omega,\k{\to}0) =
 C_\Lambda \k^2
 -\frac{\k^2}{3\pi s}
 \frac{ \sqrt{|\omega|} }{[2(\gamma_\eta{+}D)]^{3/2}}\,.
$$
The constant $C_\Lambda$ is proportional to the cutoff
and to $1/s$, but is $\omega$ and $\k$ independent.
The imaginary part of $\Sigma(\omega,\k)$ can be interpreted
as a correction to $D$, producing a frequency-dependent 
diffusion coefficient
$D(\omega) = D + C_\Lambda -\frac{1}{3\pi s}\sqrt{|\omega|}\, [2(\gamma_\eta{+}D)]^{-3/2}$.
We define the ``renormalized'' diffusion constant as
$$
  D^{\rm ren} = D + C_\Lambda\,.
$$
We need to express the physically measurable $D(\omega)$
in terms of finite $D^{\rm ren}$.
Following the standard renormalization procedure,
to this order in $1/s$ we replace $(\gamma_\eta{+}D)$ in the right-hand side
with $(\gamma_\eta^{\rm ren}{+}D^{\rm ren})$, which gives
\begin{equation}
\label{eq:D-of-omega-3dim}
  D(\omega) = D^{\rm ren} - 
  \frac{1}{3\pi s} \frac{ \sqrt{|\omega|} } {[2(\gamma_\eta^{\rm ren}{+}D^{\rm ren})]^{3/2} }\,.
\end{equation}
The renormalized diffusion constant in $d=3$ is given by the usual Kubo formula~(\ref{eq:Kubo-sigma-n}),
as $\omega\to0$ (recall that $\sigma=D\chi$),
\begin{equation}
\label{eq:nonlin-hydro-Kubo-D}
  D^{\rm ren} =
  \frac{1}{2T\chi}
  \lim_{\omega\to0} \lim_{\k\to0} \frac{\omega^2}{\k^2}
  G_{n n}(\omega,\k)\,.
\end{equation}

In $d=2$ spatial dimensions, the integral (\ref{eq:loop-n})
is logarithmically divergent at large momenta.
Introducing a large-momentum cutoff, we find
$$
  {\rm Re}\,\Sigma(\omega,\k{\to}0) = -\frac{\k^2}{16s} \frac{{\rm sign}(\omega)}{\gamma_\eta{+}D}\,,
  \ \ \ \
  {\rm Im}\,\Sigma(\omega,\k{\to}0) = \frac{\k^2}{8\pi s}
  \frac{\ln (\Lambda/|\omega|)}{\gamma_\eta{+}D}\,.
$$
The imaginary part of $\Sigma(\omega,\k)$ can be interpreted 
as a correction to $D$, producing a frequency-dependent
diffusion coefficient,
$D(\omega) = D + \frac{1}{8\pi s} \frac{1}{\gamma_\eta+D} \ln(\Lambda/|\omega|)$.
If we interpret the logarithmic divergence in the renormalization-group sense,
we can define the ``renormalized'' diffusion constant $D^{\rm ren}$
as $D(\omega)$ evaluated at some frequency scale $\omega=\mu$,
$$
  D^{\rm ren} = D + \frac{1}{8\pi s} \frac{1}{\gamma_\eta{+}D} \ln\frac{\Lambda}{\mu}\,.
$$
To this order in $1/s$ we can replace $(\gamma_\eta{+}D)$ in the right-hand side
with $(\gamma_\eta^{\rm ren}{+}D^{\rm ren})$, and express the ``bare'' diffusion constant $D$
in terms of $D^{\rm ren}$ and $\gamma_\eta^{\rm ren}$.
Note that the ``bare'' parameter $D$ knows nothing about
the arbitrary frequency scale $\mu$ at which we choose to define
the diffusion constant.
Hence $\partial D/\partial\mu=0$, 
which implies
\begin{equation}
\label{eq:RG-D}
  \mu\frac{\partial D^{\rm ren}}{\partial\mu} = 
  -\frac{1}{8\pi s} \frac{1}{\gamma_\eta^{\rm ren}{+}D^{\rm ren}}\,.
\end{equation}
In other words, in order for the frequency-dependent diffusion coefficient
$D(\omega)$ in the correlation function (\ref{eq:Gnn-1-loop}) to be independent of the arbitrary scale $\mu$, 
the renormalized diffusion constant must depend on the arbitrary frequency scale
$\mu$, at which we choose to measure it.
The diffusion coefficient is
$$
  D(\omega) = D^{\rm ren}(\mu) + \frac{1}{8\pi s}\frac{1}{\gamma_\eta^{\rm ren}(\mu) + D^{\rm ren}(\mu)}
  \ln\frac{\mu}{\omega}\,,
$$
and the correlation function of charge density (\ref{eq:Gnn-1-loop})
does not depend on $\mu$.

\subsubsection*{Scale dependence of the diffusion constant and viscosity in $d=2$}
\noindent
\begin{figure}
\begin{center}
\includegraphics[width=8cm]{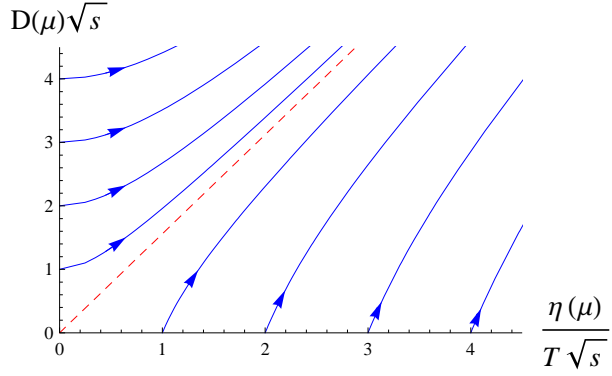}
\end{center}
\caption{
  The flow diagram for the differential equations (\ref{eq:rg-eqs-2d}), pictured
  in terms of the dimensionless variables $\eta/(T\sqrt{s})$ and $D\sqrt{s}$, where
  $\eta$ is the shear viscosity, $D$ is the diffusion constant, and $s$ is the
  equilibrium density of entropy.
  The arrows indicate the direction of decreasing $\mu$ (towards lower frequency).
  The dashed line is a straight line with a slope $(\sqrt{17}{-}1)/2$,
  indicating the asymptotics as $\mu\to0$.
}
\label{fig:rg-flow-2d}
\end{figure}%
We have found that the viscosity and the diffusion constant are running,
scale-dependent parameters in $d=2$ spatial dimensions,
satisfying renormalization-group equations (\ref{eq:RG-eta}) and (\ref{eq:RG-D}),
\begin{equation}
\label{eq:rg-eqs-2d}
  \mu\frac{\partial\gamma_\eta^{\rm ren}}{\partial\mu} = -\frac{1}{32\pi s} \frac{1}{\gamma_\eta^{\rm ren}}\,,\ \ \ \ 
  \mu\frac{\partial D^{\rm ren}}{\partial\mu} = -\frac{1}{8\pi s} \frac{1}{\gamma_\eta^{\rm ren}{+}D^{\rm ren}}\,,
\end{equation}
where $D$ is the charge diffusion constant,
$\gamma_\eta\equiv\eta/(\bar\epsilon+\bar p)$ is the shear mode damping constant,
and $\mu$ is the frequency at which we choose to measure the transport coefficients
at zero momentum.
These equations are almost identical to the equations (\ref{eq:sigma-eta-RG})
derived earlier, except for a factor of 2 in the equation for $\gamma_\eta$.
This is because the present model only keeps shear waves, but neglects
sound waves, which contributed to the renormalization of the shear viscosity in
Section~\ref{sec:loops}.
In the low-frequency limit $\mu\to0$, Eq.~(\ref{eq:rg-eqs-2d}) give the
following scale dependence of the transport coefficients:
$$
  \gamma_\eta^{\rm ren}(\mu) = \sqrt{\frac{1}{16\pi s}}\, \sqrt{\,\ln \frac{1}{\mu}}\,,\ \ \ \
  D^{\rm ren}(\mu) = \frac{\sqrt{17}-1}{2}\, \gamma_\eta^{\rm ren}(\mu)\,.
$$
The flow diagram for Eq.~(\ref{eq:rg-eqs-2d})
is shown in Figure~\ref{fig:rg-flow-2d}.
One can see that in the extreme low-frequency limit, the renormalized shear viscosity
and the diffusion constant (or charge conductivity)
cease being independent transport parameters, but are rather related by
\begin{equation}
  DT = \frac{\sqrt{17}-1}{2}\, \frac{\eta}{s}\, \approx 1.56\, \frac{\eta}{s}\,.
\end{equation}

\section{Discussion}
These lectures mainly focused on response functions of conserved densities in relativistic fluids. 
In linear response theory, hydrodynamic correlation functions may be evaluated by several methods:
the canonical approach of Sec.~\ref{sec:generalization}, 
the variational approach of Sec.~\ref{sec:variational},
or the stochastic approach of Sec.~\ref{sec:hydro}.
In non-relativistic fluids, the corresponding response functions can be measured
using light scattering~\cite{Boon-Yip}; relativistic fluids are far less common.
It would be wonderful to see relativistic hydrodynamics established experimentally as 
the effective description for many-body quantum systems with emergent Lorentz symmetry,
such as those discussed in Ref.~\cite{SS}.

There are a number of subjects related to hydrodynamic fluctuations in relativistic fluids that
we have not touched on.
These include hydrodynamic three- and four-point functions
\cite{Moore:2010bu, Arnold:2011ja},
hydrodynamics of anomalous currents \cite{Son:2009tf, Amado:2011zx, Landsteiner:2011cp, Jensen:2012jy},
hydrodynamics of relativistic superfluids \cite{Bhattacharya:2011eea, Bhattacharya:2011tr, Herzog:2011ec},
hydrodynamics of theories with broken parity \cite{ Jensen:2011xb, Bhattacharya:2011tr},
relativistic magneto-hydrodynamics \cite{Huang:2011dc},
and the relation of the noise to the closed time path formalism 
in quantum field theory~\cite{Calzetta-Hu}.

As we have seen in Section~\ref{sec:second-order}, fluctuation effects render 
second-order hydrodynamics invalid.
On the other hand, second-order hydrodynamics was introduced in the first place
in order to eliminate the acausality of the first-order hydrodynamics
associated with short-wavelength modes. 
A complete hydrodynamic theory must contain {\it both} the derivative expansion, {\it and} 
a systematic procedure to treat the fluctuations.

In these lectures, we tried to emphasize that hydrodynamics should be viewed as
more than a study of the classical hydrodynamic equations.
While many questions in hydrodynamics can be answered by viewing it as a classical field theory,
there are aspects of hydrodynamics which are more quantum field-theoretic in nature.
The derivative expansion of the constitutive relations in hydrodynamics
is not unike the derivative expansion in effective field theory, 
the hydrodynamic equations are analogous to the effective Lagrangian,
and the effects of the hydrodynamic fluctuations are analogous to 
quantum loop corrections.
A notable difference between hydrodynamics and effective field theory
is that at the moment there is no simple and systematic procedure to reformulate 
hydrodynamics as a field theory, where the derivative expansion can be applied
directly to the effective action.%
\footnote{
	Formulating an effective action for hydrodynamics appears an easier problem
	if one ignores dissipation: see for example 
	Refs.~\cite{Schutz:1970my, Brown:1992kc, Jackiw:2004nm, Dubovsky:2011sj, Banerjee:2012iz, Jensen:2012jh}.
}
A naive attempt at deriving the effective action for relativistic hydrodynamics
along the lines of Section~\ref{sec:effective-action}
leads to an action which is not particularly simple or illumminating.
We leave the problem of the effective action for dissipative relativistic hydrodynamics
including the questions of frame invariance and the derivative expansion for future work.

\vspace{0.5cm}
{\bf Acknowledgments}\\
\noindent
I would like to thank my collaborators who helped me learn  
many of the things described in these lectures. 
Thanks to L.~G.~Yaffe, D.~T.~Son, A.~O.~Starinets, C.~P.~Herzog, A.~Ritz,
G.~Moore, P.~Romatschke, K.~Jensen, and A.~Yarom.
I would like to thank the Perimeter Institute for Theoretical Physics
for hospitality and support during the completion of this paper.
This work was supported in part by NSERC of Canada.

\bibliographystyle{jhep-arxiv}
\bibliography{hydro-notes}

\end{document}